\documentclass[a4paper,11pt]{article}
\usepackage{graphicx} % Required for inserting images

\pdfoutput=1 % if your are submitting a pdflatex (i.e. if you have
\usepackage{jcappub} % for details on the use of the package, please
\usepackage[T1]{fontenc} % if needed

\usepackage{color}%to be able to use colors
\usepackage{amsmath} %Recomended for mathematical documents (use of: eqref,...)
\usepackage{amsfonts}
\usepackage{graphicx}
\usepackage{slashed}
\usepackage[dvipsnames]{xcolor}
\definecolor{refs}{RGB}{245,156,74}
\usepackage{subcaption}
\usepackage{caption}
\usepackage{float}

\usepackage{physics}

\newcommand{\be}{\begin{equation}}
\newcommand{\ee}{\end{equation}}
\newcommand{\bea}{\begin{eqnarray}}
\newcommand{\eea}{\end{eqnarray}}

\newcommand{\MPL }{M_p}
\newcommand{\mfa}{{\mathfrak a}}

\newcommand{\B}{b}
\newcommand{\C}{m}

\newcommand{\tot}{{\textnormal{tot}}}
\definecolor{tabblue}{HTML}{1f77b4}
\definecolor{taborange}{HTML}{ff7f0e}
\definecolor{tabgreen}{HTML}{2ca02c}
\definecolor{tabred}{HTML}{d62728}
\definecolor{tabpurple}{HTML}{9467bd}

\newcommand\cC{{\cal C}}

\newcommand\cG{{\cal G}}

\newcommand\cL{{\cal L}}
\newcommand\cO{{\cal O}}
\newcommand\cP{{\cal P}}
\newcommand\cR{{\cal R}}
\newcommand\cU{{\cal U}}
\newcommand\cV{{\cal V}}

\newcommand\ssB{{\scriptscriptstyle B}}
\newcommand\ssM{{\scriptscriptstyle M}}
\newcommand\ssN{{\scriptscriptstyle N}}
\newcommand\ssP{{\scriptscriptstyle P}}
\newcommand\ssQ{{\scriptscriptstyle Q}}
\newcommand\ssW{{\scriptscriptstyle W}}

\newcommand\bfv{{\bf v}}

\newcommand\bfU{{\bf U}}

\newcommand{\mfu}{\mathfrak{u}}

\newcommand\ad{{\rm ad}}
\newcommand\eff{{\rm eff}}
\newcommand\thr{{\rm th}}
\newcommand\ax{{\rm ax}}
\newcommand\ttl{{\rm tot}}

\newcommand\tphi{\tilde{\phi}}
\newcommand\tmfa{\tilde{\mfa}}

\newcommand\nn{\nonumber}
\newcommand\ol{\overline}
\newcommand\exd{{\rm d}}
\newcommand{\pref}[1]{{(\ref{#1})}}
\newcommand{\roughly}[1]{\mathrel{\raise.3ex\hbox{$#1$\kern-0.85em
\lower1ex\hbox{$\sim$}}}}
\newcommand{\lsim}{\roughly<}
\newcommand{\gsim}{\roughly>}
\newcommand\aplus{{0}}
\newcommand\aminus{{\mfa_m}}
\newcommand\aminuse{{\mfa_e}}
\newcommand\Jvec{{\bf J}}
\newcommand\vvec{{\bf v}}
\newcommand\pvec{{\bf p}}

\begin{document}

%\title{Axio-dilaton screened cosmology}
%\title{Axio-dilaton Cosmology\\ and the Big Screen}
\title{Screened Axio-dilaton Cosmology:\\ Novel Forms of Early Dark Energy}

\author[a]{Adam Smith,}
\affiliation[a]{School of Mathematical and Physical Sciences, University of Sheffield, %Hounsfield Road, 
Sheffield, %S3 7RH, United Kingdom
UK
}
\author[b]{Philippe Brax,}
\affiliation[b]{Institut de Physique Th\'eorique, Universit\'e Paris-Saclay,
%CEA, CNRS, F-91191 
Gif-sur-Yvette, % Cedex, 
 France.
}

\author[a]{Carsten van de Bruck,}

\author[c,d,e]{C.P.~Burgess}

\affiliation[c]{Department of Physics \& Astronomy, McMaster University, %1280 Main Street West, 
Hamilton ON, Canada.
}
\affiliation[d]{Perimeter Institute for Theoretical Physics, Waterloo, ON, Canada%, N2L 2Y5
}
\affiliation[e]{School of Theoretical Physics, Dublin Institute for Advanced Studies, %10 Burlington Rd., Dublin,  Co. 
Dublin, Ireland
}
\author[f]{and Anne-Christine Davis}
%\affiliation[f]{DAMTP, Cambridge University, Wilberforce Road,  Cambridge, CB3 0WA, UK.}
%\affiliation[g]{Kavli Institute of Cosmology (KICC), University of Cambridge, Madingley Road, Cambridge, CB3 0HA, UK. }
\affiliation[f]{DAMTP \& Kavli Institute of Cosmology (KICC), Cambridge University, %Madingley Road, 
Cambridge, %CB3 0HA, 
UK. 
}

\emailAdd{asmith69@sheffield.ac.uk}
\emailAdd{philippe.brax@ipht.fr}
\emailAdd{c.vandebruck@sheffield.ac.uk}
\emailAdd{cburgess@perimeterinstitute.ca}
\emailAdd{ad107@cam.ac.uk}

\date{\today}

% \begin{abstract}
\abstract{
   We study the cosmology of multi-field Dark Energy, using a well-motivated axio-dilaton model that contains the minimal number of fields to have the 2-derivative sigma-model interactions that power-counting arguments show naturally compete with General Relativity at low energies. Our analysis differs from earlier, related, studies by treating the case where the dilaton's couplings to matter are large enough to require screening to avoid unacceptable dilaton-mediated forces in the solar system. We use a recently proposed screening mechanism that exploits the interplay between stronger-than-gravitational axion-matter couplings with the 2-derivative axion-dilaton interactions to suppress the couplings of the dilaton to bulk matter. The required axion-matter couplings also modify cosmology, with the axion's background energy density turning out to resemble early dark energy. We compute the properties of the axion fluid describing the rapid oscillations of the axion field around the time-dependent minimum of its matter-dependent effective potential, extending the usual formalism to include nontrivial kinetic sigma-model interactions. We explore the implications of these models for the Cosmic Microwave Background and the growth of structure and find that for dilaton potentials of the Albrecht-Skordis form (itself well-motivated by UV physics), successful screening can be consistent with the early dark energy temporarily comprising as much as 10\% of the total density in the past. We find that increasing the dilaton-matter coupling {\it decreases} the growth of structure due to enhanced Hubble friction, an effect that dominates the usual fifth-force effects that amplify structure growth. 
% \end{abstract
}

\maketitle
\section{Introduction}
For the past twenty five years both observations and Ockham's razor have suggested regarding the late-time acceleration of the Universe's expansion as being due to a nonzero cosmological constant -- see \cite{Carroll:1991mt, Ostriker:1995su, Carroll:2000fy, Peebles:2002gy} for overviews and \cite{Peebles:2022bya} for a historical review with references. A longstanding problem with this interpretation is the small size of the result, which is not technically natural \cite{Weinberg:1988cp, Burgess:2013ara} given the much larger scales associated with most of the masses of the known particles (with the tantalizing exception of neutrinos).  

This could now be changing with, for example, the new results by the DESI experiment \cite{DESI:2024mwx, DESI:2025zgx} seeming to point towards a Universe with dynamical dark energy. Of course the jury remains out and new observations over the next few years might yet send these preliminary hints for physics beyond the standard models to join others in the realm of Hades. If they survive, we  learn something important: there are very  likely new very light (probably scalar) fields whose dynamics  generate the late time acceleration. 

  The scalars must be essentially massless on solar-system scales in order to evolve on cosmological time scales.  If so then two issues deserve great scrutiny.
\begin{itemize}
\item{\bf Why so light?}  Why are the scalars  nearly massless in the first place? It is not generic to have scalars be much lighter than the intrinsic scales of microscopic physics,\footnote{To use a condensed-matter analogy: light scalar degrees of freedom are generic at a system's critical points, which occur at pressures and temperatures that are characteristic of the underlying microscopic interactions of the constituents. What is surprising -- and can happen, but needs explanation when it does -- is the appearance of a quantum critical point at extremely low temperature and pressure relative to these microscopic underlying scales.}  unless there is a symmetry mechanism that makes it so \cite{Gildener:1976ih}. This is actually a special case of a more general issue: non-derivative interactions generically amplify quantum effects when working within a semiclassical approximation with gravity, often invalidating purely classical reasoning \cite{Burgess:2009ea}. In essence for gravity the semiclassical approximation is at heart a low-energy expansion \cite{Weinberg:1978kz,  Donoghue:1994dn, Burgess:2003jk}, where weak/strong quantum effects are assessed relative to interactions involving exactly two derivatives.\footnote{ As all of the interactions in the Einstein-Hilbert action involve two derivatives, the nonlinearities of General Relativity cannot in general be neglected at low energies.} 

So scalar fields typically only compete at all with gravity at low energies if their scalar potential is unusually very small. One well-motivated way to obtain light scalars (and suppress the interactions in their scalar potential) is as Goldstone (or pseudo-Goldstone) bosons for the spontaneous breaking of an exact (or approximate) global symmetry \cite{Weinberg:1972fn}. 

Once this is achieved the two-derivative interactions of scalars become crucial for any low-energy applications -- such as in cosmology or when testing General Relativity (GR). This observation is easily missed in the special (and most-explored) case of single-scalar models because for these models minimal coupling to gravity exhausts the possible types of two-derivative interactions.

\item{\bf Why not elsewhere?}
 The second important observation is  that any scalar light enough to be cosmologically interesting mediates a long-range force that potentially competes with gravity. The absence of observed deviations from Newton's law in gravitational tests in the solar system and in laboratory experiments therefore  tells us that couplings to any new long range forces must be suppressed to below the strength of gravity \cite{Will:2014kxa, Bertotti:2003rm,Berge:2017ovy}. 

There are two well-explored ways to suppress the coupling of a very light scalar in solar system tests. The simplest arranges for the scalar to couple weakly to ordinary matter at an atom-by-atom level (that is, to have a small coupling constant in the underlying fundamental lagrangian). The second way posits nonlinearities in the interactions that make the effective coupling of the field to a macroscopic source containing $N$ elementary particles much smaller than simply $N$ times the field's coupling to each of those elementary particles separately.\footnote{A variety of screening mechanisms of this sort have been devised for single-field models, built on properties of the scalar potential and zero-derivative matter couplings (as in the chameleon or symmetron mechanisms \cite{Pourhasan:2011sm, Khoury:2003rn, Hinterbichler:2010es, Brax:2010gi}) or large higher-order terms in field derivatives (as in K-mouflage \cite{Babichev:2009ee} or the Vainshtein \cite{Vainshtein:1972sx} mechanism). For reviews see \cite{Brax:2013ida, Koyama:2015vza, Brax:2021wcv}.}  This resembles how electromagnetic forces between electrically neutral macroscopic objects built from charged constituents can be much weaker than the forces between the constituents themselves. This second option is widely known as `screening.'  

Which of these two mechanisms arises in specific examples is model-dependent though screening mechanisms are crucial when it is difficult to have a fundamental force couple to elementary particles weaker than does gravity.  

\end{itemize}

The observation that two-derivative interactions dominate at low energies (once zero-derivative interactions have been suppressed enough to allow light scalars in the first place) makes models involving two or more scalars particularly interesting because they allow non-minimal two-derivative interactions like
\be 
{\cal L}_{\rm 2d}=- \frac{1}{2} \, \sqrt{-g} \; \cG_{ij}(\phi) \, \partial_\mu \phi^i \partial^\mu \phi^j
\ee
where $g_{\mu\nu}(x)$ is the spacetime metric and $\cG_{ij}(\phi)$ is a dimensionless positive-definite symmetric tensor that defines a metric $\exd s^2= \cG_{ij}(\phi)  \,\exd\phi^i \exd\phi^j$ on the target space.\footnote{Nonlinear sigma-model interactions like these were studied since the 60s as effective theories for low-energy pion scattering \cite{Gell-Mann:1960mvl, Weinberg:1966fm} and for Goldstone-boson interactions more generally \cite{Coleman:1969sm, Callan:1969sn}, and have been more widely considered in cosmology for inflationary applications (for reviews see \cite{Wands:2007bd, Cicoli:2023opf}).} These interactions can compete with GR at low energies without undermining the entire semiclassical approximation (unlike the higher-derivative models often studied for single-field models, to which one is driven because $\cG$ can always be eliminated by transforming to canonically normalized variables). 

Axio-dilaton models contain two scalar fields, $\phi$ and $\mfa$, that are pseudo-Goldstone bosons respectively for rigid scale transformations and for an internal shift symmetry, for which the leading target-space metric 
\be
   \exd s^2 = \exd \phi^2 + W^2(\phi) \, \exd \mfa^2 
\ee
is characterized by a single function. The phenomenology of these models have recently been explored in late-time cosmology and within the solar system \cite{Burgess:2021qti, Brax:2022vlf, Smith:2024ibv}, with four main motivations. First, they involve two scalar fields and so contain the bare minimum number needed to explore the implications of two-derivative interactions that compete optimally at low energies with the two-derivative interactions of GR. Second, they arise frequently within well-motivated UV completions of gravity (such as string vacua compactified to 4 dimensions) because of the accidental scaling and shift symmetries that are ubiquitous in these theories \cite{Burgess:2020qsc}. Thirdly, they allow new relaxation mechanisms for suppressing the size of the scalar potential \cite{Burgess:2021obw} and so make some headway on the cosmological constant problem. Lastly, axio-dilaton models also allow new screening mechanisms \cite{Brax:2023qyp} that rely on the properties of the function $W(\phi)$ (rather than higher-derivative terms, say) and so compete with gravity more easily at low energies without threatening the underlying classical approximation. 

Remarkably, the minimal two fields $\phi$ and $\mfa$ can in themselves provide a viable minimal picture for both Dark Matter and dynamical Dark Energy, with the `dilaton' $\phi$ playing the role of a quintessence-style Dark Energy and the `axion' $\mfa$ being the Dark Matter. In the simplest way of doing so \cite{Smith:2024ayu} solar system tests of gravity are evaded simply by assigning both sufficiently small couplings to ordinary matter. However, many of the most novel and interesting features of this class of models -- such as the versions explicitly produced in UV completions and potential progress on the cosmological constant problem -- predict gravitational-strength (or stronger) couplings of the scalars to ordinary matter, so in this paper we focus on the viability of late-time cosmology when couplings to matter are \emph{not} small and so when screening mechanisms are important. Exploring the implications of screening for cosmology is the main difference between the analysis we present here and the studies done in \cite{Smith:2024ibv}. 

For concreteness' sake we adopt the screening mechanism of \cite{Brax:2023qyp}, which explores two forms for the function $W^2(\phi)$: exponential and quadratic. The exponential case captures a generic runaway to large fields and embeds naturally into string-inspired models \cite{Cicoli:2023opf} (for recent examples see e.g.~\cite{Apers:2024ffe, Bernardo:2022ztc, Smith:2023oop, Poulin:2023lkg}). The quadratic case captures the generic situation near a local minimum of $W$. The basic screening occurs because the axion-matter couplings are chosen to ensure that the axion takes a value inside ordinary matter that differs from its value in the vacuum, with the resulting axion gradients driving changes in the dilaton -- due to the interaction implied by $W(\phi)$ -- that act to suppress the dilaton gradient exterior to the source (and so reduce the source's effective dilaton `charge'). 

We find that these same scalar-matter couplings can change cosmology in interesting ways, with the energetics of the matter-dependent part of the axion potential effectively introducing a Dark-Energy type equation of state in early epochs where the matter density is comparatively large. This makes it resemble early dark energy (EDE), though with important differences from the usual formulations \cite{Kaloper:2019lpl, Poulin:2018cxd, Knox:2019rjx}. At late times the axion energy density decays like a subdominant matter component and the axion dynamics is not simply driven by the behaviour of its effective potential. Indeed, the axion field performs decaying oscillations towards its matter-dependent minimum in a manner reminiscent of fuzzy dark matter in axion-like-particle models \cite{Hui:2016ltb}. But for axio-dilatons the situation is technically more intricate as both the vacuum expectation value of the axion field and its mass are time dependent. We provide a full description of the resulting axion fluid dynamics coupled to the baryons, CDM and the dilaton. The fluid description is justified by the very fast axion oscillation evolution compared to the Hubble rate. 
  
When performing this analysis we make two simple choices for the potential energy for $\phi$ and $\mfa$ in addition to the two choices (exponential or quadratic) for the form of $W$. We follow \cite{Smith:2024ibv, Smith:2024ayu} and choose the dilaton potential to be either exponential \cite{Gasperini:2001pc} or of the Albrecht-Skordis/Yoga type \cite{Albrecht:1999rm, Albrecht:2001xt, Burgess:2021obw, Burgess:2022nbx} (i.e.~the product of an exponential and a quadratic function, whose minimum serves as an attractor that traps the dilaton so that the dark energy approaches a cosmological constant at late times). The axion dynamics is driven by its scalar potential -- taken to be quadratic near its minimum -- including its coupling to matter. 

We find four very different possibilities whose physics is itself constrained by screening. The simplest possibility combines an exponential dark energy potential with a quadratic form for $W$ and nicely displays the possible interplay between screening and cosmology. Indeed, if screening is not required one can find parameter values such that the early dark energy is fairly large, temporarily making up of the order of ten percent of the Universe abundance, with a transition to decaying matter around the matter-radiation equality that ensure only small deviations of the Cosmic Microwave Background (CMB) from the predictions of $\Lambda$CDM. The growth of structure (e.g.~on\footnote{Here $f$ is the growth factor and $\sigma_8$ is the variance of the mass fluctuations within a sphere of radius $R= 8h^{-1}$Mpc.} $f\sigma_8$) increases when the coupling of the dilaton to matter increases (as is the standard expectation for scalar-tensor theories). 

But restricting the parameters to ensure sufficient screening in the late universe changes this picture significantly. First the fraction of early dark energy is bounded to be well below one percent and the kinetic energy in the axion field prevents early dark energy from playing a significant role during the epoch of matter-radiation equality. This limit on the amount of early dark energy arises because larger dark-energy densities lead to a tachyonic instability (with associated large deviations from $\Lambda$CDM). 

Larger contributions of early dark energy arise when combining the Yoga-style (Albrecht-Skordis) potential for the dilaton with the quadratic form for $W$. In this case the dilaton potential has a local minimum that attracts the dilaton at late times, leading to a cosmological-constant style equation of state. This local minimum helps tame the tachyonic instability in the screened case by preventing large field excursions around matter-radiation equality. This allows the early dark energy fraction to be as large as one percent. As the axion coupling and the coupling to matter of the dilaton pushes it away from the minimum in the absence of matter, the Hubble rate is slightly increased at redshifts when structures form. This implies that the growth of structure is hampered and we observe a \emph{decrease} of $f\sigma_8$ for increasing dilaton coupling to matter. This is the opposite of what is usually expected in scalar-tensor theories and could have appealing phenomenological consequences. 

We also consider the case of an exponential $W$ whilst keeping the Yoga potential. In this case, there is no tachyonic instability at all and so the early dark energy fraction can be increased to several percent whilst the growth of structure is still depleted for larger dilaton couplings to matter.

The rest of this paper is arranged as follows. In section \ref{first} we recap how screening works for the axio-dilaton models using the mechanism described in \cite{Brax:2023qyp}. Then we discuss the cosmological background dynamics in section \ref{two}. The axion fluid is described in section \ref{three} and the treatment of cosmological perturbations in section \ref{four}. Numerical evolution of the resulting equations are presented in section \ref{five}. Several appendices contain more technical details. 

\section{Model definition and criteria for screening}
\label{first}

This section defines the two-field axio-dilaton model to be studied and briefly recaps the discussion of \cite{Brax:2023qyp} in order to determine the parameter regime required to screen scalar-matter interactions sufficiently to evade late-time solar-system tests of gravity.  

\subsection{Action and field equations}

As stated in the introduction, we focus on two-field models whose kinetic terms are determined by the target-space metric\footnote{We denote dimensionless fields with tilde's, as in $\tphi$, reserving $\phi = \MPL \tphi$ for the dimensionful version. Notice that this convention differs from the one used in \cite{Smith:2024ayu, Smith:2024ibv}.}
\be 
\exd s^2= \exd\tphi^2 +  W^2(\tilde\phi) \, \exd\tmfa^2 ,
\ee
 and call $\tphi$ the dilaton and $\tmfa$ the axion. This is the most general two-dimensional metric consistent with a shift symmetry $\tmfa\to \tmfa+\hbox{constant}$.
As also mentioned in the introduction, motivation for this class of theories is discussed in \cite{Smith:2024ibv, Smith:2024ayu} and is partially inspired by UV extensions of the standard models of particle physics and cosmology. 

The low-energy action for the scalar fields and ordinary matter is given by
\be
   S  =   \int \exd^4 x \sqrt{-g} \left\{ \frac{\MPL^2}{2} \Bigl[ R -   (\partial \tphi)^2 -  W^2(\tphi) \, (\partial \tmfa)^2 \Bigr] - V(\tphi,\tmfa) \right\} + S_m[\tilde g_{\mu\nu}, \tmfa, \psi]   \,,   
\ee
where $R$ is the metric's Ricci scalar and the Planck mass is related to Newton's constant by $\MPL ^2= (8\pi G_\ssN)^{-1}$. $S_m$ describes the action of the matter sector, whose fields are collectively denoted $\psi$ and the dilaton only appears in this action through the Jordan-frame metric $\tilde g_{\mu\nu} := A^2(\phi) \, g_{\mu\nu}$. The model is specified once the axion-dependence of the matter action is given, together with explicit forms for the functions $V(\phi, \tmfa)$, $A(\phi)$ and $W(\phi)$. 

It is convenient when interpreting this lagrangian to shift the axion and dilaton fields so that their ambient homogeneous values during the present epoch are $\mfa = \phi = 0$. For instance, if cosmological evolution leads at late times to $\phi$ and $\mfa$ sitting at the minimum of $V(\phi,\mfa)$ then after shifting the field appropriately there is no loss in choosing the vacuum potential, $V(\phi,\mfa)$, to be minimized at $\mfa = \phi = 0$. It is also convenient to rescale the metric so that $A(0) = 1$ since this ensures Jordan frame and Einstein frame share the same units of length in the current epoch. Finally, we rescale $\phi$ and $\mfa$ so that both scalars are canonically normalized during the current epoch, which means $\phi = \MPL \,\tphi$ and $\mfa = W(0) \MPL \, \tilde\mfa$. After these choices the lagrangian is
\be
   S = \int \exd^4 x \sqrt{-g} \left\{ \frac{\MPL^2}{2} R - \tfrac{1}{2} \Bigl[  (\partial \phi)^2 + W^2(\phi) \, (\partial \mfa)^2 \Bigr] - V(\phi,\mfa) \right\} + S_m[A^2(\phi) \, g_{\mu\nu}, \mfa, \psi] \,,  \nn
\ee
and there is no loss of generality in assuming that the minimum of $V$ (if this exists) is at $\mfa = \phi = 0$ and also that $A(0) = W(0) = 1$ there.

For the dilaton-matter coupling function $A$ we take
\be \label{dilacoup}
A  = e^{-\beta \phi/\MPL },
\ee
where $\beta>0$ and we use the above-mentioned rescalings to ensure $A(0)=1$. This is motivated by the role played by $\phi$ as a pseudo-Goldstone boson for scaling symmetry (which is why this form arises so frequently from microscopic physics). This form of coupling to matter is what would arise for a Brans-Dicke scalar written in Einstein frame. When present the scalar mediates a Yukawa-type force whose coupling strength is proportional to $\beta/\MPL$. When the $\phi$ Compton wavelength is solar-system sized or larger (as is the case if it is currently cosmologically active) the agreement of GR with observations requires $\beta \lsim 10^{-3}$ \cite{Bertotti:2003rm}. It is because many microscopic theories -- and in particular the natural relaxation (Yoga) models of \cite{Burgess:2021obw} -- predict $\beta \sim \cO(1)$ \cite{Cicoli:2023opf, Brax:2023nvx} that screening mechanisms are often required. 

There are two different choices for the kinetic coupling function $W$ that are known to allow the screening described in \cite{Brax:2023qyp} and we explore them both here. The first possibility assumes $W$ has the scaling form often predicted by UV physics,
\be  \label{Wexpform}
  W^2(\phi) = e^{-\xi \phi/\MPL }  = e^{-\phi/\Lambda_\phi }  \,,
\ee
where we again rescale fields to ensure $W(0)=1$ and $\Lambda_\phi$ -- or equivalently $\xi = \MPL/\Lambda_\phi$ -- is a new coupling parameter. The other possibility assumes $W$ to be near a minimum
\be \label{Wquadform}
    W^2(\phi) \simeq W_\star^2 + \frac{(\phi-\phi_\star)^2}{2\Lambda_\phi^2} = 1 + \frac{\phi(\phi - 2\phi_\star)}{2\Lambda_\phi^2} \,,
\ee
and so is characterized by choices for the parameters $\phi_\star$ and $\Lambda_\phi$. As usual, using a strictly quadratic expression typically assumes $|\phi - \phi_\star|$ is small relative to $\Lambda_\phi$.  

It remains to specify the scalar potential in the vacuum, which we take to be additive 
\be \label{potentialdilax}
   V(\phi, \tmfa) = V_{\rm dil}(\phi) + V_{\rm ax}(\mfa)
\ee
with the dilaton contribution having the form
\be \label{VdilDef}
  V_{\rm dil} = U(\phi) \, e^{-\lambda \phi/\MPL },
\ee
where two choices are made for the prefactor $U(\phi)$. Taking $U(\phi) = V_0$ to be a constant\footnote{When $U(\phi) = V_0$ the dilaton potential has no minimum and so we instead use the freedom to shift the origin of $\phi$ to set its initial condition to a value that arranges that $\phi = 0$ at or near the current epoch.} gives a runaway dilaton model \cite{Gasperini:2001pc} for which slow roll can produce the accelerated expansion required of Dark Energy\footnote{We remark in passing that the de Sitter conjecture states that $\lambda$ should be larger than $\sqrt 2$ in order not to fall within the swampland \cite{Andriot:2024jsh}. If regarded as an important prior consideration this would preclude the existence of dynamical dark energy through classical evolution using exponential potentials. In \cite{Burgess:2021obw} $\lambda$ is taken from comparing to microscopic arguments and turns out larger than $\sqrt 2$, though the presence of a minimum due to the function $U$ also means that acceleration does not rely on slow-roll evolution in an exponential potential. In what follows we do not worry about hypothetical swampland issues.}  if $\lambda < \sqrt2$ \cite{Ferreira:1997hj}. Alternatively we also explore an Albrecht-Skordis potential \cite{Albrecht:1999rm, Albrecht:2001xt} for which the function $U(\phi)$ has a local minimum whose value sets the size of the late-time dark energy scale. In practice $U(\phi)$ is taken for simplicity to be quadratic in $\phi$ near the minimum, and once $\phi$ is shifted to ensure the minimum is at $\phi = 0$ becomes
\be\label{Uquaddef}
  U(\phi) = \tfrac12 m_\phi^2 \phi^2 + V_0 \left[ \tfrac12 \left( \frac{\lambda\phi}{\MPL}\right)^2 + \frac{\lambda\phi}{\MPL} + 1 \right]  \,,
\ee
where $V_0 = V_{\rm dil}(0) \sim \cO(H^2 \MPL^2) > 0$ and $m_\phi^2 := V''_{\rm dil}(0) \sim \cO(H^2) > 0$. The motivations for making this choice are summarized in \cite{Burgess:2021obw, Burgess:2022nbx}, and are driven by the desire  to  understand naturally the small size taken for $V_0$ (which sets the Dark Energy density). Other potentials leading to a quintessence behaviour for $\phi$ could also be considered though we do not do so. 

The axion potential in vacuum is assumed to be large enough that during the recent universe the field $\mfa$ does not stray too far from one of its minima, and we shift the axion field to arrange that this occurs at $\mfa=\aplus$. We assume the potential energy vanishes\footnote{We do so here purely by tuning, though ultimately one hopes this could also be arranged using the relaxation mechanism of \cite{Burgess:2021obw}, although at face value this would not produce a potential $V_{\rm dil}(\phi) + V_{\rm ax}(\mfa)$ of the simple additive form chosen here. Further exploration of this point is a work in progress.} at this minimum, $V(\aplus)=0$, and so\footnote{For axions arising from internal rephasing symmetries the potential is trigonometric in $\tilde \mfa$ and so choosing a quadratic form assumes $\tilde \mfa \ll 1$. The reasoning is slightly different for axions arising as the dual of a two-form Kalb-Ramond potential $B_{\mu\nu}$, however, since in this case a quadratic potential follows as the leading term in the low-energy derivative expansion \cite{Burgess:2023ifd}.}
\be \label{VaxDef}
%V_{\rm ax} \simeq \frac{m^2_\mfa}{2}(\mfa-\aplus)^2 \,.
V_{\rm ax} \simeq \tfrac12\,  m^2_\mfa\, \mfa^2 \,.
\ee
The physical mass of the axion in the vacuum is then given by $m_{\rm ax} = m_\mfa/W$ where $W$ is evaluated at the background value for $\phi$ (and so $m_{\rm ax} \to m_\mfa$ at late times when $\phi \to 0$ and $W \to 1$). It is well-known that a scalar field oscillating near the minimum of its potential behaves like non-relativistic matter, but in our analysis we do not assume the axion to be Dark Matter (and so assume the Dark Matter is part of the matter lagrangian). 

For axion-matter couplings we assume a coupling to a matter field $\psi_i$ to be proportional to the number density of matter particles, which in Jordan frame has the form $\cL_{m\mfa} \propto - \sqrt{\tilde g} \sum_i \cU_i (\mfa) \psi_i^* \psi_i$, where $\tilde g_{\mu\nu}$ is the Jordan-frame metric and the sum runs over the various species of matter fields, $\psi_i$. This amounts to giving the matter field an axion-dependent Jordan-frame mass of the form
\be \label{mivsmi0}
   m_i = m_{i0} \Bigl[ 1 + \cU_i(\mfa) \Bigr] \,,
\ee
where the function $\cU_i$ are to be specified. For definiteness we assume the axion only couples to electrons and to the particle species that makes up the cold dark matter (these choices help in evading strong bounds on the variation of masses for nuclei inside stars \cite{Brax:2023qyp}).

We furthermore assume the field $\mfa$ lies close to a local minimum of $\cU_e(\mfa)$ and $\cU_m(\mfa)$ and $V_{\rm ax}(\mfa)$ so they can all be approximated as being quadratic in $\mfa$. Here we use the subscript $m$ to refer to Dark Matter and $e$ to refer to electrons. This type of quadratic form is useful (as opposed to linear, say) because it allows many constraints on axion-matter couplings to be evaded. Higher powers of $\mfa - \mfa_i$ can be neglected by assuming $\mfa-\mfa_i$ is always small, and if so consistency requires we also drop higher powers of $\cU_i$ when these arise within the field equations (more about which below). We do not assume the minima of the coupling functions coincide with one another or with the minimum of $V_{\rm ax}$. This leads to the representation
\be \label{UmUeQuad}
  \cU_m(\mfa) \simeq \frac{(\mfa-\aminus )^2}{2\Lambda_{m}^2} \qq{and} \cU_e(\mfa)=  \frac{(\mfa-\aminuse )^2}{2\Lambda_e^2} + \dots\,,
\ee
where $\Lambda_i$ is a characteristic scale of the microscopic theory.

In cosmology and the solar system the matter sector is described by an effective fluid, with components whose (Einstein-frame) energy density, $\rho_i$, and pressure, $p_i$, are subject to an equation of state. In a regime where the matter to which the scalars are coupled is nonrelativistic the field equations for $\phi$ and $\mfa$ are obtained by substituting $\rho_i \to \rho_i(\phi,\mfa) = m(\phi, \mfa) n_i$ for each species of particle, where $n_i$ is the particle density and the field-dependent Einstein-frame mass is given by $m_i(\phi,\mfa) = A(\phi) [1 + \cU_i(\mfa)] m_{i0}$ (see \cite{Smith:2024ayu} for a detailed derivation). 

The result for the scalar field equations is 
\be \label{dileq}
\Box \phi= (WW') (\partial \mfa)^2 + \partial_\phi V_{\rm eff},
\ee
(where the prime denotes differentiation with respect to $\phi$) and 
\be \label{axieq}
\frac{1}{\sqrt{-g}}\partial_\mu \Bigl[ \sqrt{-g}W^2(\phi) \partial^\mu \mfa \Bigr] = \partial_\mfa V_{\rm eff} \,.
\ee
In these expressions the matter-dependent effective potential seen by the scalars is 
\be \label{Veffdef}
V_{\rm eff}(\phi, \mfa)  =  V(\phi, \mfa) + \sum_i m_i(\phi, \mfa) \, n_i 
%=  V(\phi, \mfa) + \sum_i \rho_i (\phi,\mfa)  
=  V(\phi, \mfa) +  \rho_{\rm nr} (\phi,\mfa)  \,,
\ee
where $\rho_{\rm nr} = \sum_i \rho_i$ is the total energy density of the nonrelativistic fluid to which the scalars couple. It follows that
\bea \label{Veffderiv}
   \partial_\phi V_{\rm eff} &=& V'_{\rm dil}(\phi) +  \left( \frac{A'}{A} \right) \sum_i \rho_i(\phi, \mfa) = V'_{\rm dil}(\phi) -  \frac{\beta \rho_{\rm nr}}{\MPL}   \\
   \qq{and} 
   \partial_\mfa V_{\rm eff} &=& V'_{\rm ax}(\mfa) + \sum_i \left( \frac{\cU_i'}{1 + \cU_i} \right) \, \rho_i(\phi,\mfa)  \simeq  V'_{\rm ax}(\mfa) + \sum_i  \cU_i' \, \rho_i(\phi,\mfa) \,, \nn
\eea
where the final, approximate, equality drops subdominant powers of $\cU_i$, for the reasons described above eq.~\pref{UmUeQuad}. With the above choices the model is fully specified.

For future purposes two things are noteworthy about eqs.~\pref{dileq} through \pref{Veffdef}. First, notice that in the presence of matter the axion field is displaced from its vacuum value $\mfa = \aplus$, moving towards the value $\aminuse$ or $\aminus$ depending on whether the electron or Dark Matter density dominates. Second, notice that in the presence of an axion gradient the $W$-dependent terms of \pref{dileq} themselves contribute as if the dilaton sees a gradient-dependent additional potential energy 
\be \label{VWdef}
   \delta V_{\ssW} = \tfrac12 W^2(\phi) (\partial \mfa)^2 .
\ee 

These are key features for the screening mechanism described below since (if large enough) the matter-dependence of the axion potential generates an axion gradient inside sufficiently large and dense macroscopic objects. But then the axion gradient makes $\delta V_\ssW$ nonzero, causing the dilaton to differ outside the source relative to the naive sum over the dilaton couplings of the source's constituents. But once the parameters for these interactions are adjusted to ensure screening occurs the same terms also change the evolution of $\mfa$ and $\phi$ in the early universe, due to the presence there of large densities of matter. 
 
\subsection{Criteria for screening and constraints}
\label{ssec:Criteria&Constraints}

We next summarize the screening mechanism of \cite{Brax:2023qyp} (those already familiar with this story can safely skip ahead to \S\ref{two}). To this end we assume the axion field is heavy enough not to mediate a long-range force, with a range much smaller than an AU ensured if $m_{\rm ax} \gsim 10^{-16}$ eV. The dilation, on the other hand, is instead assumed light enough to evolve on cosmological times and so can mediate dangerous long-range forces. The idea is to choose axion-matter couplings so that macroscopic sources generate axion gradients and then to exploit the dilaton response to these gradients to reduce (\emph{i.e.}~`screen') the source's effective `dilaton charge'. Screening occurs once two ingredients are in place: appropriate axion-matter couplings (to generate an axion gradient within or near the source) and derivative axion-dilaton couplings (so the axion gradient can modify the dilaton profile). 

\subsubsection{The Brans-Dicke problem}

In regions where the vacuum dilaton potential and gradients of the axion field are both negligible the dilaton equation \pref{dileq} becomes
\be 
\Box \phi \simeq \partial_\phi V_{\rm eff}(\phi) \simeq - \frac{\beta \rho}{\MPL} \,,
\ee
where $\beta$ is defined in \eqref{dilacoup} and the last equality uses \pref{Veffderiv}. If this equation holds both inside and outside a nonrelativistic matter source (\emph{e.g.}~a star) then the field is unscreened and the static spherically symmetric solution for the field outside the source becomes
\be \label{sol1}
\phi(r) = \phi_\infty -\frac{\beta M}{4\pi \MPL  r} \,.
\ee
Such a solution describing the field outside the Sun can be ruled out within the solar system if $\beta \gsim {\cal O}(10^{-3})$ \cite{Bertotti:2003rm}. The goal is to exploit the axion-dilaton and axion-matter couplings to reduce the size of the coefficient of $1/r$ in \pref{sol1} to acceptable levels without reducing $\beta$.

A remark about the quantity $\phi_\infty$ proves useful later. In the naive telling of the story $\phi_\infty$ is an integration constant obtained by demanding continuity with any spatially homogeneous asymptotic value taken at infinity. More generally, $\phi_\infty(t)$ can be be a function of time chosen to satisfy the asymptotic field equation far from the source, such as a cosmological solution. But in truth the approximation of spherical symmetry usually breaks down at a finite distance from the source because the residual fields from other nearby sources start to contribute appreciably. In this case $\phi_\infty$ would be chosen by matching to the value of the field at a radius close enough to be dominated by the spherically symmetric field of the nearest source, rather than further away. 

The precise value of $\phi_\infty$ often does not really matter, however, since any accidental shift `symmetry' of the equations of the form $\phi \to \phi + c$ (for constant $c$) can make $\phi_\infty$ drop out of observable quantities. In the screening mechanism described here, however, the axion-dilaton interaction breaks the dilaton shift symmetry and so ensures the energy of the field depends on $\phi_\infty$ and because of this its value in the ground state is instead determined by minimizing the energy. Agreement with ambient fields far from the source is then obtained at the cost of gradient energy, which can be much less expensive if the sources are more widely separated from one another compared to their size (see \cite{Brax:2023qyp} for more details).

\subsubsection{The screening mechanism}

As described above, for axion-matter couplings we assume  that it assigns an axion-dependence to particle masses in the Jordan frame, with $m_i = m_{i0} [1 + \cU_i(\mfa)]$ and the functions $\cU_i$ and the axion potential approximated by
\be \label{AppUVforms}
\cU_i(\mfa) \simeq \frac{(\mfa-\mfa_i )^2}{2\Lambda_i^2} 
%\qquad \hbox{and} \qquad V_{\rm ax}(\mfa) \simeq   \frac{m_\mfa^2}{2}(\mfa-\aplus)^2 
\,,
\ee
where $i = e,m$ respectively denotes electrons and Dark Matter. The axion potential in vacuum, $V_{\rm ax}$, is as given in \pref{VaxDef}.

Inside a macroscopic object (like the Sun) whose electron density dominates the ambient Dark Matter density this type of interaction implies the axion `sees' an effective matter-dependent scalar potential of the form
\be \label{U_ewrho_b}
 V_{\rm ax\, eff}(\mfa) \simeq V_{\rm ax}(\mfa) + m_{e0} \,\cU_e(\mfa) \, n_e \simeq V_{\rm ax}(\mfa) + \frac{m_{e0}}{m_\ssN}  \; \cU(\mfa) \, \rho_\ssB \,,
\ee
where electrical neutrality implies the electron and baryon number densities are equal and so we write $n_e = n_\ssB = \rho_\ssB/m_\ssN$, where $n_\ssB$ and $\rho_\ssB$ are the local baryon-number and baryon-mass densities and $m_\ssN$ is the nucleon mass. We choose parameters such that the matter-dependent part of this potential dominates the vacuum contribution for the kinds of matter densities found in the Sun. Under these circumstances we expect to find $\mfa \simeq \aminuse$ deep inside matter, with fluctuations about this value having a density-dependent mass $m_{\rm ax}(\rho_\ssB,\phi) = m_\mfa(\rho_\ssB) / W(\phi)$ where
\be  \label{mavsrho}
m_\mfa^2(\rho_\ssB) \simeq \frac{m_{e0} \, n_e}{\Lambda_e^2} \simeq \frac{\rho_e}{\Lambda_e^2} \simeq \left( \frac{m_{e0}}{m_\ssN} \right) \frac{\rho_\ssB}{\Lambda_e^2} \,,
\ee
and $\rho_e$ is the energy density in electrons.

If the axion Compton wavelength inside matter is much smaller than the distance over which $n_e$ changes appreciably then the background axion profile,  $\mfa_\ad(x)$, can be approximated by the adiabatic result that satisfies $\left. \partial_\mfa V_{\rm ax\, eff} \right|_{\ol \mfa} = 0$. If it is also true that the axion mass in vacuum is much smaller than its matter-dependent one then it can be shown that the solution for $\mfa_\ad(r)$ robustly tends to jump from $\aplus$ to $\aminuse$ within a narrow width $\ell \sim m_{\rm ax}(n_e)^{-1}$ near the source's surface (see \cite{Brax:2023qyp} for details). In these circumstances the background axion profile is well-approximated by a step function:
\be  \label{axstep}
\mfa = \ol \mfa(r) \simeq \aminuse \, \Theta(R-r) 
%+ \aplus \, \Theta(r-R)
\,,
\ee
where $R$ is the source radius and $\Theta$ is the Heaviside step-function. Although these conditions are not strictly required for screening, they do make its analysis particularly simple.

With these choices, the step in the axion profile generates an axion gradient that is localized near $r = R$, and this in turn -- by virtue of \pref{VWdef} -- generates a delta-function contribution to the potential for the dilaton, localized near $r = R$. As usual, the presence of a delta-function potential induces a jump discontinuity in the derivative of the dilaton at $r = R$, and this is how the axion gradient modifies the dilaton profile. In particular, although the value of the radial derivative $\partial_r \phi$ just inside the discontinuity is the usual one expected in the absence of screening, the jump discontinuity implies the derivative just outside is different, and if it is smaller the effective dilaton `charge' of the source has been reduced. 

To quantify this, if the dilaton profile for $r > R$ is written as
\be 
\phi = \phi_{\rm loc} - \frac{L\MPL }{r},
\ee
where $\phi_{\rm loc}$ and $L$ are integration constants, then (as discussed above) matching at the surface of the source would give $L= L_0 := 2{\beta G_\ssN M}$ in the absence of screening. But keeping track of the jump discontinuity in $\partial_r \phi$ when matching implies $L$ for the exterior solution is instead given by
\be \label{phi'invsout2}
%  \frac{L}{R} = \frac{L_0}{R} +  \Bigl( WW' \Bigr)_{r=R} \left( \frac{R}{2\ell} \right)  \frac{(\aplus - \aminuse )^2}{\MPL } \,,
  \frac{L}{R} = \frac{L_0}{R} +  \Bigl( WW' \Bigr)_{r=R} \left( \frac{R}{2\ell} \right)  \frac{\aminuse^2}{\MPL } \,,
\ee
where quantities like $WW'(\phi)$ are evaluated at $\phi(R)$. The effective dilaton charge of the source, including screening, can be defined by writing $L = 2 \beta_{\rm eff} G_\ssN M$, and if so it is $\beta_{\rm eff}$ and not $\beta$ that must satisfy solar system constraints. The dilaton charge is screened when $L$ is smaller than $L_0$, as is the case when $W'$ is negative.  

\subsubsection{Energetics}

A complication to the above screening story is that $\beta_{\rm eff}$ depends on the value taken by $\phi_{\rm loc}$, because this appears in the value $\phi(R)$ of the field where the axion gradient occurs. As argued earlier, the way to determine the value for $\phi_{\rm loc}$ is by minimizing the energy of the fields generated by the source. 

There are two important $\phi_{\rm loc}$-dependent contributions to the energy of the fields surrounding the source. The first of these comes from the gradient energy of the dilaton field exterior to the source, 
\be \label{EextEq}
   E_{\rm ext} = 4\pi  \int_R^\infty \exd r \, r^2 \; \frac{ \phi'^2}{2}= \frac{2\pi \MPL^2  L^2}{R} \,,
\ee
in which $\phi_{\rm loc}$ enters through $L$ due to the condition \pref{phi'invsout2} and for weakly gravitating sources we can compute the energy in flat space. If this were the entire story then minimizing would set $L=0$, and thereby set the dilaton charge to zero (and so perfectly screen the source from the dilaton). This is not quite what happens because there is another $\phi_{\rm loc}$-dependent contribution to the energy, given by the axion and dilaton gradients interior to the source. For weakly gravitating sources these contribute
\be
   E_{\rm in} = 2\pi \int_{0}^{R} \exd r \, r^2 \; \Bigl[\, W^2(\phi) (\mfa')^2 +  (\phi')^2 \Bigr]  \simeq \frac{\pi R^2 \aminuse^2 }{\ell}
   % (\aplus-\aminuse )^2 
  \Bigl( W^2 \Bigr)_{r=R} \,,
\ee
where the approximate equality evaluates the integral using the narrow-width profile \pref{axstep}. Any other contributions to the energy that are independent of $\phi_{\rm loc}$ do not participate in the minimization with respect to $\phi_{\rm loc}$ and so  can be dropped.

Further progress requires specifying the function $W$, so we consider two representative cases. In the case of an exponential dependence of \pref{Wexpform} we have $W^2(\phi)= \exp[-\xi \phi/\MPL] = \exp[- \phi/\Lambda_\phi]$ and ref.~\cite{Brax:2023qyp} finds the effective dilaton charge evaluated at the field that minimizes the energy is 
\be  \label{betaeffexp}
 \beta_{\rm eff}  \simeq \frac{R}{2  \xi G_\ssN M} \qquad\qquad \hbox{(exponential $W^2(\phi)$)}.
\ee
On the other hand, when $W^2$ has a minimum, as in \pref{Wquadform}, we have $W^2 = W^2_\star + \frac12(\phi - \phi_\star)^2/\Lambda_\phi^2$ 
%it turns out that the minimization of the energy gives
%
%\be 
%\phi_{\rm loc}=\phi_\star \,,
%\ee
%
and the resulting effective dilaton coupling turns out in this case to be 
\be \label{betaeffquad}
%   \beta_{\rm eff} \simeq \frac{\beta }{1+  (R/\ell) \frac{(\aplus-\aminuse )^2}{4\Lambda_\phi^2} } \qquad\qquad \hbox{(quadratic $W^2(\phi)$)}\,.
   \beta_{\rm eff} \simeq \frac{\beta }{1+  (R/\ell) \frac{\aminuse^2}{4\Lambda_\phi^2} } \qquad\qquad \hbox{(quadratic $W^2(\phi)$)}\,.
\ee
This can also be much smaller than $\beta$ if the denominator is large, as can happen if 
%$(\aplus - \aminuse)^2/\Lambda_\phi^2 \gg 1$ 
$\aminuse^2/\Lambda_\phi^2 \gg 1$ and/or the width $\ell$ of the axion profile is sufficiently narrow relative to the source radius $R$.   

\subsubsection{Benchmark values}\label{Benchmark values}

This section establishes benchmark parameters that satisfy the above simplifying assumptions. When making these estimates we use the convention described below eqs.~\pref{dilacoup} and \pref{Wexpform} that the present-day value for $A$ and $W$ is $A(0) = W(0) = 1$. 

For applications to screening we assume the present-day axion mass in matter for typical solar densities, $n_e \sim n_\odot$, to be ({\it c.f.}~eq.~\pref{mavsrho}) 
\be
  m_{\rm ax}(n_\odot)  \sim \left( \frac{m_e n_\odot}{\Lambda_e^2} \right)^{1/2} \sim 2\times10^{-12} \; \hbox{eV} \,,
\ee
so that its Compton wavelength is of order $\ell \sim 100$ km (much smaller than the solar radius). For a representative solar electron density of $n_{\odot} \sim 10^{24}$/cm${}^3 \sim 10^{10}$ eV${}^3$ -- and so $\rho_{e\odot} \sim m_e n_\odot \sim 10^{16}$ eV${}^4$ -- this size of an axion mass in matter determines the coupling scale to be $\Lambda_e \sim 10^{11}$ GeV. We take the vacuum axion mass to be a thousand times smaller (to ensure the step occurs near the solar surface), and so take 
\be \label{mabenchmark}
   m_{\rm ax} = m_\mfa \sim  2\times10^{-15} \; \hbox{eV}. 
\ee
This choice also ensures the axion does not mediate a long-range force in the solar system and that its mass is larger than the Hubble scale during post nucleosynthesis cosmology.

The electron-axion coupling also implies a field-dependence to the electron mass and so one must also check that the gradients in the axion field required by screening are not ruled out by what we know about electron properties on Earth and in the Sun. This could show up in precision measurements of atomic properties, such as through a position-dependence it predicts for the electron/nucleon mass ratio (see \cite{Sherrill:2023zah} for recent constraints). To evade these bounds we require that $\delta m_e/m_e = \cU_e$ remain smaller than $10^{-15}$, which given the form \pref{AppUVforms} implies we require 
\be \label{emassstability}
%  \frac{(\aplus - \aminuse)^2}{\Lambda_e^2} \lsim 10^{-15} \,. 
  \frac{  \aminuse^2}{\Lambda_e^2} \lsim 10^{-15} \,. 
\ee
For $\Lambda_e \sim 10^{11}$ GeV this implies 
%$W|\aplus - \aminuse| \lsim 10^{3}$
$|\aminuse| \lsim 3\times 10^{3}$ GeV.

Next we ask for sufficient screening in the solar system: $\beta_{\rm eff}/\beta \lsim 10^{-3}$ as suggested by the Cassini bound  \cite{Bertotti:2003rm} if $\beta \sim \cO(1)$. For exponential $W^2$ eq.~\pref{betaeffexp} shows this is a condition on $\Lambda_\phi$ alone. Using that $G_\ssN M/R \sim 10^{-6}$ for the Sun shows sufficient screening implies
\be 
  \xi \gsim 10^9  \qquad \hbox{and so} \qquad
 \Lambda_\phi \lsim 10^9 \; {\rm GeV} \qquad \hbox{(exponential $W$)}.
\ee
The screening criterion in the case of quadratic $W^2$ requires 
%$(R/\ell) [(\aplus - \aminuse)^2/\Lambda_\phi^2] \gsim 10^{3}$
$(R/\ell) [ \aminuse^2/4\Lambda_\phi^2] \gsim 10^{3}\beta$. Taking the radius of the sun to be $7\times10^5\;\text{km}$, the above choice for $m_{\rm ax}(n_\odot)$ implies $R/\ell \sim 7\times 10^3$ and so  
\be 
%  \frac{(\aplus-\aminuse )^2}{\Lambda_\phi^2} \gsim 10 \,.
  \Lambda_\phi \lsim \frac{\mfa_e}{\sqrt{\beta}} \qquad \hbox{(quadratic $W$)}\,.
\ee
Combined with the above upper bound on 
%$W|\aplus - \aminuse|$ 
$W| \aminuse|$ this implies $W_\star \Lambda_\phi \lsim 3\times10^3$ GeV. 
We remark in passing that the constraint on $\beta_{\rm eff}(n_\odot)$ can be slightly stronger ($\sim 10^{-4}$) if for the parameters chosen the sourcing of the dilaton by planets is not also screened, since in this case the constraints -- such as coming from the precession of perihelia -- involve the product $\beta_{\rm eff}(n_\odot) \, \beta_{\rm eff}(n_\ssP)$ rather than $\beta_{\rm eff}^2(n_\odot)$. It is also true that partial screening erodes the protection that Brans-Dicke models otherwise have against tests of the equivalence principle, making these more dangerous than they might otherwise have been (see \cite{Brax:2023qyp} for more details).
 
A final concern comes from general bounds on axion decay constants, coming from energy loss from supernovae and massive stars \cite{ParticleDataGroup:2024cfk}. Because these involve axion emission they are very sensitive to couplings to matter that are linear in the axion field. This makes them less constraining for quadratic couplings (see for example \cite{Olive:2007aj, Beadle:2023flm}) like the ones considered here, because for these radiation must occur by emitting two particles simultaneously whenever the background axion sits at the minimum $\ol \mfa = \aminuse$. There can be linear emission from regions where $\mfa_\ad$ deviates from $\aminuse$ but in our case these regions are restricted to the surface of the star, where temperatures and densities are much smaller. Although these constraints have not yet been worked out in detail for the specific models of interest here, the preliminary estimates in \cite{Brax:2023qyp} suggest that the above choices for $W \Lambda_e$ remain viable.

\section{Axion evolution and Early Dark Energy}
\label{two}

We now turn to the cosmological implications of the screening mechanism discussed above. It is known that the axion-dilaton interaction plays a central role in the cosmological dynamics of axio-dilaton models \cite{Burgess:2021obw, Smith:2024ayu, Smith:2024ibv}, and the axion-matter interactions required for screening can change things even more because of the matter-dependent axion potential they generate. In particular, we show here how these axion-matter interactions can provide a robust new production mechanism for a form of early dark energy.\footnote{We highlight this as a novel production and removal mechanism for early epochs of Dark Energy, with no claim that the result resolves the Hubble tension (as was Early Dark Energy's initial motivation \cite{Poulin:2018cxd}).} 

We start with the study of background evolution for the coupled matter/axio-dilaton system, deferring the discussion of their cosmological perturbations to the next sections. For the axion the background evolution also splits up into two types of motion: slow evolution on timescales similar to the Hubble scale superimposed on rapid oscillations about the slower evolution. This section describes each of these in turn.

\subsection{Adiabatic axion evolution}
\label{sec:adiabatic}

Much of the background axion dynamics can be simply understood using the time-dependent effective scalar potential \pref{Veffdef} seen by the axion (including matter-dependent terms). This potential is time-dependent because it depends on both $\phi$ and the particle densities $n_i$ of the nonrelativistic matter to which the axion couples. When the evolution is adiabatic the axion's central value simply traces out the time-dependent minimum of this potential. 

\subsubsection{Matter-dependent axion potential}
\label{sec:abar}

With the choices outlined in the previous sections the effective potential experienced by the scalars is given by 
\be  \label{Veffdef2}
%   V_{\rm eff}(\phi, \mfa)   =   V_{\rm dil}(\phi) + \tfrac12 m_\mfa^2 (\mfa - \aplus)^2  +  A(\phi) \, m_{e0} \, n_e \left[ 1 + \frac{(\mfa - \aminuse)^2}{2\Lambda_e^2} \right] +  A(\phi) \,  m_{m0} \, n_m \left[ 1 + \frac{(\mfa - \aminus)^2}{2\Lambda_m^2} \right]  \,,
   V_{\rm eff}(\phi, n_i, \mfa)   =   V_{\rm dil}(\phi) + \tfrac12 m_\mfa^2 \mfa^2  +  A(\phi) \, m_{e0} \, n_e \left[ 1 + \frac{(\mfa - \aminuse)^2}{2\Lambda_e^2} \right] +  A(\phi) \,  m_{m0} \, n_m \left[ 1 + \frac{(\mfa - \aminus)^2}{2\Lambda_m^2} \right]  \,,
\ee 
of which the axion-dependent part can be written
\be  \label{Veffdef2ax}
%   V_{\mfa\,{\rm eff}}(\phi, \mfa)  =   \tfrac12 m_\mfa^2 \left[ (\mfa - \aplus)^2  + \frac{\rho_{e0}}{\rho_{e\,{\rm th}}}  (\mfa - \aminuse)^2  + \frac{\rho_{m0}}{\rho_{m\,{\rm th}}}  (\mfa - \aminus)^2 \right] \,,
   V_{\mfa\,{\rm eff}}(\phi, n_i, \mfa)  =   \tfrac12 m_\mfa^2 \left[ \mfa^2  +\cC_e  (\mfa - \aminuse)^2  + \cC_m (\mfa - \aminus)^2 \right] \,,
\ee 
which defines 
\be \label{cCidefs}
    \cC_i(t) := \frac{A(\phi) \, m_{i0} \, n_i(t)}{m_\mfa^2 \Lambda_i^2} = \frac{\rho_{i0}(t)}{\rho_{i\,\thr}} 
\ee
where $\rho_{i0} := A(\phi) \, m_{i0} n_i$ differs from the corresponding (Einstein-frame) matter energy density, $\rho_i$, by the difference between $m_{i0}$ and the full mass $m_i = m_{i0}[1 + \cU_i(\mfa)]$ given in \pref{mivsmi0}. This difference is not important in practice in the regime where $\cU_i(\mfa) \ll 1$, as is required for the approximate equality in the axion equation in \pref{Veffderiv} to be valid. 

The threshold value
\be \label{rhothbench}
   \rho_{i\,{\rm th}} := m_\mfa^2 \Lambda_i^2 = \left(100 \; \hbox{eV}\right)^4 \left( \frac{m_\mfa}{10^{-15} \; \hbox{eV}} \right)^2 \left( \frac{\Lambda_i}{10^{10} \; \hbox{GeV}} \right)^2 \,,
\ee
denotes the density above which the matter-dependent terms dominate the vacuum term, and the above representative values are chosen motivated by the benchmarks used when discussing screening (see the discussion surrounding \pref{mabenchmark}).  Notice that these give threshold densities that are achieved by the Hot Big Bang plasma during the interval after nucleosynthesis and before recombination.
  
Taking two derivatives defines the density-dependent axion mass parameter
\be \label{maeffdef}
m^2_{\mfa\,{\rm eff}} :=  \partial_\mfa^2 V_{\rm eff} 
%= m_\mfa^2 +  \frac{A(\phi) \, m_{e0} \, n_e }{\Lambda_e^2}  + \frac{A(\phi) \,  m_{m0} \, n_m }{\Lambda_m^2} 
=    m_\mfa^2 \Bigl( 1  + \cC_e  +\cC_m  \Bigr).
\ee
Notice that this implies the axion mass asymptotes to $m_{\mfa\,\eff} \to m_\mfa$ in the future (once $\rho_{i0} \ll \rho_{i\,\thr}$) but falls with the matter density, $m_{\mfa\,\eff} \propto a^{-3/2}$, in the remote past (when $\rho_{i0} \gg \rho_{i\,\thr}$).

Denoting the minimum of the potential \pref{Veffdef2ax} by $\mfa_\ad(t)$, one finds  
\be \label{abarform}
%\bea
%    \ol \mfa(t)   &=& \frac{m_\mfa^2}{m^2_{\mfa\,{\rm eff}}}  \left[\aplus  + \frac{\rho_{e0} }{\rho_{e\,{\rm th}}} \, \aminuse   + \frac{\rho_{m0} }{\rho_{m\,{\rm th}}} \, \aminus  \right] \nn\\
%    &=& \aplus + \frac{m_\mfa^2}{m^2_{\mfa\,{\rm eff}}} \left[\frac{\rho_{e0} }{\rho_{e\,{\rm th}}} \,  (\aminuse - \aplus) +\frac{\rho_{m0} }{\rho_{m\,{\rm th}}} \, (\aminus - \aplus) \right] \,.
%\eea
    \mfa_\ad(t)    = \frac{m_\mfa^2}{m^2_{\mfa\,{\rm eff}}}  \Bigl( \cC_e \, \aminuse   + \cC_m \, \aminus  \Bigr) = \frac{\cC_e \, \aminuse + \cC_m \, \aminus}{1 + \cC_e + \cC_m} \,.
\ee  
%
%
%\bea
%    \ol \mfa(t)  %&=& \frac{m_\mfa^2 \aplus + (A m_{e0} n_e \aminuse/\Lambda_e^2) + (A m_{m0} n_m \aminus/\Lambda_m^2)}{m^2_{\mfa\,{\rm eff}}} \nn\\
%     &=& \frac{1}{m^2_{\mfa\,{\rm eff}}} \left[m_\mfa^2 \aplus +A(\phi) \left( \frac{m_{e0} n_e \aminuse}{\Lambda_e^2}  + \frac{m_{m0} n_m \aminus}{\Lambda_m^2} \right) \right] \nn\\
%    &=& \aplus + \frac{A(\phi)}{m^2_{\mfa\,{\rm eff}}} \left[ \frac{m_{e0} n_e}{\Lambda_e^2} (\aminuse - \aplus) + \frac{m_{m0} n_m}{\Lambda_m^2}(\aminus - \aplus) \right] \,.
%\eea 
%
This is a function of time in a cosmological context because $\phi$, $n_e$ and $n_m$ are time-dependent, with both particle densities falling monotonically $n_i \propto a^{-3}$ where $a(t)$ is the metric's scale factor. At sufficiently late times the vacuum contribution dominates $V_\eff$ and so eventually $\mfa_\ad \to \aplus$ , as in simpler axion cosmologies. At sufficiently early times, on the other hand, it is the matter contributions that dominate. Assuming the electron contribution is much smaller than the Dark Matter contribution -- as is very likely given the strong constraint \pref{emassstability} -- then $\mfa_\ad(t) \to \aminus$ at very early times.

The axion part of the potential evaluated at its minimum is 
\be \label{Vaeffmin}
  \ol{V}_{\mfa\,{\rm eff}}(\phi,n_i) := V_{\mfa\,{\rm eff}}[\phi, n_i, \mfa_\ad(\phi,n_i)] = \tfrac12 \, m_\mfa^2 \left[ \frac{\cC_e \aminuse^2 + \cC_m \aminus^2 + \cC_e \cC_m (\aminuse - \aminus)^2}{1 + \cC_e + \cC_m}  \right]
\ee
which depends on time as well as $\phi$ because of the appearance of $n_i(t)$ within the $\cC_i$'s. Notice that although the potential $V_{\mfa\,\eff}(\phi, n_i, \mfa)$ given in \pref{Veffdef2ax} satisfies
\be
  V_{\mfa\,\eff}(\phi, n_i, \mfa) = V_{\mfa\,\eff}(\phi, \bar n_i, \mfa) + \sum_{i=e,m}\frac{A(\phi) \, m_{i0} }{2\Lambda^2_i}\,(n_i - \ol n_i) (\mfa - \mfa_i)^2 \,,
\ee
the same is {\it not} true of \pref{Vaeffmin} due to the change in $\mfa_\ad$ that a shift $\ol n_i \to n_i$ generates.

At very late times $\ol{V}_{\mfa\,{\rm eff}} \to 0$ -- and so $\ol{V}_{{\rm eff}} \to V_{\rm dil}(\phi)$ -- because $n_i \to 0$ implies $\cC_e , \, \cC_m \to 0$ and so $\mfa_\ad \to \aplus$. For $\cC_e \ll \cC_m \ll 1$ eq.~\pref{Vaeffmin} implies $\ol V_{\mfa\,{\rm eff}} \simeq \frac12 m_\mfa^2 \cC_m \mfa_m^2$ and so falls with the universal expansion proportional to $a^{-3}$.  At very early times, however, the matter-dependent terms dominate and we instead have $\cC_i \gg 1$ and so $\ol{V}_{\mfa\,{\rm eff}} \to  \ol{V}_{\mfa\,{\rm in}}(\phi)$ -- implying $\ol{V}_{\rm eff} \to V_{\rm dil}(\phi)  + \rho_{e0} + \rho_{m0} + \ol{V}_{\mfa\,{\rm in}}(\phi)$ -- with a large-$\cC$ expansion of \pref{Vaeffmin} giving
%
%\bea
%    V_{\mfa\,{\rm in}} &=& \frac{\cC_e \cC_m }{2(\cC_e + \cC_m)} (\aminus - \aminuse)^2 
%    +\frac{m_\mfa^2 \Bigl[\cC_e(\aminuse-\aplus) + \cC_m(\aminus-\aplus) \Bigr]^2 }{2(\cC_e+\cC_m)^2} + \cdots \nn\\
%    &=& \tfrac12 m_\mfa^2 \left\{\frac{\rho_{e0}}{\rho_{e\,{\rm th}}} \frac{(\aminus - \aminuse)^2}{1+\cR}  +  \frac{\Bigl[(\aminus-\aplus) + \cR (\aminuse-\aplus) \Bigr]^2 }{(1+\cR)^2} + \cdots \right\}
%\eea
\bea \label{Vaindef}
   \ol  V_{\mfa\,{\rm in}} &=& \tfrac12 m_\mfa^2 \left[ \frac{\cC_e \cC_m }{ (\cC_e + \cC_m)} (\aminus - \aminuse)^2 
    +\frac{ ( \cC_e \aminuse + \cC_m \aminus )^2 }{ (\cC_e+\cC_m)^2} + \cdots \right]\nn\\
    &=& \tfrac12 m_\mfa^2 \left\{ \left( \frac{\rho_{e0}}{\rho_{e\,{\rm th}}} \right) \frac{( \aminus - \aminuse)^2}{1+\cR}  +  \frac{( \aminus + \cR  \aminuse )^2 }{(1+\cR)^2} + \cdots \right\}
\eea
where the second line defines the time-independent ratio
\be \label{ratiodef}
  \cR := \frac{\cC_e}{\cC_m} =  \left( \frac{m_{e0}}{m_{m0}} \right) \left( \frac{n_e}{n_m} \right) \left( \frac{\Lambda_m}{\Lambda_e} \right)^2 \,.
\ee

Notice that the leading term in \pref{Vaindef} is proportional to $n_e$ and so falls like $1/a^3$, but it also vanishes in the limit either $\cC_e$ or $\cC_m$ vanish (or if $\aminus = \aminuse$). This makes it in practice very small because of mass-variation contraints like \pref{emassstability} that require $\cC_e$ to be small. This allows the first subleading term to dominate, 
\be
%  V_{\mfa\,{\rm in}} \simeq \frac{m_\mfa^2\Bigl[(\aminus-\aplus) + \cR (\aminuse-\aplus) \Bigr]^2 }{2(1+\cR)^2} \,,
  V_{\mfa\,{\rm in}} \simeq \frac{m_\mfa^2 (\aminus + \cR \aminuse )^2 }{2(1+\cR)^2} \,,
\ee
which is time-independent because it depends on $\phi$ and $n_i$ only through the small time-independent ratio\footnote{Notice that $\cR$ can depend on position once inhomogeneities are included to the extent that the Dark Matter and electron distributions differ from one another.} $\cR$. $V_{\mfa\,{\rm in}}$ becomes 
%$\tfrac12   m_\mfa^2(\aminus-\aplus)^2$
$\tfrac12   m_\mfa^2 \, \aminus^2$ if $\cR$ is negligible, as is intuitive because in this limit the dominance of the Dark Matter contribution in \pref{abarform} pulls $\mfa_\ad \to \aminus$ and this nulls out the Dark Matter part of $V_{\mfa\,{\rm eff}}$, allowing the vacuum contribution $\frac12 m_\mfa^2 \mfa_m^2$ dominate.

For later purposes it is useful to isolate the evolution of the vacuum part of the axion potential, evaluated using the solution (\ref{abarform}):
\begin{eqnarray}\label{EDEcontribution}
    \ol V_{\mfa\,{\rm ax}}(\phi, n_i) := \tfrac{1}{2}m_\mfa^2 \mfa_\ad^2(t) \simeq \tfrac{1}{2}m_\mfa^2\left(\frac{\cC_m \mfa_m}{1+\cC_m}\right)^2,
\end{eqnarray}
where the approximate equality again drops the contributions involving $\cC_e$. As described above, this is what dominates $\ol V_{\mfa\,{\rm eff}}$ at early times, leading in the $\cC_m \gg 1$ limit to the constant $\ol V_{\mfa\,{\rm eff}} \simeq \ol V_{\mfa\,{\rm ax}} \simeq \frac12 m_\mfa^2 \mfa_m^2$. By contrast, once $\cC_m$ becomes much smaller than unity -- {\it i.e.}~once $\rho_{m0} \ll \rho_{m\,\thr}$ -- we instead have
\begin{equation}
   \ol V_{\mfa\,{\rm ax}}(\phi, n_i)  \simeq \tfrac12  m_\mfa^2 \cC_m^2 \mfa_m^2 \propto \frac{1}{a^6}.
\end{equation}
It is the vacuum axion potential $\ol V_{\mfa\,{\rm ax}}$ that is plotted with the label `axion' in {\it e.g.}~Fig.~\ref{fig:cosmological constraints}.

\subsubsection{Interpretation as Early Dark Energy}

The above arguments show how the background axion energy behaves like a novel form of Early Dark Energy (EDE) inasmuch as it remains approximately constant until eventually falling off to zero once ${\rho_{m0} \lsim \rho_{m\,\thr}}$ (or $\cC_m \lsim 1$). Unlike standard EDE models this energy turns itself off without any need for vacuum phase transitions, because this is automatically accomplished by the changing matter density once $\rho_{m0}$ falls below $\rho_{m\,\thr}$. 

Eqs.~(\ref{abarform}) and (\ref{Vaeffmin}) show that the maximal fraction of energy associated with this type of early dark energy occurs at the epoch $t_{\star}$ defined by $\rho_{m0}(t_\star) = \rho_{m\,\thr}$ (as may also be seen in our later numerical evaluation, such as shown in the first panel of Fig.~\ref{fig:cosmological constraints}). Denoting the total energy at this time by $\rho_{\ttl \,\star} = \rho_{\ttl}(t_{\star})$, we find the maximum fraction of EDE ({\it i.e.}~in the background axion) is given by 
\begin{eqnarray}\label{ede frac}
    f_{\textnormal{EDE}} = \left. \frac{\ol V_{\mfa\,{\rm eff}}(\phi, n_i)}{\rho_{\tot}}\right|_{\rho_{m0} = \rho_{m\,\thr}} \simeq \frac{m_\mfa^2\mfa_m^2}{4\rho_{\tot\,\star}} \,,
\end{eqnarray}
where the last approximate equality comes from evaluating \pref{Vaeffmin} at $\cC_e \simeq 0$ and $\cC_m = 1$.

In the normal telling of the EDE story, to have the desired effect on recombination physics requires this maximal fraction to arise around the era of matter-radiation equality, $t_{\star} \simeq t_{\rm eq}$. The value of $\Lambda_m$ required to ensure this is true is found by remarking that $t \simeq t_{\rm eq}$ implies $\rho_{m} + \rho_{b}\simeq \rho_r$ (where $\rho_r$ and $\rho_b$ are the energy density of radiation and baryons respectively) and so $\rho_m + \rho_b \simeq \frac{1}{2}\rho_{\ttl}$. Neglecting $\rho_b$ relative to $\rho_m$ and choosing $t_\star \simeq t_{\rm eq}$ then implies
\begin{align}
    \rho_{\ttl\,\star} \simeq  2\rho_m = 2\rho_{m0}\left[1+ \cU_m(\mfa_\ad)\right] 
    \simeq 2\rho_{m\,\thr},
\end{align}
where the last equality uses the definition of $t_\star$ as well as $\cU_m(\mfa_\ad) \ll 1$. Plugging this into (\ref{ede frac}), we therefore evaluate the axion's maximum EDE fraction as
\begin{eqnarray}
    f_{\textnormal{EDE}} \simeq   \frac{m_\mfa^2\mfa_m^2}{8\rho_{m\,\thr}} \simeq  \frac{\mfa_m^2}{8\Lambda_m^2} = \tfrac{1}{4}\, \cU_m(\mfa = 0).
\end{eqnarray}

\subsubsection{Criteria for adiabatic evolution}

The minimum configuration $\mfa_\ad(t)$ matters to the extent that it provides an accurate solution to the axion equation of motion in a cosmological context, which is true within the adiabatic approximation. To see why, notice that in a cosmological setting the field equation satisfied by homogeneous configurations, $\ol\phi(t)$, $\ol\mfa(t)$ and $\ol n_i(t)$, becomes
\be \label{axioneqcosmo}
   \ddot{\ol\mfa} + 3 H_\mfa \dot{\ol\mfa} + \frac{\ol m_{\mfa\,{\rm eff}}^2}{\ol W^2} \Bigl[ \ol\mfa- \mfa_\ad(t) \Bigr] = 0,
\ee
which uses the notation $\ol W := W(\ol \phi)$ and $\ol m_{\mfa\,\eff} := m_{\mfa\,\eff}(\ol\phi,\ol n_i)$ and
\be 
   H_\mfa := \tfrac{1}{3}\partial_t \ln \ol W^2 + H \,,
\ee
where $H = \dot a/a$ is the Hubble scale. In \pref{axioneqcosmo} $\mfa_\ad(t)$ is the configuration introduced above that is defined to satisfy $\partial_\mfa V_{\rm eff} = 0$. So long as $\ol\phi$ also only varies significantly over Hubble times, we have $H_a={\cal O}(H)$. The adiabatic solution to \pref{axioneqcosmo} assumes the hierarchy
\be \label{adiabaticcondition}
\ol m_{{\rm ax}\,{\rm eff}}(t) = \frac{\ol m_{\mfa\,{\rm eff}}}{\ol W}\gg H_\mfa \,, 
\ee
since in this case the axion follows the minimum of the effective potential, $\ol\mfa(t)\simeq \mfa_\ad$. The growth of the axion mass as $a \to 0$ in the regime where $\rho_{i0} \gg \rho_{i\,\thr}$ makes the adiabatic approximation better and better the earlier in the universe's history we go. In what follows we work in a regime where this is a good approximation in all post-BBN epochs. This assumption is also consistent with the choices for $m_\mfa$ that allow the screening of the dilaton in the solar system.  

The properties of $\mfa_\ad$ explored in \S\ref{sec:abar} imply that an adiabatically evolving axion at the background level interpolates between a cosmological constant early in the Universe and an extra component of Dark Matter at late times.  As noted earlier, no phase transitions are required to have the EDE turn off and not dominate the universe perpetually since the transition is instead achieved through the matter-dependence of the effective scalar potential. 

There is considerable latitude in precisely how large $V_{\mfa\,{\rm in}}$ can be in such a scenario because this depends on the relative size of the vacuum and matter-dependent parts of $V_{\rm eff}$. The larger the vacuum component, $V_{\rm ax}$, the earlier $V_{\mfa\,{\rm in}}$ transitions from constant to something falling proportional\footnote{An interesting possibility is to have the axion itself be Dark Matter. This can be achieved by placing the EDE transition at or before nucleosynthesis, since in this case the axion will have already transitioned to the minimum of its vacuum potential by nucleosynthesis, recovering the model described in \cite{Smith:2024ibv} (with an additional coupling between the axion-as-CDM and electrons). As will be shown below, imposing screening in this case however causes there to be an extremely strong interaction between the axion and dilaton, as seen in (\ref{full dilaton friedmann}). This coupling destabilises the dynamics of the dilaton if the axion CDM is ever a dominant component so we do not pursue this option further here.} to $n_i$. The maximum total fraction of the universal energy density invested in EDE occurs at this transition given by (\ref{ede frac}) with $\Lambda_m$ controlling when $\rho_{m0} \sim \rho_{m\,\thr}$. To get an idea of what is involved, if we choose this to occur at the epoch of radiation-matter equality (as discussed above) then we find that taking $\rho_{m\,\thr} = m_\mfa^2 \Lambda_m^2 \simeq 1$ eV${}^4$ in \pref{rhothbench} implies $\Lambda_m \sim 10^{6}$ GeV if we use the benchmark value $m_\mfa \sim 10^{-15}$ eV discussed above, in the screening section. Notice that this choice together with those made for the screening benchmarks around eq.~\pref{mabenchmark} imply that all of the factors appearing in \pref{ratiodef} are small, justifying the choice $\cR \ll 1$. 

\subsection{Beyond adiabatic evolution}

The speed of convergence of nonadiabatic homogeneous evolution $\ol\mfa(t)$ towards the adiabatic solution $\mfa_\ad$ can be studied by defining $\delta \ol\mfa= \ol\mfa - \mfa_\ad$ and rewriting the axion field equation as
\be \label{axionapproacheq}
\partial_t\Bigl[ \ol W^2 a^3  \partial_t \delta \ol \mfa \Bigr] + a^3 \ol m_{\mfa\,{\rm eff}}^2 \delta \ol\mfa= -\partial_t\Bigl[ \ol W^2 a^3 \dot \mfa_\ad  \Bigr],
\ee
where $\dot \mfa_\ad = \partial_t \mfa_\ad$. A particular integral of this equation is
\be \label{particint}
\delta \ol\mfa_p = -\frac{1}{a^3 \ol m^2_{\mfa\,{\rm eff}}}\partial_t\Bigl[ \ol W^2 a^3 \dot \mfa_\ad \Bigr] \,,
\ee
in the approximation $\ol m_{\mfa\,{\rm eff}} \gg H$ -- assuming all the background quantities satisfy $\dot f/f \sim {\cal O}(H)$ -- because the first term on the left-hand side of \pref{axionapproacheq} is subdominant in this limit. This particular solution is suppressed relative to $\mfa_\ad$ itself by powers of $H/\ol m_{\mfa\,{\rm eff}}$ and so becomes a negligible correction in the strict adiabatic limit.

To understand  fully the approach to $\mfa_\ad$ we require the general solution to \pref{axionapproacheq}, which is obtained by adding to \pref{particint} a general solution to the homogeneous equation with no source term: $\delta \ol\mfa = \delta \ol\mfa_p + \delta \ol\mfa_h$. The homogeneous solution is given at leading order in $\ol m_{\mfa\,{\rm eff}}/H$
by
\be \label{bgrho}
\delta \ol\mfa_h \sim \frac{\sqrt{2C}}{\sqrt{a^3 \ol  W  \ol m_{\mfa\,{\rm eff}}}}\cos \left[\int m(t) \, \exd t-S_0 \right],
\ee
where $C$ and $S_0$ are integration constants and $m(t) :=\ol  m_{\rm ax~eff}(t) = \ol m_{\mfa\,{\rm eff}}/\ol W$ is the physical matter-dependent mass first seen in \pref{adiabaticcondition} that approaches the vacuum mass $m_{\rm ax}$  -- defined below \pref{VaxDef} -- as $\ol n_i \to 0$. This solution describes the well-known damped oscillations whose amplitude decreases with time as the axion reverts to the adiabatic solution $\mfa_\ad$. The adiabatic solution is an attractor for nearby solutions within a finite basin of attraction. 

The energy density of these homogeneous oscillations is
\be 
\bar \rho_{\rm osc}= \tfrac{1}{2} \ol W^2 (\delta \dot{\ol \mfa})^2 + \tfrac12 \ol m^2_{\mfa\,{\rm eff}} \delta \ol\mfa^2 
\,,
\ee
which for $\ol m_{\mfa\,{\rm eff}} \gg H$ evaluates to
\be 
\bar \rho_{\rm osc} = \frac{C m(t)}{a^3}  = \frac{C \ol m_{\mfa\,{\rm eff}}}{\ol W a^3} ,
\ee
and so mimics the background matter density of a fluid comprised of particles of mass $m(t)$ and conserved number density 
\be  \label{OscillationEvolution}
n_\mfa(t) := \frac{\bar \rho_{\rm osc}}{m(t)}= \frac{C}{a^3}.
\ee

The existence of rapid oscillations with frequency much larger than the Hubble scale complicates the numerical generation of cosmological solutions to the field equations, since reliably calculating in the regime $\ol m_{\mfa\,{\rm eff}} \gg H$ requires simulations with an enormous dynamic range. We deal with this issue below by integrating out the fast oscillations to obtain an effective fluid description of the much slower transfer of energy over Hubble timescales (for both background and fluctuations). These techniques were developed long ago in \cite{Madelung:1927ksh, Spiegel:1980ykb, Turner:1983he} (and revived more recently \cite{Chavanis:2011zi, Suarez:2011yf, Uhlemann:2014npa, Marsh:2015daa, Hui:2016ltb}) and our purpose in the next sections is to extend these to include the multifield axio-dilaton case. 

\subsection{Axion-driven dilaton instability}\label{Dilaton Tachyonic Instability}

Before treating the axion fluid, we first digress to discuss one consequence of axion evolution for the dilaton that provides a useful guide when interpreting our later numerical calculations. Axion evolution affects how the dilaton evolves due to the presence of the derivative dilaton-axion interaction. This is much like in the screening mechanism, though there is an important change of sign for time-dependent axions compared with spatially varying axion profiles. 

To see why, recall that the equation for a homogeneous dilaton, $\bar \phi$, in a cosmological background spacetime is
\be  \label{homophieq}
  \ddot{\bar\phi} + 3H \dot{\Bar{ \phi}}- \dot{\mfa}_\ad^2 \ol W(\bar\phi) \ol W'(\bar\phi)= - \partial_\phi V_{\rm eff}(\Bar{\phi}),
\ee
where $\ol W' := \partial_\phi W(\bar \phi)$ and we assume the axion field is homogeneous and evolves adiabatically: $\ol\mfa = \mfa_\ad$. (Fluctuations are described more fully in later sections.) For the present purposes the important term is the last one on the left-hand side, which for given $\mfa_\ad$ has the effect of adding a new term to the effective potential seen by the dilaton, of the form (see \emph{e.g.}~eq.~\pref{VWdef})
\be
  \delta V_\ssW(\phi) = - \tfrac12 \, \dot{\mfa}_\ad^2 W^2(\phi)  \,.
\ee

What is important about this contribution is that it is negative definite\footnote{The unusual sign is required to properly capture motion along target-space geodesics (in the absence of other forces) \cite{Brax:2023now} and has been recognized to allow unusual cosmologies as these models started being explored in more detail \cite{Burgess:2021obw, Brax:2023tls, Smith:2024ayu}.} and so it favours $\phi$ seeking out regions that maximize the size of $W$ (which in turn tends to suppress axion interactions). Whether this is a catastrophe or not depends on what other forces are pushing on $\phi$ and whether they can successfully compete. We find below, for example, that for $W \propto e^{-\xi \phi/\MPL }$ -- as in \pref{Wexpform} for instance -- the tendency for this term to drive $\phi$ to smaller values is countered by the tendency of the potential \pref{VdilDef} to push it to larger values. 

In the case of quadratic kinetic coupling $W^2(\phi)$ eq.~\pref{homophieq} becomes
\be 
\ddot{\Bar{ \phi}} + 3H \dot{\Bar{ \phi}}- \frac{\dot{\mfa}_\ad^2(\rho)}{2\Lambda_\phi^2}(\Bar{\phi}-\phi_\star)= - \partial_\phi V_{\rm eff}(\Bar{\phi}) \,,
\ee
showing how axion evolution induces a tachyon-like mass term for $\phi$, which competes with the dynamics driven by the effective potential. Using \pref{maeffdef} in \pref{abarform} and differentiating gives in the limit 
%
%\bea \label{abardot}
%    \dot{\ol \mfa}   &=& - \Bigl( \beta \dot \phi + 3 H \Bigr) \left[ \frac{\rho_{e0} }{\rho_{e\,{\rm th}}} \,  (\aminuse - \aplus) + \frac{\rho_{m0} }{\rho_{m\,{\rm th}}} \, (\aminus - \aplus) \right]/\left(1 +   \frac{\rho_{e0} }{\rho_{e\,{\rm th}}}   + \frac{\rho_{m0} }{\rho_{m\,{\rm th}}}  \right)^2 \nn\\
%    &\simeq&  \Bigl( \beta \dot \phi + 3 H \Bigr) \, \Bigl( \aplus - \aminus \Bigr) \frac{ \rho_{m0} /\rho_{m\,{\rm th}}}{(1  + \rho_{m0} / \rho_{m\,{\rm th}})^2} \qquad \hbox{(if $\rho_{e0}/\rho_{e\,{\rm th}} \simeq 0$)}\,.
%\eea 
\bea \label{abardot}
    \dot{\mfa}_\ad   &=& - \Bigl( \beta \dot \phi + 3 H \Bigr) \left( \frac{\rho_{e0} }{\rho_{e\,{\rm th}}} \,   \aminuse   + \frac{\rho_{m0} }{\rho_{m\,{\rm th}}} \,  \aminus   \right)/\left(1 +   \frac{\rho_{e0} }{\rho_{e\,{\rm th}}}   + \frac{\rho_{m0} }{\rho_{m\,{\rm th}}}  \right)^2 \nn\\
    &\simeq& - \Bigl( \beta \dot \phi + 3 H \Bigr)   \aminus \, \frac{ \rho_{m0} /\rho_{m\,{\rm th}}}{(1  + \rho_{m0} / \rho_{m\,{\rm th}})^2} \qquad \hbox{(when $\rho_{e0}/\rho_{e\,{\rm th}}$ is negligible)}\,.
\eea 
we see that the tachyonic term has the largest effect when $\rho_{m0} \sim \rho_{m\,{\rm th}}$; \emph{i.e.}~at the axion's transition between the early dark energy to the late matter-like behaviour. 

In this transition regime we have $\rho_{m0} \sim \rho_{m\,{\rm th}}$ and so
%
%\be 
%\frac{\dot{\bar \mfa}^2}{\Lambda_\phi^2} \simeq  \frac{1}{16} \Bigl( \beta \dot \phi + 3 H \Bigr)^2 \, \frac{( \aplus - \aminus )^2}{\Lambda_\phi^2} \,.
%\ee
\be 
\frac{\dot{\mfa}_\ad^2}{\Lambda_\phi^2} \simeq   \Bigl( \beta \dot \phi + 3 H \Bigr)^2 \, \frac{\aminus^2}{16\Lambda_\phi^2} \,.
\ee
Instabilities can be important if this dominates other contributions to the dilaton equation, which for cosmological evolution are typically $\cO(H^2)$ in size. This indicates that problems with runaway solutions need not be an issue if %$|\aplus - \aminus| \ll \Lambda_\phi$
$| \aminus| \ll \Lambda_\phi$, but could become problems otherwise. We return to this observation in \S\ref{five} below, where we present numerical analyses of models with various choices for $W$ and $V$.

\section{Sigma-model interactions and the axion fluid}
\label{three}

This section describes more fully how to handle the slow evolution in the energy of the fast oscillations around $\mfa_\ad$. This can be done cleanly in the limit $m_{{\rm ax}\,{\rm eff}} \gg H$ because then the oscillations are much faster than the time scale governing the evolution of the other cosmologically evolving fields. Our purpose is to extend the techniques of \cite{Madelung:1927ksh, Spiegel:1980ykb, Turner:1983he, Chavanis:2011zi, Suarez:2011yf, Uhlemann:2014npa, Marsh:2015daa, Hui:2016ltb} to include the multifield axio-dilaton case. We do so in enough generality that the results of this section can also be used for non-cosmological applications, such as when testing GR in the solar system. 

We have seen that the amplitude of homogeneous oscillations around the background decreases with the Universal expansion leading to an energy density that behaves like Dark Matter. Once inhomogeneous fluctuations are considered axion oscillations behave like a fluid with its own background and perturbations. The dynamics of this axion fluid are obtained by integrating out the fast modes (\emph{i.e.}~the oscillations) to obtain the effective description of the much slower evolution (\emph{i.e.}~the slow variation of the oscillation amplitude and the initial phase). 

We here set up the fluid description that is evolved numerically in later sections, doing so fairly explicitly because previous treatments did not have the two-derivative scalar-scalar interactions (and because the behaviour of axio-dilaton systems can be of wider interest than just cosmology).

\subsection{Integrating out rapid oscillations}

The decoupling process averages over the fast oscillations \cite{Turner:1983he}, leaving a low-frequency effective field  theory. From the point of view of averaging over fast axion motion we can regard the quantities $g_{\mu\nu}(x)$, $\phi(x)$ and $n_i(x)$ as specified background configurations that vary only slowly in space and time. At a later point we specialize these to small fluctuations about a homogeneous background, $\phi = \ol\phi(t) + \delta \phi(x)$, $n_i = \ol n_i(t) + \delta n_i(x)$ and so on, but we need not do so yet here with the exception of  the frequency of the rapid background oscillations. 

To this end we change variables from $\mfa$ to $\psi$ where
\be 
\mfa (x) = \ol \mfa(x) + \frac{1}{\sqrt{2} \,m(t)}\left[ e^{-i\int_0^t \exd t'\, m(t')}\psi(x) + e^{i\int_0^t \exd t'\, m(t')}\psi^\star(x)\right] ,
\ee
where  -- as in \pref{bgrho} -- for brevity of notation we define the physical mass evaluated at the background dilaton and matter densities, 
\be \label{mdef}
m^2(t) :=  \frac{m^2_{\mfa\,{\rm eff}}(\ol\phi, \ol n_i) }{W^2(\ol\phi)} =  \frac{\ol m^2_{\mfa\,{\rm eff}} }{\ol W^2}  \quad \hbox{as opposed to} \quad
 m^2_{{\rm ax}\,{\rm eff}}(x) = \frac{m^2_{\mfa\,{\rm eff}}(\phi,  n_i) }{W^2(\phi)}\,.
\ee
Notice we do not (yet) similarly refer $\ol\mfa$ to the homogeneous background, and so it remains the local minimum of the potential even as $\phi$ and $n_i$ vary in position. 

The evolution of the slowly varying fields $\ol\mfa$ and $\psi$ are found by inserting this ansatz into the axionic action and coarse-graining the result by performing the integration over a region $M$ of spacetime much larger than the oscillation frequency and wavelength, but much smaller than the scales over which $g_{\mu\nu}$, $\phi$ and $n_i$ vary in cosmology. This leads to the effective lagrangian density for `slow' evolution of the form 
\be 
  \Bigl\langle \cL_{\rm ax} \Bigr \rangle := - \frac{1}{\cV_\ssM} \int_M \exd^4 x \sqrt{-g}\Bigl[ \tfrac12 \, W^2(\phi) g^{\mu\nu}   \partial_\mu \mfa \, \partial_\nu \mfa   +  V_{\mfa\,{\rm eff}}(\mfa)  \Bigr] 
  %= -   \sqrt{-g}\Bigl[ \tfrac12 \, W^2(\phi) g^{\mu\nu} \langle \partial_\mu \mfa \, \partial_\nu \mfa \rangle + \langle V_{\mfa\,{\rm eff}}(\mfa) \rangle \Bigr]
  \,,
\ee
where $\cV_\ssM$ is the volume of $M$ and the effective potential is given by \pref{Veffdef2ax}, which for the present purposes we write as
%
%\be 
%V_{\mfa\,{\rm eff}} =  \ol{V}_{{\rm eff}}(\phi,t) + \tfrac12 \, m_{\mfa\,{\rm eff}}^2(\phi,\ol n_i) \, (\mfa-\ol \mfa)^2 + A(\phi) \sum_i \frac{m_{i0}}{2\Lambda_i^2}  (n_i -\ol n_i) (\mfa-\mfa_i)^2 .
%\ee
\be 
V_{\mfa\,{\rm eff}} (\mfa) =  \ol{V}_{\mfa\,{\rm eff}}(\phi,n_i) + \tfrac12 \, m_{\mfa\,{\rm eff}}^2(\phi, n_i) \, (\mfa- \mfa_\ad)^2  \,,
\ee
at each spacetime point $x$. Here $m_{\mfa\,{\rm eff}}$ and $\mfa_\ad$ are as defined in eqs.~\pref{maeffdef} and \pref{abarform} and $\ol{V}_{\mfa\,{\rm eff}}(\phi,t) = V_{\mfa\,{\rm eff}}(\mfa = \mfa_\ad)$, as defined in eq.~\pref{Vaeffmin}. 
%In all three cases $n_i$ is evaluated at a homogeneous background value $\ol n_i$ and the third term corrects the result when the relevant matter densities do not coincide with this background value.

We evaluate the action using the derivatives
%
%\bea
%  \partial_\mu \mfa (x) &=&  \frac{1}{\sqrt{2 m(t)}}\left[ e^{-i\int_0^t \exd t\, m(t)} \partial_\mu \psi(x) + e^{i\int_0^t \exd t\, m(t)} \partial_\mu \psi^\star(x)\right]  \\
%  &&\; + \delta_\mu^t \left[ \dot{\bar \mfa}  - \left( \frac{\dot m}{2m} + i m \right) \frac{\psi}{\sqrt{2 m(t)}}  e^{-i\int_0^t \exd t\, m(t)}   - \left( \frac{\dot m}{2m} - i m \right) \frac{\psi^\star}{\sqrt{2 m(t)}}   e^{i\int_0^t \exd t\, m(t)}  \right] ,\nn
%\eea
%
\bea
  \dot \mfa   &=& - \frac{i}{\sqrt{2}}\left[ e^{-i\int_0^t \exd t'\, m(t')}   \psi(x) - e^{i\int_0^t \exd t'\, m(t')}   \psi^\star(x)\right]  \\
  &&\; +  \dot{\bar \mfa}  +  \frac{1}{\sqrt{2} m}\left[ e^{-i\int_0^t \exd t'\, m(t')} \left( \dot \psi - \frac{\dot m}{m}  \psi \right) + e^{i\int_0^t \exd t'\, m(t')} \left( \dot \psi^\star - \frac{\dot m}{m}  \psi^\star \right)  \right]   ,\nn
\eea
where in the regime of interest the first line is systematically larger than the second line. Spatial derivatives are all parametrically small and are given by
\be 
  \nabla \mfa  =  \nabla{\bar \mfa} + \frac{1}{\sqrt{2} m}\Bigl[ e^{-i\int_0^t \exd t'\, m(t')}  \nabla \psi   +  e^{i\int_0^t \exd t'\, m(t')}  \nabla \psi^\star \Bigr]   \,.
\ee 

Performing the coarse graining eliminates terms with unequal powers of $\psi$ and $\psi^\star$ and so gives
\bea
  \Bigl\langle V_{\mfa\,{\rm eff}}  \Bigr \rangle &=&  \ol{V}_{\mfa\,{\rm eff}}(\phi,n_i)  + \tfrac12 \, m_{\mfa\,{\rm eff}}^2(\phi, n_i) \, (\ol\mfa- \mfa_\ad)^2 + \tfrac12 \, \left( \frac{ m^2_{\mfa\,{\rm eff}}}{m^2} \right)  \, \psi^\star \psi  \nn\\
  &=&  \ol{V}_{\mfa\,{\rm eff}}(\phi,n_i)   + \tfrac12 \, m_{\mfa\,{\rm eff}}^2(\phi, n_i) \, (\ol\mfa- \mfa_\ad)^2 + \tfrac12 \, \ol W^2 \,\left( \frac{m^2_{\mfa\,{\rm eff}}}{\ol m^2_{\mfa\,{\rm eff}}} \right) \psi^\star \psi  \,,  
\eea
%
%
%\be
%  \Bigl\langle V_{\mfa\,{\rm eff}}  \Bigr \rangle =  V_{\mfa\,{\rm in}}(t) + \tfrac12 \, W^2(\phi,\ol n_i) \, \psi^\star \psi  + A(\phi) \sum_i \frac{m_{i0}}{2\Lambda_i^2}  (n_i -\ol n_i) \left[ (\mfa_\ad-\mfa_i)^2  + \frac{1}{m^2} \, \psi^\star \psi \right] \,,  
%\ee
%
and\footnote{We drop mixed time and space derivatives because in later sections we study metrics for which $g_{ti} = 0$.}
\be
  \Bigl\langle g^{\mu\nu} \partial_\mu \mfa \, \partial_\nu \mfa \Bigr\rangle  =   g^{\mu\nu} \partial_\mu \ol \mfa \, \partial_\nu \ol \mfa + g^{tt} \psi^\star \psi + \frac{g^{tt}}{m} \Bigl( i \psi^\star  \dot \psi -i \dot{\psi}^\star \psi \Bigr)    + \frac{g^{ij}}{m^2} \partial_i \psi^\star \, \partial_j \psi   + \cdots \nn
\ee
and so
\bea \label{Lfluid0}
  \Bigl\langle \cL_{\rm ax} \Bigr \rangle &=&  -   \sqrt{-g}\Bigl[ \tfrac12 \, W^2 \Bigl\langle g^{\mu\nu} \partial_\mu \mfa \, \partial_\nu \mfa \Bigr\rangle +  \Bigl\langle V_{\mfa\,{\rm eff}}  \Bigr \rangle  \Bigr] \nn\\
  &=&  -   \sqrt{-g}\left[ \tfrac12 W^2 g^{\mu\nu} \partial_\mu \ol \mfa \, \partial_\nu \ol \mfa  +  \ol{V}_{\mfa\,{\rm eff}}(\phi,n_i) + \tfrac12 \, m_{\mfa\,{\rm eff}}^2(\phi, n_i) \, (\ol\mfa- \mfa_\ad)^2  \phantom{\frac12} \right.  \\
  && \quad\left.  +\tfrac12 \left( W^2 g^{tt}+ \frac{\ol W^2 m^2_{\mfa\,{\rm eff}}}{\ol m^2_{\mfa\,{\rm eff}}} \right) \psi^\star \psi  +\frac{W^2}{2m}  g^{tt} \Bigl( i \psi^\star  \dot \psi -i \dot{\psi}^\star \psi \Bigr)  
   +  \frac{W^2}{2m^2} g^{ij} \partial_i \psi^\star \, \partial_j \psi+ \cdots \right]. \nn
\eea
The $\ol\mfa$-dependent terms of this action describe the corrections to the adiabatic approximation, $\ol\mfa = \mfa_\ad + \delta \ol\mfa$, such as \pref{particint} in the full theory before coarse-graining. The $\psi$-dependent terms describe the slow evolution of the oscillating part of the field. 

The action $S_{\mfa\,{\rm eff}} = \int \exd^4x \langle \cL_{\rm ax} \rangle$ built from this lagrangian can be varied with respect to $g_{\mu\nu}$, $\phi$ and $\psi$ to give the contribution of axions to the metric, dilaton and background axion evolution once the fast axion oscillations are integrated out. For instance
\be  \label{AxionFluidTmn}
T^{\mu\nu}_{\mfa\,{\rm eff}} = W^2(\phi)  \Bigl\langle \partial^\mu \mfa \, \partial^\nu \mfa - \tfrac12 \, g^{\mu\nu} \, (\partial \mfa)^2 \Bigr\rangle - g^{\mu\nu} \Bigl\langle V_{\mfa\,{\rm eff}}(\mfa)  \Bigr\rangle = \frac{2}{\sqrt{-g}} \, \frac{\delta S_{\mfa\,{\rm eff}}}{\delta g_{\mu\nu}} \,,
\ee
captures the axion-dependence of the Einstein equations, including  the axion-dependence of the masses on  the cosmological nonrelativistic fluids. Differentiating \pref{Lfluid0} with respect to $\phi$ similarly reproduces the coarse-grained version of the right-hand side of the dilaton equation \pref{dileq}:
\be
    (WW') \Bigl\langle (\partial \mfa)^2 \Bigr\rangle + \Bigl\langle \partial_\phi V_{\rm eff} \Bigr\rangle = - \frac{1}{\sqrt{-g}} \, \frac{\delta S_{\mfa\,{\rm eff}}}{\delta \phi} 
    %= (WW') \Bigl\langle (\partial \mfa)^2 \Bigr\rangle  +  V'_{\rm dil}(\phi) -  \frac{\beta \rho_{\rm nr}}{\MPL} 
    \,.
\ee

Inhomogeneous fields are then described by expanding the resulting field equations about their homogeneous solutions, as for any other fluid, and (at leading order) dropping terms that do not contribute to the evolution of linear fluctuations. Before doing so, however, it is useful to re-express the axion evolution in terms of an approximate conservation law since this is convenient when deriving the evolution equations for the cosmological fluid. It also suggests recasting the variable $\psi$ in a more directly physical way.  

\subsubsection{Axion dynamics and Madelung variables}

Since averaging over fast oscillations removes all terms involving unequal powers of $\psi$ and $\psi^\star$ the lagrangian \pref{Lfluid0} enjoys an emergent approximate global symmetry, $\psi \to e^{i \alpha} \psi$. Part of the information in the field equation for $\psi$ can be traded for the conservation of the corresponding Noether current, in much the same way as the field equation for $\mfa$ can be regarded as expressing Noether's theorem for the current for the underlying axionic shift symmetry.  

The conserved current density for this symmetry is $J^\mu = i \left[ \psi^\star ({\delta S}/{\delta \partial_\mu \psi^\star}) - ({\delta S}/{\delta \partial_\mu \psi}) \psi \right]$ and so the charge and spatial current densities
\be
  J^t  = - \sqrt{-g}\, g^{tt} \frac{W^2 }{m} \psi^\star \psi \qquad \hbox{and} \qquad
  J^i =  \sqrt{-g}\, g^{ij} \frac{W^2}{2m^2}  \Bigl[ i (\partial_j \psi^\star )\psi  -i\psi^\star (\partial_j \psi ) \Bigr]  \,,
\ee
satisfy 
\be \label{conseq}
  \partial_t J^t + \partial_i J^i = 0 \,. 
\ee
This suggests defining the fluid-like Madelung variables \cite{Madelung:1927ksh}
\be \label{Madelung}
\psi(x) :=  \sqrt{\varrho_\mfa(x)} \; e^{iS(x)} \,,
\ee
where $\varrho_\mfa$ parameterizes the axion fluid density and spatial gradients of $S$ will turn out to define the fluid velocity. With these definitions the current components become 
\be \label{JvsrhoS}
  J^t  = - \sqrt{-g}\, g^{tt} \frac{W^2 }{m}  \varrho_\mfa \qquad \hbox{and} \qquad
  J^i =  \sqrt{-g}\, g^{ij}\frac{W^2 \varrho_\mfa}{m^2} \, \partial_j S  \,.
\ee

In terms of the new variables the lagrangian density \pref{Lfluid0} becomes
\bea \label{Lfluid}
  \Bigl\langle \cL_{\rm ax} \Bigr \rangle   &=&  -   \sqrt{-g}\left[ \tfrac12 W^2 g^{\mu\nu} \partial_\mu \ol\mfa \, \partial_\nu \ol \mfa  +  \ol{V}_{\mfa\,{\rm eff}}(\phi,n_i) + \tfrac12 \, m_{\mfa\,{\rm eff}}^2(\phi, n_i) \, (\ol\mfa- \mfa_\ad)^2  - \frac{W^2 \varrho_\mfa}{m}  g^{tt} \dot S \phantom{\frac\Phi2} \right. \nn \\
  && \quad \left.  +\tfrac12 \left( W^2 g^{tt}+ \frac{\ol W^2 m^2_{\mfa\,{\rm eff}}}{\ol m^2_{\mfa\,{\rm eff}}} \right) \varrho_\mfa 
   +  \frac{W^2}{2m^2} g^{ij} \left( \varrho_\mfa \partial_i S \, \partial_j S + \frac{\partial_i \varrho_\mfa \, \partial_j \varrho_\mfa}{4\varrho_\mfa} \right) + \cdots \right] , 
\eea
so the field equation obtained by varying $S$ is just the conservation law \pref{conseq} once \pref{JvsrhoS} is used. The equation obtained by varying $\varrho_\mfa$ on the other hand gives the `Hamilton-Jacobi' equation 
\be \label{SdotEq}
  \frac{W^2}{m} g^{tt} \dot S =\tfrac12 \left( W^2 g^{tt}+ \frac{\ol W^2 m^2_{\mfa\,{\rm eff}}}{\ol m^2_{\mfa\,{\rm eff}}} \right) + \frac{W^2}{2m^2} g^{ij} \partial_i S \, \partial_j S - \frac{W^2}{8m^2 \varrho_\mfa^2} g^{ij } \partial_i \varrho_\mfa \, \partial_j \varrho_\mfa - D_i \left( \frac{W^2}{4m^2} g^{ij} \, \frac{\partial_j \varrho_\mfa}{\varrho_\mfa} \right)
\ee
from which the evolution of the axion fluid velocity $\vec v_\mfa$ is obtained below. Here $D_i$ denotes the covariant derivative built using the Christoffel symbols of the spatial metric $g_{ij}$.

\subsection{A useful class of metrics}

For practical applications -- such as to cosmology and to the solar system -- we now restrict to a special class of metrics. In later sections we specialize further to small perturbations around cosmological spacetimes. Consider therefore a metric of the following form: 
\be  \label{MetricForm}
\exd s^2= - \exd t^2 \Bigl[1+2\Phi(x) \Bigr] + a^2(t) \Bigl[ 1-2\Psi(x)  \Bigr] \delta_{ij} \, \exd  x^i \, \exd  x^j \,,
\ee
In our later applications we take $\Phi, \, \Psi \ll 1$ and explore perturbations about a spatially flat FRW metric, but we do not do so immediately. For solar-system applications we take $a = 1$.

\subsubsection{Current conservation}

For these choices the current \pref{JvsrhoS} becomes
%
%\be \label{noether0}
% J^t  = a^3 (1-\Phi-3\Psi) \frac{W^2 }{m}  \varrho_\mfa \qquad \hbox{and} \qquad
%  \vec J =  a(1+\Phi-\Psi)\frac{W^2 \varrho_\mfa}{m^2} \, \nabla S \,,
%\ee
%
%
\be \label{noether0}
 J^t  = \frac{a^3 (1+\Phi-3\Psi) W^2  \varrho_\mfa}{m(1+2\Phi)}  \qquad \hbox{and} \qquad
  \Jvec =  \frac{a(1+\Phi-3\Psi) W^2 \varrho_\mfa}{m^2(1-2\Psi)} \, \nabla S \,,
\ee
where spatial indices are now raised, lowered and contracted using the flat metric $\delta_{ij}$. Current conservation then implies the axion `number density' and fluid velocity
\be \label{navadefs}
  n_\mfa := \frac{W^2\varrho_\mfa }{m} \qquad \hbox{and} \qquad
    \vvec_\mfa :=  \frac{ \nabla S}{a m} \,,
\ee
satisfy
\be \label{FluidCons}
\partial_t\left[ \left( \frac{ 1+\Phi-3\Psi }{1+2\Phi}\right) n_\mfa \right] + 3  H  n_\mfa + \frac{1}{a} \nabla \cdot \left[  \left( \frac{ 1+\Phi-3\Psi }{1-2\Psi}\right)  n_\mfa \vvec_\mfa \right] = 0,
\ee
where (as usual) $H = \dot a/a$ and we use the expressions for $J^\mu$ implied by eqs.~\pref{JvsrhoS} and \pref{navadefs}. The definitions \pref{navadefs} are motivated by the observation that they make eq.\pref{FluidCons} resemble the conservation equation for a fluid with a conserved number density $n_\mfa$ and velocity $\vvec_a$. 

Specialized to a homogeneous configuration $\Phi = \Psi = 0$ and $n_\mfa = \ol m_\mfa(t)$ the conservation law \pref{FluidCons} implies the background density satisfies 
\be \label{olnaevo}
  \ol n_\mfa = \frac{C}{a^3} \,,
\ee
for constant $C$. In the next section we compute the energy density for this axion fluid and show that its evolution agrees with the behaviour derived microscopically in (\ref{bgrho}) for the energy density in homogeneous oscillations, showing how these are captured in the fluid language by the background evolution. 

For weak gravitational fields eqs.~\pref{noether0} can be linearized in $\Phi$ and $\Psi$ to become
\be \label{noether i}
  J^t  \simeq a^3 (1-\Phi-3\Psi) \, n_\mfa \quad \hbox{and} \quad
 \Jvec  \simeq %a^2(1+\Phi-\Psi) \frac{W^2}{m}   \varrho_\mfa \vec v_{\mfa }  = 
% a^2(1+\Phi-\Psi) \, n_\mfa \vec v_{\mfa } \qquad \hbox{(weak gravity)} \,.
 a^2(1+\Phi-\Psi) \, n_\mfa \vvec_{\mfa } \qquad \hbox{(weak gravity)} \,
\ee
where the metric perturbations contribute to a small change in current and density. 
\subsubsection{Euler equation}

Using the metric \pref{MetricForm} in \pref{SdotEq} (after multiplying through by $m g_{tt}/W^2$) gives the evolution equation
\bea \label{HamiltonJacobi}
   \dot S &=& - m\left[ \Phi + \tfrac12 (1+2\Phi)  \left( \frac{\ol W^2 m^2_{\mfa\,{\rm eff}}}{W^2  \ol m^2_{\mfa\,\eff}} - 1\right) \right] \\
   && \qquad  - \frac{ 1}{2ma^2} \left( \frac{ 1+2\Phi}{1-2\Psi} \right) \left\{ \nabla S \cdot \nabla S   - \frac{\nabla \cdot \left[ W^2 \sqrt{(1+2\Phi)(1-2\Psi)}  (\nabla \sqrt{\varrho_\mfa}) \right] }{ W^2\sqrt{\varrho_\mfa}\sqrt{(1 + 2\Phi)(1-2\Psi)}} \right\} \,,\nn
\eea
where we recall the definition $m = \ol m_{\mfa\,\eff}/\ol W$ made in eq.~\pref{mdef}. Eq.~\pref{HamiltonJacobi} allows $\dot S$ to be eliminated in future expressions. The explicit expression for $m^2_{\mfa\,\eff}/\ol m^2_{\mfa\,\eff}$ appearing here is 
\be
 \frac{  m^2_{\mfa\,{\rm eff}}}{  \ol m^2_{\mfa\,{\rm eff}}} = \frac{1 + (\rho_{e0}/\rho_{e\,\thr}) + (\rho_{m0}/\rho_{m\,\thr})}{1 + (\ol\rho_{e0}/\rho_{e\,\thr}) + (\ol\rho_{m0}/\rho_{m\,\thr})}  \,,
\ee
with $\rho_{i0} := A(\phi) \, m_{i0} n_i$ as defined just below eq.~\pref{Veffdef2ax} and we recall $\rho_{i\,\thr} = m_\mfa^2 \Lambda_i^2$. In appendix \ref{thephase} we use this to show the relationship between the phase of the axion field and the energy of axion particles recovers the standard expression expected for particles in a galactic halo.

Taking the gradient of \pref{HamiltonJacobi} gives the Euler equation governing the evolution of the axion fluid velocity, $\vvec_\mfa$. Written in terms of the fluid momentum 
\be 
\pvec_\mfa = m(t) \, \vvec_\mfa \,,
\ee
this reads
\be \label{AxionEuler}
  \partial_t \pvec_\mfa + H \, \pvec_\mfa = - \frac{m}{a} \nabla \left\{ \Phi + \tfrac12 (1 + 2\Phi) \left[\left( \frac{\ol W^2 m^2_{\mfa\,{\rm eff}}}{W^2 \ol m^2_{\mfa\,{\rm eff}}} \right)- 1 \right] +  \tfrac12 v_\mfa^2  \left( \frac{ 1+2\Phi}{1-2\Psi} \right)  + \Phi_\ssQ \right\} \,,
\ee
where $v_\mfa$ is the modulus of $\vvec_\mfa$ and we define the `quantum pressure' $\Phi_\ssQ$ by
\be \label{PhiQdef}
 \Phi_\ssQ  :=  - \frac{ 1}{2m^2a^2} \left( \frac{ 1+2\Phi}{1-2\Psi} \right)  \frac{\nabla \cdot \left[ W^2 \sqrt{(1+2\Phi)(1-2\Psi)}  \; (\nabla \sqrt{\varrho_\mfa}) \right] }{ W^2 \sqrt{\varrho_\mfa}  \sqrt{(1+2\Phi)(1-2\Psi)}}  \,.
\ee

\subsubsection{Axion stress-energy}\label{axionstressenergy}

We next collect expressions for the axion fluid's stress energy -- {\it c.f.}~\pref{AxionFluidTmn} --- in terms of the fluid variables. Writing (as above) $\langle \cL_{\rm ax} \rangle = \sqrt{-g} \; \cP$ and specializing \pref{Lfluid} to a homogeneous background $\ol\mfa(t)$ and to the metric \pref{MetricForm} gives
\bea \label{cPfluid}
 \cP 
  &=&    \frac{W^2 \dot{\ol\mfa }^2}{2(1+2\Phi)}  -  \frac{W^2 (\nabla{\ol\mfa })^2}{2(1-2\Psi)}  -  \ol{V}_{\mfa\,{\rm eff}}(\phi,n_i) - \tfrac12 \, m_{\mfa\,{\rm eff}}^2(\phi, n_i) \, (\ol\mfa- \mfa_\ad)^2  - \frac{W^2 \varrho_\mfa}{m (1+2\Phi)}   \dot S  \nn  \\
  && \quad   +\tfrac12 \varrho_\mfa  \left( \frac{W^2  }{1+2\Phi} -  \frac{\ol W^2 m^2_{\mfa\,{\rm eff}}}{\ol m^2_{\mfa\,{\rm eff}}} \right)
   -  \frac{W^2}{2m^2a^2 (1- 2\Psi)}  \left( \varrho_\mfa \nabla S \cdot \nabla S + \frac{\nabla \varrho_\mfa \cdot \nabla \varrho_\mfa}{4\varrho_\mfa} \right)  \nn\\
 &=&    \frac{W^2 \dot{\ol\mfa }^2}{2(1+2\Phi)}  -  \frac{W^2 (\nabla{\ol\mfa })^2}{2(1-2\Psi)}  -  \ol{V}_{\mfa\,{\rm eff}}(\phi,n_i) - \tfrac12 \, m_{\mfa\,{\rm eff}}^2(\phi, n_i) \, (\ol\mfa- \mfa_\ad)^2      \\
  && \qquad\qquad\qquad\qquad\qquad\qquad\qquad\qquad\qquad    -  \frac{ \nabla \cdot \left[ W^2 \sqrt{(1+2\Phi)(1-2\Psi)}  \;  \nabla \varrho_\mfa  \right]}{ 4m^2a^2 \sqrt{(1 + 2\Phi)(1-2\Psi)^3}}    \,,\nn
\eea
where the second equality eliminates $\dot S$ using \pref{HamiltonJacobi} and $\ol V_{\mfa\,\eff}(\phi, n_i)$ is given by \pref{Vaeffmin}. 

This expression is useful when evaluating the energy density, $\rho_\ax :=   - {T_t^{t}}_{\mfa(\rm eff)}=   - g_{tt}T^{tt}_{\mfa(\rm eff)}$, of the axion fluid. Using \pref{AxionFluidTmn} this becomes
\bea \label{axionenergy}
\rho_\ax  &=& - g^{tt} \left\{ W^2 \dot{\bar \mfa}^2 + W^2 \psi^\star \psi  + \frac{W^2}{2m} \left[ i \psi^\star  \left( \dot \psi - \frac{\dot m}{m}  \psi \right) + \hbox{h.c.} \right] \right\}   -{\cal P} \nn\\
&=& \frac{1}{1+2\Phi} \left\{ W^2 \dot{\bar \mfa}^2 + W^2 \varrho_\mfa  - \frac{W^2\varrho_\mfa}{m} \dot S\right\}   -\cP \\
 &=&    \frac{W^2 \dot{\ol\mfa }^2}{2(1+2\Phi)}  + \frac{W^2 (\nabla{\ol\mfa })^2}{2(1-2\Psi)} +  \ol{V}_{\mfa\,{\rm eff}}(\phi,n_i) + \tfrac12 \, m_{\mfa\,{\rm eff}}^2(\phi, n_i) \, (\ol\mfa- \mfa_\ad)^2  + \frac{W^2  \varrho_\mfa v_\mfa^2 }{2 (1- 2\Psi)}     \nn  \\
  && \quad + \frac{\varrho_\mfa W^2(1+\Phi)}{(1+2\Phi)} + \tfrac12 \, \varrho_\mfa W^2  \left[ \left( \frac{\ol W^2 m^2_{\mfa\,{\rm eff}}}{W^2\ol m^2_{\mfa\,{\rm eff}}} \right) - 1\right] 
  + \frac{W^2\nabla \varrho_\mfa \cdot \nabla \varrho_\mfa}{8m^2a^2\varrho_\mfa (1- 2\Psi)}   \nn\,.
\eea
The first four terms of this expression describe the energy of the background evolution $\ol\mfa$ while the rest capture the energy of the fluid describing fast axion oscillations. Notice in particular that for homogeneous backgrounds (for which gradients and $\vvec_\mfa$ vanish) in the absence of a gravitational field the fluid part of the energy becomes
\be \label{fluidbackgroundE}
   \ol\rho_{f} = \ol \varrho_\mfa \ol W^2  = m(t) \, \ol n_\mfa  = \frac{C m(t)}{a^3} \,,
\ee 
where the second equality uses \pref{navadefs} for $n_\mfa$ and the last equality uses current conservation in the form given in \pref{olnaevo}. As advertised, this precisely captures the energy of the underlying homogeneous oscillation given in \pref{OscillationEvolution}. Notice also that \pref{fluidbackgroundE} implies the usual matter-like dependence $\ol\rho_f \propto a^{-3}$ in the late universe when the axion mass is time-independent, but instead implies $\ol\rho_f \propto a^{-9/2}$ in the earlier universe when $m \propto \sqrt{\rho_{m0}} \propto a^{-3/2}$. 
 
The other components of the energy momentum-tensor are obtained in a similar way. Eq.~\pref{AxionFluidTmn} reveals the energy flux/momentum density is given by
\be 
  U^i_\ax :=  T^{ti}_{\mfa(\rm eff)} = W^2  g^{tt} g^{ij} \Bigl\langle \dot\mfa \, \partial_j \mfa \Bigr\rangle =   W^2   g^{tt} g^{ij} \left[  \dot{\ol\mfa} \, \partial_j \ol\mfa +  \frac{1}{2m} \Bigl(i \psi^\star \partial_j \psi + \hbox{h.c.} \Bigr) \right]  \,.
\ee
The fluid part of this expression takes a familiar form once expressed in terms of the potentials $\Phi$ and $\Psi$ and the fluid variables:
\be
   \bfU_\ax = \frac{W^2}{a^2(1+2\Phi)(1-2\Psi)} \left( \frac{ \varrho_\mfa \nabla S}{m} \right) =  \frac{W^2\varrho_\mfa \vvec_\mfa}{a(1+2\Phi)(1-2\Psi)}    \,.
\ee

The shear tensor similarly becomes 
\bea  \label{Shear}
T^{ij}_{\mfa\,{\rm eff}} &=& W^2 g^{ik} g^{jl}   \left[ \partial_k \ol \mfa \, \partial_l \ol \mfa + \frac{1}{2m^2} \Bigl( \partial_k \psi^\star \, \partial_l \psi + \hbox{h.c.} \Bigr) \right] + g^{ij} \cP \nn\\
&=& \frac{W^2}{a^4(1-2\Psi)^2}  \left[ \partial^i \ol \mfa \, \partial^j\ol \mfa + \frac{1}{m^2} \left(\varrho_\mfa \, \partial^i S \, \partial^j S + \frac{\partial^i \varrho_\mfa \,  \partial^j \varrho_\mfa}{4\varrho_\mfa^2} \right)    \right] + \frac{\cP\delta^{ij}}{a^2(1-2\Psi)}  \,,\\
&=& \frac{W^2}{a^4(1-2\Psi)^2}  \left[ \partial^i \ol \mfa \, \partial^j \ol \mfa + a^2 \varrho_\mfa v_{\mfa}^i  v_{\mfa}^j + \frac{\partial^i \varrho_\mfa \,  \partial^j \varrho_\mfa}{4m^2\varrho_\mfa^2}     \right] + \frac{\cP\delta^{ij}}{a^2(1-2\Psi)}  \,,\nn
\eea
with its characteristic dependence on $\varrho_\mfa v^i_\mfa v^j_\mfa$ when the fluid is incompressible.

\subsection{Linear perturbations}

We next delve into the wonderous world of cosmological perturbations by linearizing the above expressions about a homogeneous cosmological background solution. To this end we regard both gravitational potentials $\Phi$ and $\Psi$ to be small and restrict to dilaton and matter configurations that are similarly linearized around homogeneous backgrounds
\be \label{phinifluct}
   \phi = \ol\phi(t) + \delta \phi(x) \qquad \hbox{and} \qquad n_i = \ol n_i(t) + \delta n_i(x) \,.
\ee

We perform a similar split for the slowly evolving axion oscillations, $\psi = \ol\psi + \delta \psi$, by choosing 
\be
   n_\mfa = \ol n_\mfa + \delta n_\mfa \qquad \hbox{and so} \qquad
    \varrho_\mfa = \ol\varrho_\mfa(t) \, \Bigl[ 1 + \delta_\mfa(x) \Bigr] 
\ee
and assuming the axion fluid velocity $\vvec_\mfa$ vanishes in the background. If interested in post-Newtonian solar-system applications a natural choice for the size of the fluid speed is $v_\mfa^2 \sim \Phi$ but in cosmology we instead drop $v_\mfa^2$ terms relative to linearized fluctuations. (In the remainder of this section we keep the $v_\mfa^2$ terms to keep our discussion general, but drop them in our later numerical evolution.)  

We assume that the background evolution is well-described by adiabatic evolution -- as is justified because subdominant terms in $H/\ol m_\ax$ are negligible -- and so $\ol \mfa = \mfa_\ad$ everywhere. This implies that $\ol\mfa$ inherits the fluctuation structure \pref{phinifluct} of the fields $\phi$ and $n_i$, with
\bea \label{adiabaticgrad}
  \ol\mfa(x) = \mfa_\ad(x) & =&  \ol \mfa_\ad(t) + \frac{  \frac{\delta \rho_{e0} }{\rho_{e\,\thr} } \left[ \aminuse + \frac{\ol\rho_{m0}}{\rho_{m\,\thr}}(\aminuse - \aminus) \right] + \frac{\delta \rho_{m0}}{\rho_{m\,\thr}} \left[ \aminus + \frac{\ol\rho_{e0}}{\rho_{e\,\thr}}( \aminus - \aminuse) \right]    }  {\left(1 +   \frac{\ol\rho_{e0} }{\rho_{e\,{\rm th}}}   + \frac{\ol\rho_{m0} }{\rho_{m\,{\rm th}}}  \right)^2 }\nn\\
    &\simeq&  \ol \mfa_\ad(t) + \frac{  \left( \frac{\delta \rho_{m0}}{\rho_{m\,\thr}} \right)  \aminus     }  {\left(1   + \frac{\ol\rho_{m0} }{\rho_{m\,{\rm th}}}  \right)^2 } \qquad \hbox{(when $\rho_{e0}/\rho_{e\,{\rm th}}$ is negligible)}\,,
\eea   
where we recall $\rho_{i\,\thr} = m_\mfa^2 \Lambda_i^2$ and note that $\delta \rho_{i0}$ is given in terms of $\delta \phi$ and $\delta n_i$ by
\be \label{deltarhoi}
   \delta \rho_{i0} = \left( - \beta \delta \phi + \frac{\delta n_i}{\ol n_i} \right) \ol \rho_{i0} \,.
\ee
The same is also true for the axion mass, which becomes
\bea
 \frac{  m^2_{\mfa\,{\rm eff}}}{  \ol m^2_{\mfa\,{\rm eff}}} &=& \frac{1 + (\rho_{e0}/\rho_{e\,\thr}) + (\rho_{m0}/\rho_{m\,\thr})}{1 + (\ol\rho_{e0}/\rho_{e\,\thr}) + (\ol\rho_{m0}/\rho_{m\,\thr})} \simeq 1 + \frac{\delta \rho_{e0}}{\rho_{e\,\thr}} + \frac{\delta \rho_{m0}}{\rho_{m\,\thr}} + \cdots \nn\\
 &\simeq& 1  + \frac{\delta \rho_{m0}}{\rho_{m\,\thr}}  \qquad \hbox{(when $\rho_{e0}/\rho_{e\,{\rm th}}$ is negligible)} \,,
\eea

The expansion of the current \pref{noether0} then is $J^\mu = \ol J^\mu + \delta J^\mu$ with
\be
   \ol J^t = a^3 \ol n_\mfa \,, \quad %\hbox{and} \quad
   \delta J^t = a^3 \Bigl[ \delta n_\mfa - \ol n_\mfa (\Phi + 3\Psi) \Bigr] \,,
\ee
and
\be
   \ol J^i = 0 \,, \quad %\hbox{and} \quad 
   \delta J^i =  a^2  \, \ol n_\mfa  v^i_{\mfa }\,,
\ee
to linear order in the fluctuations. Current conservation therefore implies both $\ol n_\mfa \propto a^{-3}$ and
\be
   \partial_t \Bigl\{  a^3 \Bigl[ \delta n_\mfa - \ol n_\mfa (\Phi + 3\Psi) \Bigr]\Bigr\} + a^2 \ol n_\mfa \nabla \cdot \vvec_\mfa = 0 \,.
\ee 

Keeping lowest nontrivial order in the fluctuations in the Hamilton-Jacobi equation \pref{HamiltonJacobi} gives
\be\label{dotS}
 \dot S \simeq - m(t) \Bigl[  \tfrac12 v_\mfa^2 + \Phi + \Phi_\phi + \Phi_\rho + \Phi_\ssQ \Bigr] \,,
\ee
where the time-dependent mass appearing here is
\be
  m(t) = \frac{\ol m_{\mfa\,\eff}}{\ol W} = \frac{m_\mfa}{\ol W} \left( 1 + \frac{\ol \rho_{e0}}{\rho_{e\,\thr}} + \frac{\ol \rho_{m0}}{\rho_{m\,\thr}} \right) \,,
\ee
and we define the generalized potentials
\be 
  \Phi_\phi :=  -\frac{W' }{W} \delta \phi\,, \quad
  \Phi_\rho := \tfrac12  \left(  \frac{\delta \rho_{e0}}{\rho_{e\,\thr}} + \frac{\delta \rho_{m0}}{\rho_{m\,\thr}} \right) \,,
\ee
and the quantum pressure $\Phi_\ssQ$ defined in \pref{PhiQdef} linearizes to:
\be  \label{PhiQlin}
 \Phi_\ssQ  \simeq    - \frac{1}{2m^2a^2 }  \left[ \frac{\nabla^2 \sqrt{\varrho_\mfa}}{\sqrt{\varrho_\mfa}}    \right]   =   \frac{1}{4m^2a^2 }   \left[- \frac{\nabla^2 \varrho_\mfa}{\varrho_\mfa} + \frac{(\nabla \varrho_\mfa)^2}{2\varrho_\mfa^2}    \right]    \,.
\ee

In terms of these potentials the relationship between $\delta n_\mfa$ and $\delta_\mfa$ obtained from \pref{navadefs} is
\be
   \delta n_\mfa = \ol n_\mfa \left( \delta_\mfa + \frac{\delta W^2}{\ol W^2}  \right) = \ol n_\mfa \left( \delta_\mfa - 2 \Phi_\phi \right) 
\ee
and the evolution of the axion fluid velocity is given by the linearized Euler equation \pref{AxionEuler}, which states
\be \label{AxionEulerlin}
 \partial_t  \pvec_\mfa + H \, \pvec_\mfa  = - \frac{ m}{a}  \nabla \Bigl[ \Phi + \tfrac12   v_\mfa^2 +  \Phi_\phi + \Phi_\rho + \Phi_\ssQ \Bigr]    \,,
\ee
for $\pvec_\mfa = m(t) \,\vvec_\mfa$. 

The quantities relevant to the Einstein equations are the linearized energy density and stress energy, given explicitly in eqs.~\pref{axionenergy} and \pref{Shear}. For the homogeneous background these give the background pressure 
\be
  \ol\cP_\ax  =    \tfrac12 \ol W^2 \dot{\ol\mfa }^2   -  \ol{V}_{\mfa\,{\rm eff}}(\ol\phi, \ol n_i) - \tfrac12 \, m_{\mfa\,{\rm eff}}^2(\ol\phi, \ol n_i) \, (\ol\mfa- \mfa_\ad)^2   \simeq   -  \ol{V}_{\mfa\,{\rm eff}}(\ol\phi, \ol n_i)    \,,
\ee
which only receives contributions from $\ol\mfa$ and not from the oscillatory fluid (as expected for a matter-type equation of state). The approximate equality here uses the adiabatic approximation that drops subdominant powers of $H/m$ and for which $\ol\mfa \simeq \mfa_\ad$. 

The background energy density similarly is
\be  \label{axionenergybgnd}
  \ol \rho_\ax    =     \tfrac12  \ol W^2 \dot{\ol\mfa }^2  +  \ol{V}_{\mfa\,{\rm eff}}(\ol\phi, \ol n_i) + \tfrac12 \, m_{\mfa\,{\rm eff}}^2(\ol\phi, \ol n_i) \, (\ol\mfa- \mfa_\ad)^2        + \ol\varrho_\mfa \ol W^2 \simeq     \ol{V}_{\mfa\,{\rm eff}}(\ol\phi, \ol n_i)   + \ol\varrho_\mfa \ol W^2  \,,
\ee
with the approximate equality again assuming adiabatic evolution for $\ol\mfa$. The last term of this expression can be recognized as the background fluid energy density $\ol\rho_f = \ol \varrho_\mfa \ol W^2$ given in \pref{fluidbackgroundE}. 

The fluctuation in energy density in $\ol\mfa$ and in the oscillatory fluid is obtained by linearizing \pref{axionenergy} about the background. In the adiabatic approximation the contribution from the fluctuations in $\ol\mfa = \mfa_\ad$ inherited from fluctuations in $\phi$ and $n_i$ are  
\be \label{axionenergydeltabar}
 \delta \rho_\ad   \simeq   \partial_\phi  \ol{V}_{\mfa\,{\rm eff}} \, \delta \phi + \partial_{n_i} \ol{V}_{\mfa\,\eff} \, \delta n_i \,,
\ee
The fluctuation in the fluid energy density is similarly
\be \label{axionenergydeltaf}
\delta_f := \frac{\delta \rho_f}{\ol\rho_f}  
% =      \tfrac12   v_\mfa^2  - \Phi  + \delta_\mfa - 2\Phi_\phi + \Phi_\phi + \Phi_\rho
 =      \tfrac12   v_\mfa^2 + \delta_\mfa  - \Phi  -  \Phi_\phi  + \Phi_\rho  \,.
\ee
 
We can now work at the level of linear perturbations and obtain the Euler equation in terms of the effective energy density of the axion. Using (\ref{FluidCons}) the conservation equation is
\be 
\dot{\delta}_f  -3 \dot{\Psi}= \dot{\Phi}_\phi+ \dot{\Phi}_\rho- \frac{1}{a}\Theta_a,
\label{non-con}
\ee
where $\Theta_a := \nabla \cdot \bfv_\mfa$ is the axion fluid's velocity expansion and the two terms $\dot{\Phi}_\phi+ \dot{\Phi}_\rho$ on the right-hand side have their origins in the perturbations of the axion mass: $\partial_t\left(\frac{\delta m}{m}\right)$. This reproduces the usual result in the traditional case when the potentials $\Phi_\phi$ and $\Phi_\rho$ are absent. 

We now turn to the Euler equation which reads 
\be 
{ {\dot \bfv}_\mfa}+ \left(H+ \frac{\dot m}{m}\right)\bfv_\mfa  = -\frac{1}{a} \nabla (\Phi +\Phi_\phi + \Phi_Q+\Phi_\rho) \,,
\ee
the divergence of which then gives
\be 
\dot{\Theta}_\mfa +\left({H}+ \frac{\dot{m}}{m}\right) \Theta_\mfa= -\frac{1}{a} \nabla^2(  \Phi+ \Phi_\phi + \Phi_Q).
\ee
This, together with the conservation equation, gives the growth equation for $\delta_f$ in the quasi static approximation
\be \label{axion fluid growth}
\ddot{\delta}_f +\left({H}+ \frac{\dot{m}}{m}\right) \dot{\delta}_f -\frac{1}{a}\nabla^2 ( \Phi+ \Phi_\phi + \Phi_Q+\Phi_\rho)=0,
\ee
which can be expanded once the Poisson equation and the Klein-Gordon equation have been used. Notice in particular that the Hubble friction is enhanced when the axion mass varies. Moreover gravity is modified, with the three new potentials $\Phi_\rho$, $\Phi_\phi$ and $\Phi_Q$ supplementing the gravitational potential $\Phi$. 

%qqqq
\section{Cosmological Perturbations }
\label{four}

In the previous section the focus was mainly on the axion fluid, but we now collect all the components of the Universe and their interactions within the axio-dilaton models. The full description of the cosmology is in particular obtained by specifying the coupling between the different universal fluids. In particular, since dark matter and the baryons couple to the axion fluid we wish to investigate their mutual exchange of energy and momentum.

The exchange of energy-momentum between the components of the cosmological fluid means each component is not separately conserved on its own. But for the fluids we do not have an action formulation for the equations of motion and so cannot as easily read off how their energy density changes as a function of what the scalar fields are doing. This is most simply derived by computing the rate with which the scalar sector loses or emits energy as a function of the fluids and then using the overall conservation of stress energy, as guaranteed by the Bianchi identities as applied to the Einstein equation
\be \label{Einstein eq}
G_{\mu\nu}= 8 \pi G (T^\phi_{\mu\nu}+ T^\mfa_{\mu\nu}+ T^{\ttl}_{\mu\nu}),
\ee
to infer how much the fluid sector absorbs or emits, as we now argue in detail. 

The right-hand side of \pref{Einstein eq} separates out the parts of the slowly moving scalar energy-momentum tensors that are independent of the presence of matter, with
\be
T^\phi_{\mu\nu}= \partial_\mu \phi \, \partial_\nu \phi -g_{\mu\nu}\Bigl[ \tfrac12 \, (\partial \phi)^2 +V_{\rm dil} (\phi)\Bigr],
\ee
and
\be 
T^\mfa_{\mu\nu}= W^2(\phi) \partial_\mu \mfa \, \partial_\nu \mfa -g_{\mu\nu} \Bigl[ \tfrac12 \, W^2(\phi) \, (\partial \mfa)^2 + V_{\rm ax}(\mfa) \Bigr] \,,
\ee
where $V_{\rm dil}$ and $V_{\rm ax}$ are as defined in and below eq.~\pref{potentialdilax}. The quantity $T^\ttl_{\mu\nu}$ denotes the rest of the total stress energy including both the axion fluid and the matter-dependent couplings with the axion and dilaton. Because the scalar stress energies exclude the scalar couplings to matter they are not covariantly conserved when evaluated at the solutions to the equations of motion: neither $D^\mu T^\phi_{\mu\nu}$ or $D^\mu T^\mfa_{\mu\nu}$ vanish in general, though their nonzero values are easily computed using the scalar field equations. But the Bianchi identity $D^\mu G_{\mu\nu} = 0$ ensures the total stress energy is covariantly conserved and so $D^\mu T^\ttl_{\mu\nu} = - D^\mu T^\phi_{\mu\nu} - D^\mu T^\mfa_{\mu\nu}$, and this allows the fluid response to the scalar fields to be computed. 

\subsection{Einstein Equations}

We start by writing down the background and perturbed Einstein equations in the presence of the  coupled axion fluid, starting from (\ref{Einstein eq}). 

The background Einstein equation is the Friedmann equation for the background metric, 
\be  \label{BackgroundHubble0}
3H^2 \MPL^2 =  \frac{1}{2}\left[ \dot{\Bar{\phi}}^2+W^2(\bar\phi)\dot{\Bar{\mfa}}^2 \right] +V_{\rm dil}(\Bar{\phi})+ V_{\rm ax}(\bar{\mfa}) +\tfrac{1}{2}\Bar{\rho}_{f0}  +\Bar{\rho}_{\tot},
\ee
where $V_{\rm dil}$ and $V_{\rm ax}$ are respectively defined in \pref{VdilDef} and \pref{VaxDef}, repeated here for convenience: 
\begin{equation}
 %   V(\bar{\mfa}) = \frac{m_\mfa^2}{2}(\Bar{\mfa}-\aplus)^2.
   V_{\rm ax}(\bar{\mfa}) = \tfrac12 m_\mfa^2 \Bar{\mfa}^2,
   \qquad\hbox{and}\qquad
   V_{\rm dil}(\bar\phi) = U(\bar\phi) \, e^{-\lambda \bar\phi/\MPL }\,,
\end{equation}
and $\Bar{\rho}_\tot = \Bar{\rho}_m + \Bar{\rho}_b +\Bar{\rho}_r$ is the sum of the energy density of dark matter, baryonic matter, and additional relativistic species also present during the late time cosmology, respectively.

The quantity $\ol\rho_{f0}$ appearing in \pref{BackgroundHubble0} is not quite the same as the axion fluid density $\rho_f$ given in \pref{fluidbackgroundE}, differing because here we do not  include the matter coupling in the axion energy momentum tensor (lumping it instead into the energy density of matter). As a result it involves the mass $m_\mfa$ rather than the matter-dependent mass $\ol m_{\mfa\,{\rm eff}}(t)$:
\be
\ol\rho_{f0} :=  \Bigl\langle \tfrac12 \ol W^2 \dot\mfa^2 + \tfrac12 \, m_\mfa^2 \mfa^2 \Bigr\rangle   \,.
\ee
By contrast $\ol\rho_f$ contains the energy density of the rapid axion oscillations, with
\be
    \Bigl\langle \tfrac12 \ol W^2 \dot\mfa^2 \Bigr\rangle = \Bigl\langle \tfrac12 \, \ol m_{\mfa\,{\rm eff}}^2(t) \, \mfa^2 \Bigr\rangle = \tfrac12 \, \ol \rho_f = \tfrac12 \ol W^2 \ol \varrho_\mfa =\tfrac12  m(t) \, \ol n_\mfa \,,
\ee
and so
\be
 \ol\rho_{f0} =  \tfrac12 \, \ol \rho_f \left[1 + \frac{m_\mfa^2}{m_{\mfa\,{\rm eff}}^2(t)} \right] =  \tfrac12 \, \ol \rho_f \left(1+\frac{1}{1+\cC_e + \cC_m}\right) ,
\ee
where the last equality uses \pref{maeffdef} to evaluate $m_\mfa^2/m_{\mfa\,{\rm eff}}^2(t)$.

At the level of linear perturbations the 00--component of the Einstein equations is 
\bea
    &&\left[\frac{k^2}{a^2}\Psi+3H\dot{\Psi}\right]\MPL ^2  +\tfrac{1}{2}\dot{\Bar{\phi}}\,\delta\dot{\phi} + \tfrac{1}{2}\left\{ 2\Phi\left[ V(\phi)+V(\Bar{\mfa})+\left( W^2+\frac{m_\mfa^2}{m^2(t)}\right)\frac{\varrho_\mfa}{2}\right]+V,_\phi\delta\phi\right\} \nonumber
    \\
    && \qquad +\tfrac{1}{2}WW,_{\phi} \delta\phi \,\dot{\Bar{\mfa}}^2
    +\tfrac14 a^2W^2\dot{\delta\varrho_\mfa} \left[1+\frac{k^2}{2m^2(t)a^2}+\frac{m_\mfa^2}{m^2_{\mfa\,{\rm eff}}(\rho_m)}\right] \nn\\
    && \qquad\qquad\qquad = -\tfrac{1}{2}\left(\delta\rho_\tot+2\Phi\Bar{\rho}_\tot +\frac{W^2\delta\rho_m}{2m^2(t)\Lambda_{m}^2}\Bar{\varrho}_\mfa\right),
\eea
and the 0i--component reads
\begin{align}
    k^2(H\Phi + \dot{\Psi})\MPL ^2= \tfrac12 k^2 \dot{\Bar{\phi}} \,\delta\phi+\tfrac12 aW^2 \Bar{\varrho}_\mfa\Theta_\mfa + \tfrac12 a \Bar{\rho}_\tot\Theta_\tot.
\end{align}
These agree with the result obtained by starting from the perturbed axio-dilaton equations in \cite{Burgess:2021obw, Smith:2024ayu,  Smith:2024ibv} and averaging terms. Here $\Theta_\tot = \partial_i v^i_{\tot}$ is the divergence of the total velocity field, $\bfv_{tot}$, of all other fluids. These equations are used to solve for the evolution of cosmological perturbations. 

\subsection{Coupled axion-matter dynamics}

As mentioned earlier, the Bianchi identity $D^\mu G_{\mu \nu}=0$ allows us to derive the non-conservation equations for each fluid (with details of the calculations given in appendix \ref{the bianchi identity}). For nonrelativistic matter we can neglect the pressure and so write
\be 
T^{b}_{\mu\nu}=\rho_{\B}u_{\mu}^{\B}u_\nu^{\B} \quad \hbox{and} \quad
T^{m}_{\mu\nu}=\rho_{\C}u_{\mu}^{\C}u_\nu^{\C},
\ee
where $b$ and $m$ respectively denote ordinary matter (the combined baryon/electron fluid) and CDM. The fluid 4-velocities separately satisfy $u^2_{\B}=u^2_{\C}=-1$ and so also $u^\mu_{\B}D_\nu u_{\mu, \B}=u^\mu_{\C}D_\nu u_{\mu, \C}=0$. 
 
For Dark Matter the arguments of the appendix teach us that the Dark Matter fluid satisfies
\be \label{fluidevoeq}
\dot{\rho}_{\C} u_\nu^{\C} + 3 h_{\C} \rho_{\C} u_\nu^{\C}+ \rho_{\C}  \dot{u}_\nu^{\C} = \left(
 -\frac{\beta}{\MPL }  \partial_\nu \phi +\left\langle \frac{\partial \cU_m(\mfa)}{\partial \mfa}\partial_\nu \mfa\right\rangle \right)  T^{\C}~,
\ee
where $T^m := g^{\mu\nu} T^m_{\mu\nu}$ and we drop the axion-electron coupling $\cU_e$ in comparison to the axion-CDM coupling $\cU_m$. Here the overdot denotes $ \dot{\rho_a}=  u_a^\mu D_\mu \rho_a$ and lower-case $h_a$ denotes the local Hubble rate $3h_{a} =  D_\mu u^\mu_{a}$ for $a = m$ or $b$, emphasizing that each fluid has an individualized time derivative and experiences its own Hubble flow. The brackets $\langle \,\cdots \rangle$ denote the average over fast axion oscillations. 

Because we work in an approximation where the baryon fluid moves in lock-step with the electron fluid, the baryons satisfy an equation identical to \pref{fluidevoeq} but with $u^m_\mu \to u^b_\mu$, $\rho_m \to \rho_b$, $h_m \to h_b$ and $T^m \to T^b$ but $\cU_m \to m_e\,\cU_e/m_\ssN$ (see eq.~\pref{U_ewrho_b} for why the electron/nucleon mass ratio appears).

Contracting \pref{fluidevoeq} with $u^\nu_{\C}$ gives the rate of energy change due to the Dark Matter fluid's interaction with the scalars. For Dark matter this becomes
\be \label{DMeq1}
 \dot{\rho}_{\C}  + 3 h_{\C} \rho_{\C}= -\frac{\beta}{\MPL } \rho_{\C}  \dot{\phi} +u^\nu_{\C}\left\langle \frac{\partial \cU_m(\mfa)}{\partial \mfa}\partial_\nu \mfa\right\rangle  a\rho_{\C} \,.
\ee
Projecting \pref{fluidevoeq} onto the directions orthogonal to the fluid evolution can be done with the tensor
\be 
  h^{\C}_{\mu\nu}= g_{\mu\nu}+ u^{\C}_\mu u^{\C}_\nu
\ee
leading to the Euler equations
\be \label{DMeq2} 
 \dot{u}^\mu_{\C}- \frac{\beta  \dot{\phi}}{\MPL }u^\mu +u^\nu_{\C}\left\langle \frac{\partial \cU_m(\mfa)}{\partial \mfa}\partial_\nu \mfa\right\rangle  u^\mu_{\C}= \frac{\beta}{\MPL }  \partial^\mu \phi -\left\langle \frac{\partial \cU_m(\mfa)}{\partial \mfa}\partial^\mu \mfa\right\rangle .
\ee
The average over the rapid oscillations in these equations gives the following expression in terms of the slow axion fluid variables
\be \label{DMeq0}
  \left\langle \frac{\partial \cU_m(\mfa)}{\partial \mfa}\partial_\mu \mfa\right\rangle = \frac{\bar \mfa -\aminus }{\Lambda_{m}^2}\partial_\mu{\bar\mfa}  + \frac{1}{2\Lambda_{m}^2}  \partial_\mu{ \left( \frac{\rho_\mfa}{m^2(t)}\right)},
\ee
where the total density is involved through the axion adiabatic solution's dependence on $\rho_m$ and $\rho_e$. 

Again an almost identical line of argument goes through for the baryon fluid, with the counterparts to \pref{DMeq1} and \pref{DMeq2} obtained by making the substitutions $u^m_\mu \to u^b_\mu$, $h_m \to h_b$, $\rho_m \to \rho_b$ and $\cU_m \to  m_e \, \cU_e/m_\ssN$. Once this is done the analogue of \pref{DMeq0} then is
\be \label{DMeq0e}
  \left\langle \frac{\partial \cU_e(\mfa)}{\partial \mfa}\partial_\mu \mfa\right\rangle = \frac{\bar \mfa -\aminus }{\Lambda_e^2}\partial_\mu{\bar\mfa}  + \frac{1}{2\Lambda_e^2}  \partial_\mu{ \left( \frac{\rho_\mfa}{m^2(t)}\right)} \,.
\ee

The above equations -- together with the Einstein and scalar-field equations -- are the fully relativistic equations of the baryon-CDM-axion fluid system. 

\subsection{Linear perturbations}

We next linearize these equations to compute the evolution of small inhomogeneous fluctuations around a homogeneous background. At the background level both $h_{\B}$ and $h_\C$ are simply given by the background Hubble scale $H$ and time derivatives for the two fluids coincide with the cosmic time derivatives, leading to 
\be 
\dot{\bar\rho}_{\C}+ 3H \bar \rho_{\C}= \left[ -\frac{\beta}{\MPL } \dot{\bar\phi}+ \frac{\bar \mfa-\mfa_m}{\Lambda_{m}^2}  \dot{\bar\mfa} + \frac{1}{2\Lambda_{m}^2} \partial_t{ \left( \frac{\bar \varrho_\mfa}{m^2(t)}\right)}\right] \bar\rho_{\C},
\ee
and 
\begin{equation} \label{baryon continuity}
\dot{\bar\rho}_{\B}+ 3H \bar \rho_{\B}= \left\{ -\frac{\beta}{\MPL } \dot{\bar\phi}+\frac{m_e}{m_\ssN} \left[ \frac{\bar \mfa-\mfa_m}{\Lambda_{e}^2}  \dot{\bar\mfa} + \frac{1}{2\Lambda_{e}^2} \partial_t{ \left( \frac{\bar \varrho_\mfa}{m^2(t)}\right)} \right]\right\} \bar\rho_{\B},
\end{equation}
where we assume $m_e/m_\ssN \ll 1 $. We note in passing that these equations are unusual inasmuch as they depend on $\Bar{\mfa}(t) = \Bar{\mfa}(\Bar{\rho}_m,\Bar{\rho}_e)$, that depends only on the cold dark matter density and the electron density since the axion does not couple directly to protons and neutrons. In principle there are three conservation equations, one each for CDM, for the electrons and for the baryons, but in practice -- and in the numerical work -- we work in the limit where the baryons and electrons are strongly coupled to one another and so do not distinguish between the electrons and the nuclei. The Boltzmann code we use below also does not separate the different components of the baryons. 

At the perturbative level, the increased number of terms makes for more work. But nothing can stop us now! So using 
$ 
h_{\B,\C}= H-H \Phi -\dot \Phi + \frac{\Theta_{\B,\C}}{\mfa}
$
where
$
\Theta_{\B,\C}= \partial_i v^i_{\B,\C}
$
and writing the 4-velocity $u^\mu_{\B,\C}$ as $\{u^0_{\B,\C}, u^i_{\B,\C} \} = \{ 1-\Phi, a^{-1}v^i_{\B,\C}\}$ in Newtonian gauge, we find that the Dark Matter conservation equation reads
\begin{eqnarray} 
\dot{\delta \rho}_{\C} + &&3H \delta \rho_{\C} -3  \dot{\Phi} \bar \rho_{\C}+ \frac{1}{a}\partial_i v^i_{\C}\bar \rho_{\C}\nonumber \\ &&=-\frac{\beta}{\MPL }\bar \rho_{\C}\dot{\delta \phi} +\left[-\frac{\beta}{\MPL } \dot{\bar\phi} +
\frac{\bar \mfa-\mfa_m }{\Lambda_{m}^2} \dot{\bar\mfa}+ \frac{1}{2\Lambda_m^2} \partial_t{ \left( \frac{\bar\varrho_\mfa}{m^2(t)}\right)}\right]\delta \rho_{\C}+F_\C\nonumber ,\\
\end{eqnarray}
while conservation of the baryon fluid is
\begin{eqnarray} 
\dot{\delta \rho}_{\B} + &&3H \delta \rho_{\B} -3  \dot{\Phi} \bar \rho_{\B}+ \frac{1}{a}\partial_i v^i_{\B}\bar \rho_{\B}\nonumber \\ &&=-\frac{\beta}{\MPL }\bar \rho_{\B}\dot{\delta \phi} +\left[-\frac{\beta}{\MPL } \dot{\bar\phi} + \frac{m_e}{m_\ssN} \left( 
\frac{\bar \mfa-\mfa_m }{\Lambda_{e}^2} \dot{\bar\mfa}(\Bar{\rho})+ \frac{1}{2\Lambda_e^2} \partial_t{ \left( \frac{\bar\varrho_\mfa}{m^2(t)}\right)}\right)\right]\delta \rho_{\B}+F_\B\nonumber ,\\
\end{eqnarray}
where the $F$ terms involve the variation of the axion source term with respect to the total density
\be
F_\C = \frac{1}{2\Lambda_{m}^2} \partial_t\left(  \frac{\delta\rho_\mfa}{m^2(t)}\right)\bar \rho_{\C} \quad\hbox{and} \quad
F_\B = \frac{m_e}{2m_\ssN\Lambda_{e}^2} \partial_t\left(  \frac{\delta\rho_\mfa}{m^2(t)}\right)\bar \rho_{\B} . 
\ee

Now when we define $\delta_{ \B,\C}= {\delta \rho_{\B,\C}}/{\bar \rho_{\B,\C}}$, we find that the source terms almost cancel and we get 
\be 
\dot{\delta}_{\C}- 3 \dot{\Phi}+ \frac{\Theta_{\C}}{a}=  -\frac{\beta}{\MPL} \dot{\delta\phi}+\partial_t\left(\frac{\delta\varrho_\mfa}{2\Lambda_{m}^2m^2(t)}\right),
\quad 
\ee
and
\be
\dot{\delta}_{\B}- 3 \dot{\Phi}+ \frac{\Theta_{\B}}{a}= -\frac{\beta}{\MPL} \dot{\delta\phi}+ \partial_t\left(\frac{m_e\delta\varrho_\mfa}{2m_\ssN\Lambda_{e}^2m^2(t)}\right).
\ee
Notice the new term from the exchange of energy between matter and the axion fluid. This term mimics a similar term in $-\partial_0 \Phi_\rho$ appearing  in (\ref{non-con}).  This confirms that the matter and axion  fluids exchange energy and interact via a term in $\rho_{m}\rho_\mfa/2m^2(t) \Lambda_{m}^2$. Notice too that in the axion fluid equation an extra term in $\partial_0 \Phi_\phi$ implies that some of the energy flows into the dilaton perturbations too. 

Finally using the time derivative along the fluid flow
\be 
\dot{u}_{\B,\C}^{i}=  \dot{v}_{\B,\C}^{i}+\partial^i \Phi+ H \frac{v_{\B,\C}^i}{a},
\ee
we find the Euler equations for the matter fluids
\begin{eqnarray}
\dot{v}_{\C}^{i} + \Biggl[H-\frac{\beta {\dot{\bar\phi}}}{\MPL }+\frac{\bar \mfa -\aminus }{\Lambda_{m}^2} \dot{\bar \mfa}(\Bar{\rho}) &&+ \frac{1}{2\Lambda_{m}^2} \partial_t{ \left( \frac{\bar\varrho_\mfa}{m^2(t)}\right)}\Biggr]v_{\C}^i\nonumber \\
&&=\frac{1}{a}\left[-\partial^i\Phi+ \frac{\beta}{\MPL }\partial^i \phi  - \frac{1}{2\Lambda_{m}^2} \partial^i \left( \frac{\varrho_\mfa}{m^2(t)}\right)\right],\nonumber \\
\end{eqnarray}
and
\begin{eqnarray}
\dot{v}_{\B}^{i} + \Biggl\{H-\frac{\beta {\dot{\bar\phi}}}{\MPL }+\Biggl[\frac{\bar \mfa -\aminus }{\Lambda_{e}^2} \dot{\bar \mfa}(\Bar{\rho}) &&+ \frac{1}{2\Lambda_{e}^2} \partial_t{ \left( \frac{\bar\varrho_\mfa}{m^2(t)}\right)}\Biggr]\frac{m_e}{m_\ssN}\Biggr\}v_{\B}^i\nonumber \\
&&=\frac{1}{a}\left[-\partial^i\Phi+ \frac{\beta}{\MPL }\partial^i \phi  - \frac{1}{2\Lambda_{e}^2} \partial^i \left( \frac{\varrho_\mfa}{m^2(t)}\right)\frac{m_e}{m_\ssN}\right].\nonumber \\
\end{eqnarray}
Taking the divergence of the Euler equations we have
\begin{eqnarray}
\dot{\Theta}_{\C} + \Biggl[H-\frac{\beta \dot{\phi}}{\MPL }+\frac{\bar \mfa -\aminus }{\Lambda_{m}^2}\dot{\bar \mfa}&& + \frac{1}{2\Lambda_{m}^2}\partial_t { \left( \frac{\bar\varrho_\mfa}{m^2(t)}\right)}\Biggr] \Theta_{\C}\nonumber \\
&&=-\frac{\Delta\Phi}{a}+ \frac{\beta}{\MPL }\frac{\Delta \phi}{a}  - \frac{1}{2a\Lambda_{m}^2} \Delta \left( \frac{\varrho_\mfa}{m^2(t)}\right),\nonumber \\
\end{eqnarray}
and 
\begin{eqnarray}
\dot{\Theta}_{\B} + \Biggl[H-\frac{\beta \dot{\phi}}{\MPL }+\Biggl[\frac{\bar \mfa -\aminus }{\Lambda_{e}^2}\dot{\bar \mfa}&& + \frac{1}{2\Lambda_{e}^2}\partial_t { \left( \frac{\bar\varrho_\mfa}{m^2(t)}\right)}\Biggr]\frac{m_e}{m_\ssN}\Biggr]\Theta_{\B}\nonumber \\
&&=-\frac{\Delta\Phi}{a}+ \frac{\beta}{\MPL }\frac{\Delta \phi}{a}  - \frac{1}{2a\Lambda_{e}^2} \Delta \left( \frac{\varrho_\mfa}{m^2(t)}\right)\frac{m_e}{m_\ssN}.\nonumber \\
\end{eqnarray}
The last term in both of these equations is a quantum pressure term from the axion fluid in the evolution of matter. This parallels the quantum pressure term in the axion fluid equations.

\subsection{Dilaton evolution}

The final ingredient is the dilaton equation in the presence of the coupled axion fluid, which reads
\begin{equation} \label{full dilaton friedmann}
 \ddot{\Bar{\phi}} + 3H \dot{\Bar{\phi}}- WW,_\phi\left(\dot{\bar \mfa}^2+ \bar\varrho_\mfa\right)= - \partial_\phi V_{\rm eff}(\Bar{\phi}).
\end{equation}
at the background level 
{whilst its perturbed counterpart is given by} 
\begin{align}
    \ddot{\delta\phi}+3H\dot{\delta{\phi}}+\left[\frac{k^2}{a^2}-\left(W,_{\phi}^2+W,_{\phi\phi}W\right)\dot{\Bar{\mfa}}^2 + \left(W,_{\phi}^2-W,_{\phi\phi}W\right)\Bar{\varrho}_\mfa+ V_{,\phi\phi}\right]\delta\phi \nonumber\\
     -\dot{\Bar{\phi}}\left(\dot{\Phi}+ 3\dot{\Psi}\right)
    -WW,_\phi\left(1-\frac{k^2}{2m^2(t)}\right)\delta\rho_\mfa+2\Phi \left(V,_\phi-WW,_\phi\Bar{\rho}_\mfa \right)\nonumber\\
    = \beta\Bar{\rho}_\C(\delta_\C+2\Phi)+\beta\Bar{\rho}_\B(\delta_\B+2\Phi)+WW,_\phi\frac{\delta\rho_m}{m^2\Lambda_{m}^2}\Bar{\rho}_\mfa.
\end{align}
We next turn to integrating these background and perturbation equations, specializing to the choices mentioned at the beginning of this paper for both $W(\phi)$ and $V(\phi)$. 

\section{Numerical case studies}
\label{five}

In this section, we will apply the results of the previous sections to two different choices for the coupling functions $W^2(\phi)$ and dilaton potential, $V_{\rm dil}(\phi)$. We choose $W^2$ either to be exponential -- as in eq.~\pref{Wexpform} -- or near a minimum (and so quadratic) -- as in eq.~\pref{Wquadform}. For the dilaton potential we make two similar choices for the prefactor $U(\phi)$ appearing in \pref{VdilDef}: it is either a constant -- leading to a pure exponential form for $V_{\rm dil}$ -- or quadratic -- making $V_{\rm dil}$ an Albrecht-Skordis potential \cite{Albrecht:1999rm, Albrecht:2001xt}, such as arises in the RG stabilization mechanism \cite{Burgess:2022nbx} used in Yoga models \cite{Burgess:2021obw}. For the matter-axion couplings we take the form \pref{UmUeQuad}, and in all cases we fix $\Lambda_e = 10^{11}\,\text{GeV}$.

The phenomenology is sufficiently varied that we will devote separate subsections to the four possible cases.

\subsection*{Numerical Implementation}

The  effects of the axio-dilaton on the background cosmology and the linear perturbations are derived using the formalism derived above as implemented in  a modified version of CLASS \cite{Diego_Blas_2011}. This includes
\begin{itemize}
    \item The dynamics of multiple interacting scalar fields, specialised to the case of the coupled axio-dilatons studied in this paper. 
    \item The interactions between such scalar fields and ordinary matter species, baryons and CDM. %This means that for the numerical analysis, the matter density $\rho$ differs from the conserved  $\rho_m$, and the numerical computation of the early dark energy fraction.
    \item The modifications to cosmological perturbations and structure growth arising from such couplings.
    \item The dynamics of additional fluid species and their possible interactions with other species, such as  the effects of the axion fluid.
\end{itemize}
We begin by exploring the cosmological implications of the presence of early dark energy as described above and  the effects caused by imposing solar system constraints. Finally we explore the  Yoga models and show how their features enable the axion to act as non-negligible early dark energy.

 In all cases we choose parameters to ensure $H_0 = 100h\; \rm km/s\;\hbox{Mpc}^{-1}$ with $h = 0.6756$, $\Omega_\B h^2 = 0.022$, and $\Omega_{\rm CDM} h^2 = 0.12$. For the perturbations, we assume adiabatic Gaussian initial conditions for the power spectrum from the 2018 Planck LCDM best fit \cite{Planck:2018vyg}, and set the spectral index $n_s = 0.966$, the pivot scale $k_{piv} = 0.05\;\rm M_{pc}^{-1}$ and scalar amplitude $A_s = 2.10\times10^{-9}$. 

For the background densities we plot the non-conserved CDM and baryon energy density $\rho_{\C,\B}(\chi, \mfa, x)$ with the axion and dilaton couplings included, along with the matter-independent axion energy density. For the density parameters, we show both the axion-dependent and axion independent density fractions, along with the corresponding matter-independent axion potential, $V_\mfa$, and matter-dependent effective potential, $V_{\mfa\,\text{eff}}$, to properly account for interaction energies. The axion dependent and independent matter density parameters are denoted by  $\Omega_i$ and $\Omega_{0\,i}$ respectively. 

 For the sake of comparison between the different scenarios, in all cases we choose parameters to ensure the axion's transition between the minimum of $\cU_m(\mfa)$ and $V_\mfa(\mfa)$ occurs just before the onset of matter-radiation equality, at $\rho_{m\,th} \approx 1 \, \text{eV}^4$. 
 
 A key limiting factor on the amount of early dark energy present in the cosmology is the model building bound on the axion-matter coupling, $\cU_m(\mfa) \ll 1$, which also forces us to consider $f_{\textnormal{EDE}} \ll 1$ through the relation (\ref{ede frac}). This means that while in principle one could push the axion transition back further than matter radiation equality to keep the effects on the matter and angular power spectra minimal, this would correspond to the matter coupled to the axion making up a smaller percentage of the total energy density, causing the total fraction of early dark energy to be reduced, as seen in (\ref{ede frac}). In order to raise the fraction of the total matter content taken up by early dark energy again one would need to consider larger and larger $f_{\textnormal{EDE}}$ as the axion transition is pushed further and further back, which would require $\cU_m(\mfa) > 1$ and so we do not pursue such a scenario further here.

\subsection{Constraints}

We next describe the main constraints these models face, starting with those coming from cosmology and continuing to combine them with constraints coming from solar-system tests.

\subsubsection{Cosmological Constraints}\label{cosmological constraints}

The most stringent cosmological constraints on early time solutions to the Hubble tension come from their effects on the extremely well-constrained second and third peaks of the angular power spectrum. With the view of understanding the effects caused by the axionic early dark energy on these features, we begin by ignoring the constraints required by solar system screening and focus on the simplest cosmological scenario (but return to these issues below). 

Here we take 
\begin{align}
    V(\phi) =V_0e^{-\lambda\phi/M_{\rm p}},\qq{and}  W^2(\phi) = W_*^2+\frac{(\phi-\phi_*)^2 }{2\Lambda_\phi^2}.
\end{align}
For the first two cases considered here we use the freedom to rescale the value of the dilaton due to its exponential potential to set $\phi_* = 0$ and $\phi_{ini} \approx 0$ with $V_0$ phenomenologically re-scaled to give the present day dark energy density. We also take $\lambda = 0.1$ unless stated otherwise.

\begin{figure}[hbt!]
    \centering
     \begin{subfigure}[b]{\textwidth}
         \centering
         \includegraphics[width=0.93\textwidth]{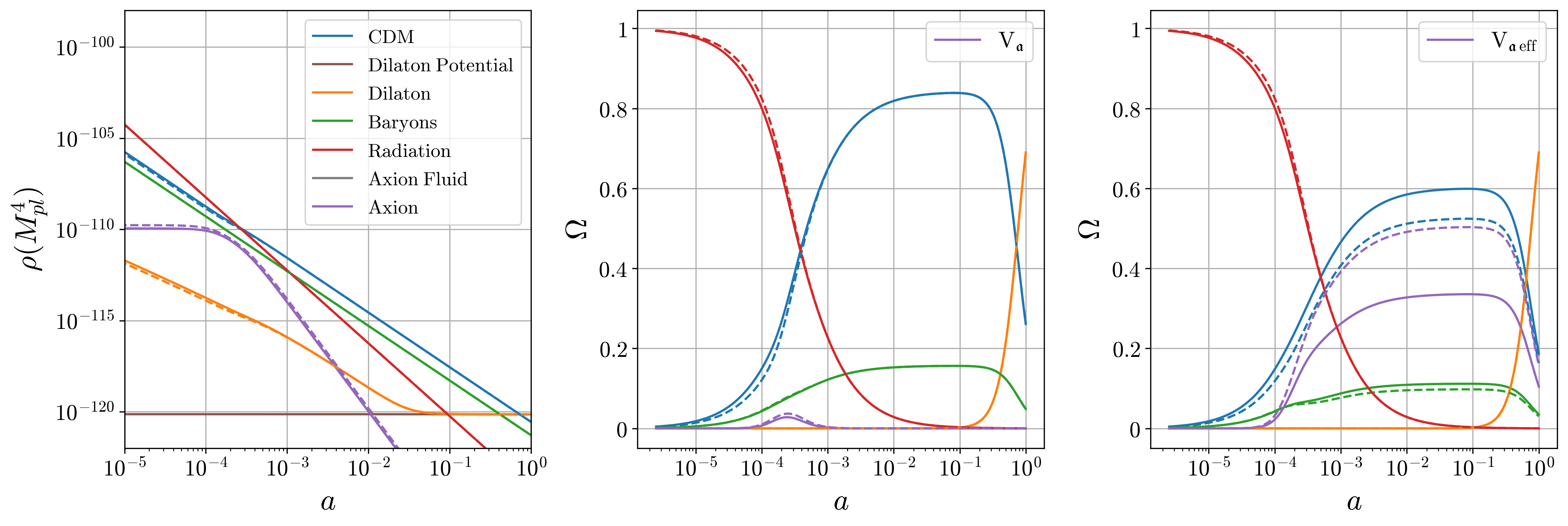}

     \end{subfigure}
     \hfill
     \begin{subfigure}[b]{\textwidth}
         \centering
         \includegraphics[width=0.93\textwidth]{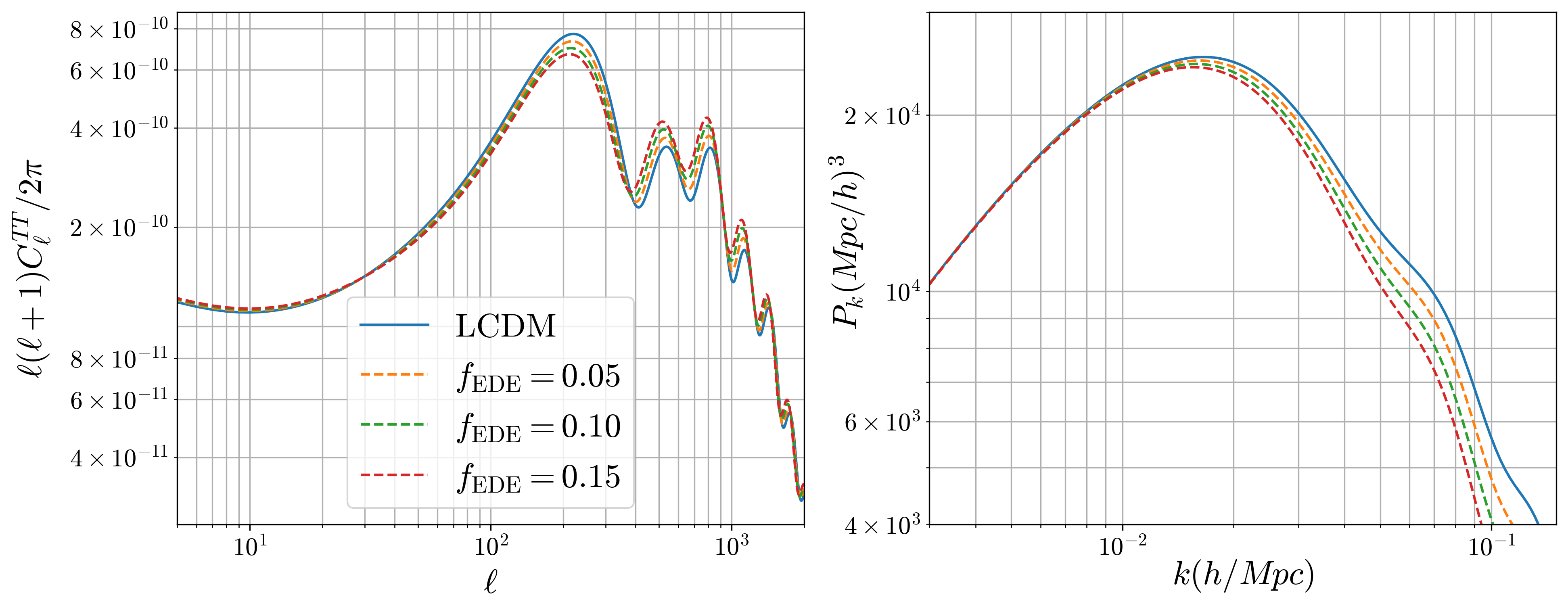}
     \end{subfigure}
     \hfill
     \begin{subfigure}[b]{\textwidth}
         \centering
         \includegraphics[width=0.93\textwidth]{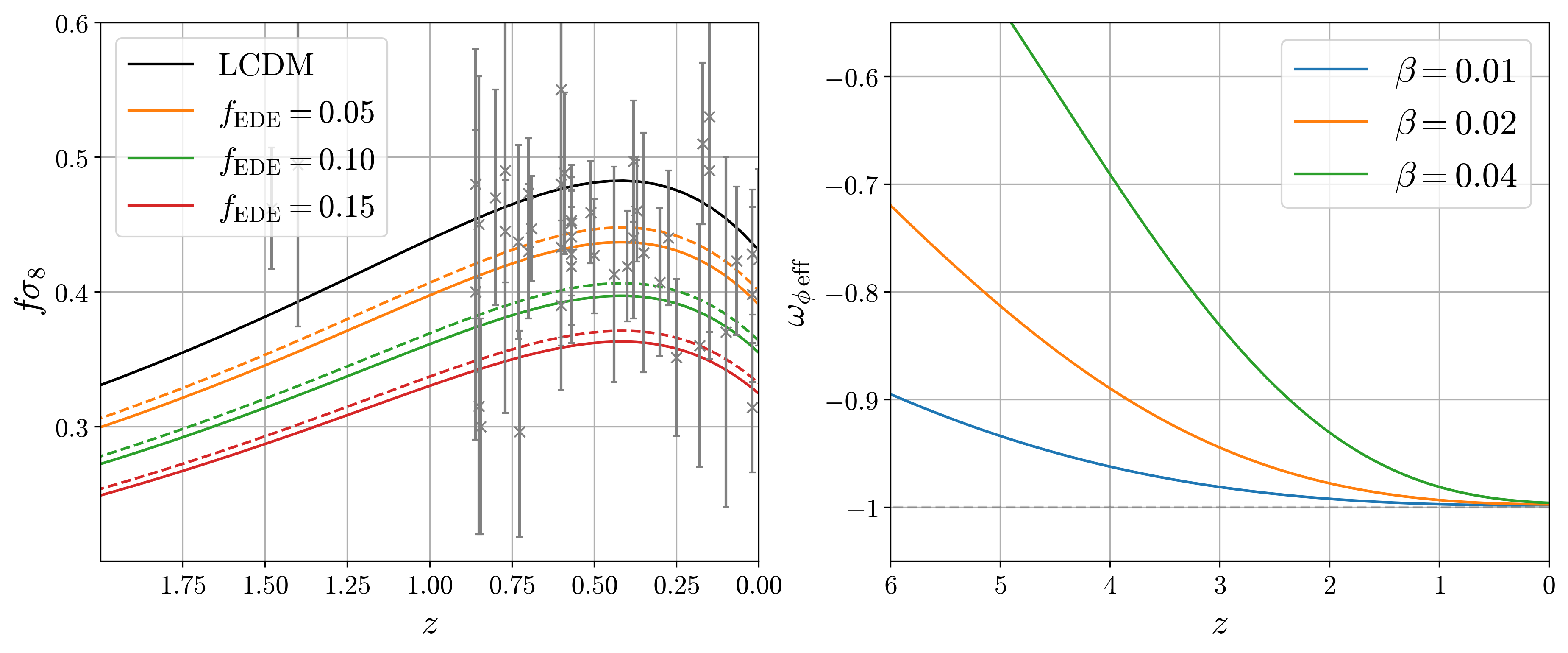}
     \end{subfigure}

        \caption{The case of an exponential dilaton potential  and a quadratic coupling  $W^2(\phi)$. Top row shows background evolution for $m_\mfa = 2\times10^{-15}  \, \rm eV$, $\Lambda_{\phi} = 10^{10}  \, \rm GeV$, $\Lambda_{m} = 5\times10^{5} \,  \rm GeV$, 
        %$\aplus-\aminus
        $\aminus  = 5.5\times10^{14} \, {\rm GeV}$, and $4.5\times10^{14}\,\rm GeV$ in solid and dashed respectively. Middle row shows matter and angular power spectra for 
        %$\aplus-\aminus  
        $\aminus  = 5.5\times10^{5}  \, \rm GeV,\;4.5\times10^{5}  \, \rm GeV,\;\rm and \;3.2\times10^{5}  \, \rm GeV$ in orange, green and red respectively. The bottom left plot shows $f\sigma_8$ for the corresponding early dark energy fractions with $\beta = 10^{-2}$ in solid and $\beta = 4\times10^{-2}$ in dashed, while the bottom right plot shows the evolution of the effective dark energy equation of state for $\mfa_m = 5.5\times 10^{5}  \, \text{GeV}$ and for various coupling strengths. In all other plots $\beta  = 10^{-2}$. The axion fluid has decayed away so fast that it is not visible in the background plots. Finally, in all cases $\Lambda_e =10^{11}\text{GeV}$.}
        \label{fig:cosmological constraints}
\end{figure}

The results for this scenario are shown in Fig.~\ref{fig:cosmological constraints}. The top row depicts the background evolution of the relevant fluids and fields where the axion's vacuum potential contributes an early dark energy fraction of approximately $3\%$ and $5\%$. As argued in \S \ref{axionstressenergy}, placing the transition of the axion from the minimum of the matter coupling to the minimum of its bare potential at around matter-radiation equality ensures the axion fluid dilutes away between initial conditions set at BBN and this epoch, leading to it playing a negligible role on the background dynamics. 

As the axion transitions just before recombination, its evolution induces changes in the masses of the coupled baryons and CDM, which can be seen as a jump in the density of both species before recombination. As we raise the value of
%$\aplus-\aminus $,
$\aminus $, in turn raising the dark energy fraction, the axion must cover a greater distance in the field space within the same transition window, inducing greater axion velocities and hence greater variations in particle masses. This in turn has effects on the angular and matter power spectra, causing deviations away from the $\Lambda$CDM best fit at large $\ell$ for a dark energy fraction greater than $f_{\text{EDE}}\sim5\%$. We see that as the early dark energy fraction is raised, correspondingly raising the coupling between the axion and matter through the relation (\ref{ede frac}), there is a net reduction in power for small scales in the matter power spectrum and a shifting of power in the first peak of the angular power spectrum into the higher peaks. This is caused by the couplings between the axion early dark energy and matter fields leading to matter-radiation equality occurring later and later as the coupling strength is raised.

The net reduction in small scale power in the matter power spectrum leads a net reduction in clustering at late times. As can be seen in the bottom plot of figure \ref{fig:cosmological constraints}. Here we plot the observable parameter $f\sigma_8$ as a function of redshift. Where $f_i$ is the linear growth rate of each matter species

\begin{equation}
    f_i(z, k) := \frac{1}{\mathcal{H}}\frac{\delta_i'(z,k)}{\delta_i(z,k)} \, ,
\end{equation}
and $\sigma_8$ is the variance of the mass fluctuations within a sphere of radius $R=8 h^{-1}$Mpc, defined by
\begin{equation}
   \sigma_8^2 = \int \frac{dk}{k} |{\cal W}(kR)|^2 \Delta^2(k) \,,
\end{equation}
where ${\cal W}(kR)$ denotes the Fourier transform of the real--space top-hat window function and $\Delta^2(k)$ is the dimensionless power spectrum defined by $\Delta^2(k) = k^3P(k)/2\pi^2$. 

The product of these two parameters is a direct observable associated with redshift space distortions \cite{Song:2008qt}. It is given by
\begin{equation}
    f\sigma_8 = \frac{\sigma_8(z,k_{\sigma8})}{\mathcal{H}}\frac{\delta_{\text{matter}}'(z,k_{\sigma8})}{\delta_{\text{matter}}(z,k_{\sigma8})} \, ,
\end{equation}
where $k_{\sigma8} = 0.125h~$Mpc$^{-1}$ and $\delta_{\text{matter}} = (\delta \rho_\ssB +\delta \rho_\C)/(\rho_B+\rho_\C)$. The coupling of the dilaton to matter however has the opposite effect. As the dilaton's Brans-Dicke coupling is raised $f\sigma_8$ goes back up due to the additional fifth forces mediated by the dilaton, re-accelerating structure growth. 

In view of the DESI results \cite{DESI:2024mwx, DESI:2025zgx} Fig.~\ref{fig:cosmological constraints} also shows the effective dilaton equation of state $\omega_{\phi\, \text{eff}}$ one would infer from this model if one were to try to describe its predictions in terms of vanilla Dark Matter (falling like $1/a^3$) and a single-field quintessence model with equation-of-state parameter $\omega_{\phi\,{\rm eff}}$, doing so for several different coupling strengths to matter. Following \cite{Das:2005yj, Brax:2011qs, vandeBruck:2020fjo, Khoury:2025txd}, it is given by 
\begin{equation}
    \label{exp eos}\omega_{\phi\,\text{eff}} = \frac{\omega_\phi(\phi)}{1+\left[e^{\beta(\phi_0-\phi)} - 1\right]\frac{\rho_{\C,\textnormal{tod}} + \rho_{\B,\textnormal{tod}}}{a^3\rho_\phi}},
\end{equation}
where $\omega_\phi(\phi) = [{\dot{\phi}^2-2V(\phi)}]/[{\dot{\phi}^2 + 2V(\phi)}]$ is the actual equation of state parameter for the dilaton field and $\phi_0$, $\rho_{\C,\textnormal{tod}}$ and $\rho_{\B,\textnormal{tod}}$ are respectively the values of the dilaton, the dark matter density and the baryon energy density today. The point of this comparison is that this analysis is effectively what DESI does when interpreting its data using a $w_0w_a$CDM model, and that this kind of analysis can easily give $\omega_{\phi\,{\rm eff}} < - 1$ (in the phantom regime) even though the actual evolution always satisfies $-1 < \omega_\phi < 1$ (as required by energy constraints). From this point of view finding a phantom equation of state can be seen as evidence both for time-dependent Dark Energy and a coupling between Dark Energy and Dark Matter that ruins the Dark Matter's naive $1/a^3$ evolution.

In the distant past the mean value of the dilaton is smaller than at present as it sits further up its exponential potential causing the denominator of (\ref{exp eos}) to be larger, and this causes the effective equation of state for the dilaton to be significantly greater than $- 1$, and the figure shows how the precise value depends on choices made for the dilaton-matter coupling parameter $\beta$.

The gradual nature of the axion transition necessitates a full data analysis to understand if such a period of early dark energy can play a role in reducing the Hubble tension. The variation of particle masses expected around the axion transition also has visible effects on the background dilaton evolution, whose coupling to matter is proportional to the total matter energy density. The dangerous tachyonic instability discussed in \S\ref{Dilaton Tachyonic Instability} when imposing solar system screening constraints is avoided here by ensuring a hierarchy between the axion and dilaton cutoff scales, $\Lambda_\phi\gg \Lambda_\mfa = \Lambda_m$ making the dilaton less sensitive to axionic evolution.

\subsubsection{When $\beta$ satisfies solar system constraints}

We next focus on the additional effects arising from requiring the matter-dilaton coupling to be small enough to evade constraints coming from tests of GR within the solar system, which we impose by choosing $\beta = 10^{-3}$. We explore the cosmology of this choice while continuing to use the case where $W^2$ is quadratic. 

Restating our parameter choices from \S\ref{ssec:Criteria&Constraints} here for convenience, this means our parameter choices are 
\begin{align}
    \Lambda_\phi\lesssim \frac{\aminuse}{\sqrt{\beta}},
\end{align}
which enforces $\Lambda_\phi \lesssim 10^{4}$ GeV for $\sqrt{\beta} \sim 10^{-2}$ as we impose $|\aminuse| \lsim 3\times 10^{3}$ GeV to adhere to constraints on the variation of the electron mass on Earth, as discussed in \S\ref{Benchmark values}. We also take
\begin{equation}
    m_\mfa\sim 10^{-3} m_\mfa(\rho_\odot)\sim 2\times10^{-15} \,{\rm eV},
\end{equation}
so that the axion field varies over scales much smaller than the radius of the Sun for efficient screening, which enforces $\Lambda_{m} \sim 5\times10^5\,\rm GeV$ to keep the axion transition at around the correct redshift, when $\rho_{m\,th} = m_\mfa^2\Lambda_m^2 \sim 1\;\text{eV}^4$ from (\ref{rhothbench}).  These parameter constraints on the local screening therefore kill any chance of a hierarchy between $\Lambda_{m}$ and $\Lambda_\phi$ to keep the dilaton's tachyonic instability at bay.

\begin{figure}[hbt!]
     \begin{subfigure}[b]{\textwidth}
         \centering
         \includegraphics[width=\textwidth]{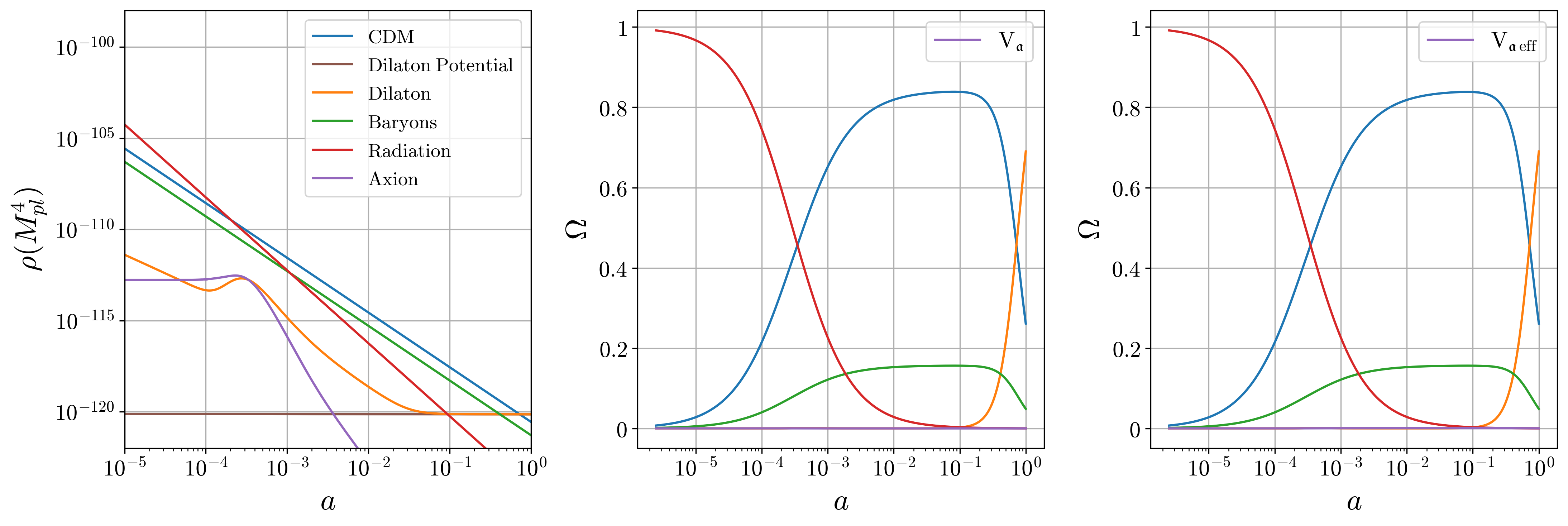}
     \end{subfigure}
     \hfill
     \begin{subfigure}[b]{\textwidth}
         \centering
         \includegraphics[width=0.49\textwidth]{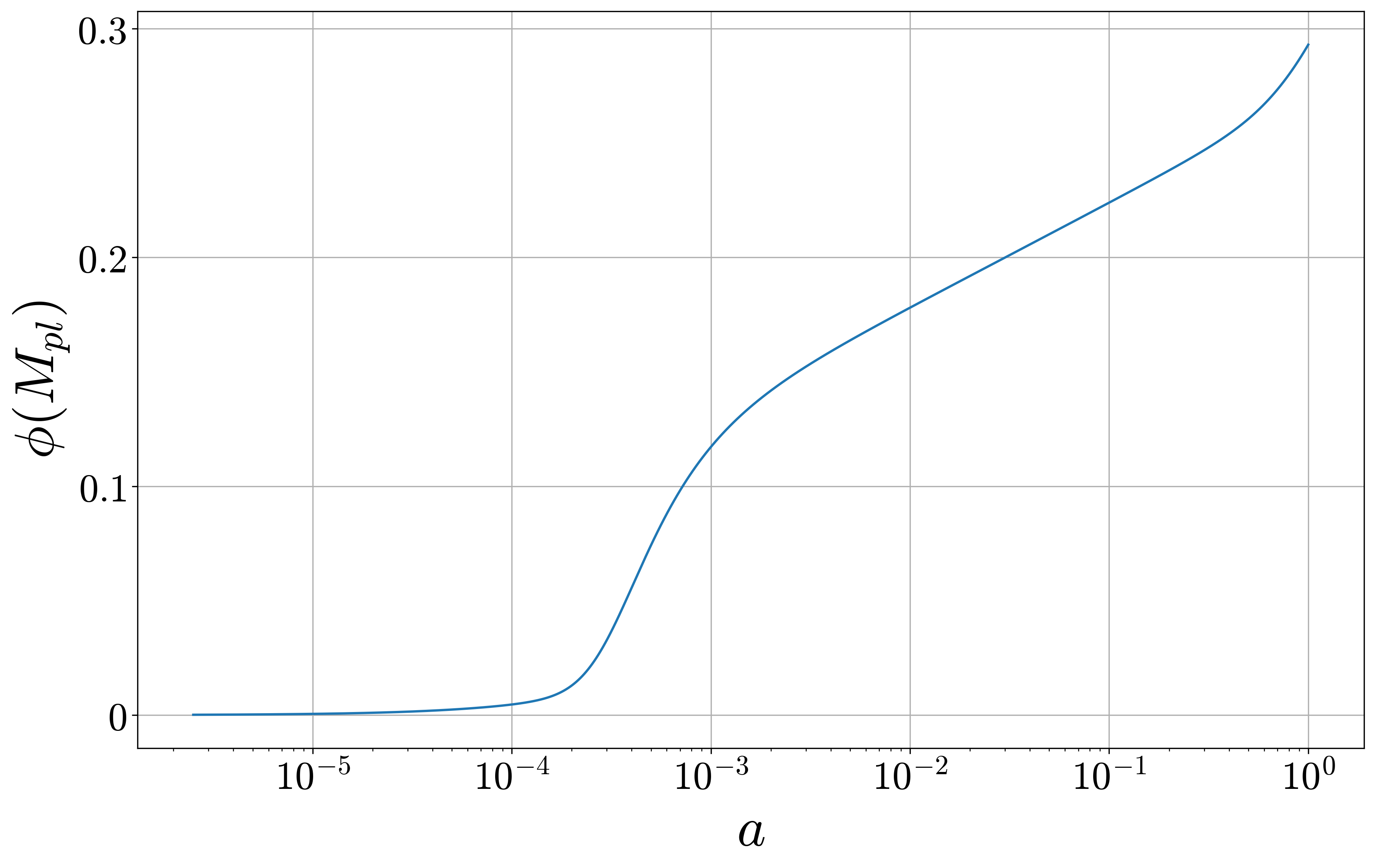}
     \end{subfigure}

        \caption{The case of an exponential dilaton potential $V$ and a quadratic $W$ which satisfies screening. The top row shows the background evolution for $m_\mfa = 2\times 10 ^{-15}\,\rm eV, \Lambda_{m} = 5\times10^{5}\,\rm GeV, \Lambda_\phi = 3.2\times10^{3}\,\rm GeV, \beta = 10^{-2},  
        %\aplus-\aminus 
        \aminus  = 1.7\times10^4\,\rm GeV$. The bottom plot shows the dilaton field evolution for the same parameters. Finally, in all cases $\Lambda_e = 10^{11} \text{GeV}$.}
        \label{fig:solar system constraints}
\end{figure}

The effects are shown in Fig.\,\ref{fig:solar system constraints}, where the tachyonic instability induces a large increase in dilaton velocity around the transition, shown in the bottom panel. This in turn leads to an increase in the axionic kinetic energy which would then dominate the total axion energy density, caused by a rapid increase in the $W$ function as the dilaton evolves, as depicted as an extra bump in axion energy density in the top left panel around the axion's transition. This produces an axion equation of state with $\omega_\mfa > -1$ during its transition, ruining its ability to act as early dark energy.

\subsection{Yoga-type quadratic well}\label{Yoga_quadratic_section}

To illustrate one possible route to take to stop the tendency of a dilaton runaway and so be able to have the axion be interpretable as early dark energy we next add additional structure to the dilaton's potential, using the Albrecht-Skordis form with a mild quadratic potential well parameterized as 
\be
    V(\phi) =V_0\Bigl(1-u_1\phi+ \tfrac12 u_2 \phi^2\Bigr) e^{-4\zeta\phi/\MPL} \qquad \hbox{and} \qquad W^2(\phi) = W_*^2+\frac{(\phi-\phi_*)^2 }{2\Lambda_\phi^2}.
\ee
For concreteness, here we take the dilaton potential used in the Yoga mechanism \cite{Burgess:2021obw, Smith:2024ayu}, which chooses $\zeta = \sqrt{\frac23}$ and $V_0 \approx \left(\frac{1}{500}\right)^4\rm \MPL ^4$. The parameters $u_1$ and $u_2$ are chosen to ensure the dilaton's potential has a minimum around $\phi\sim 74~\rm{ \MPL}$, which in turn ensures that the dark energy scale and the electroweak hierarchy are well described. We then start the dilaton's evolution off at $\phi_{ini} \approx 74~ \rm \MPL$, and hence take the local asymptotic value of the dilaton to be $\phi_* = 74~\rm \MPL$.

\begin{figure}[hbtp!]
     \begin{subfigure}[b]{0.99\textwidth}
         \centering
         \includegraphics[width=\textwidth]{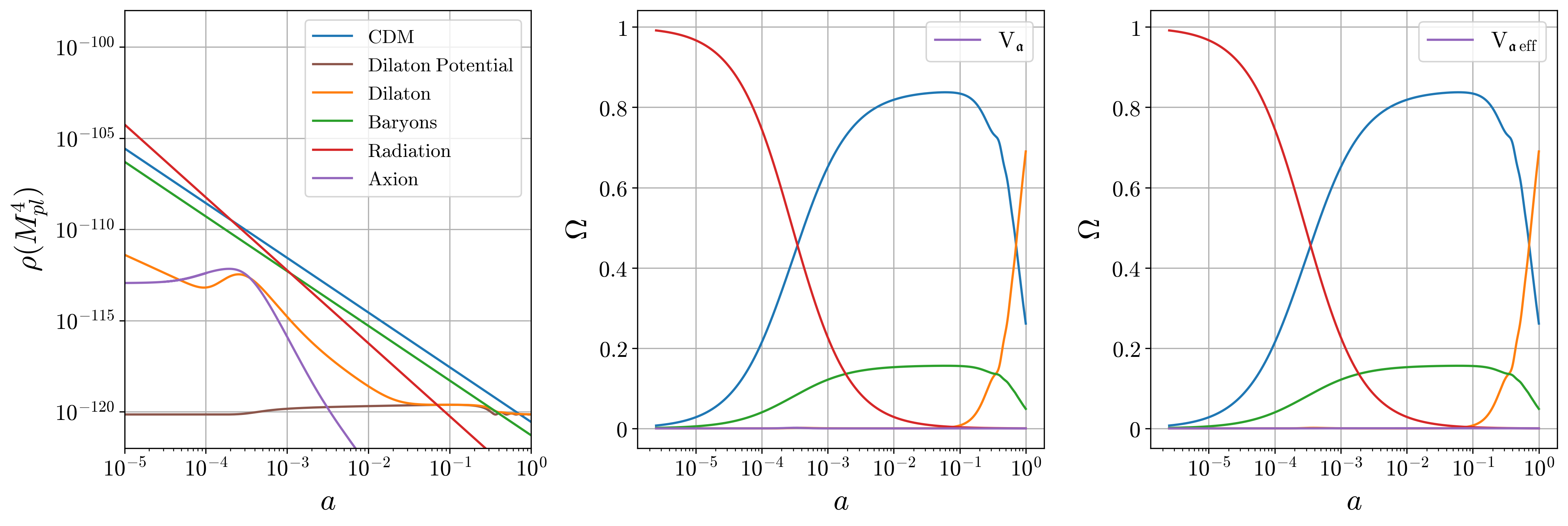}
     \end{subfigure}
     \hfill
     \begin{subfigure}[b]{0.99\textwidth}
         \centering
         \includegraphics[width=\textwidth]{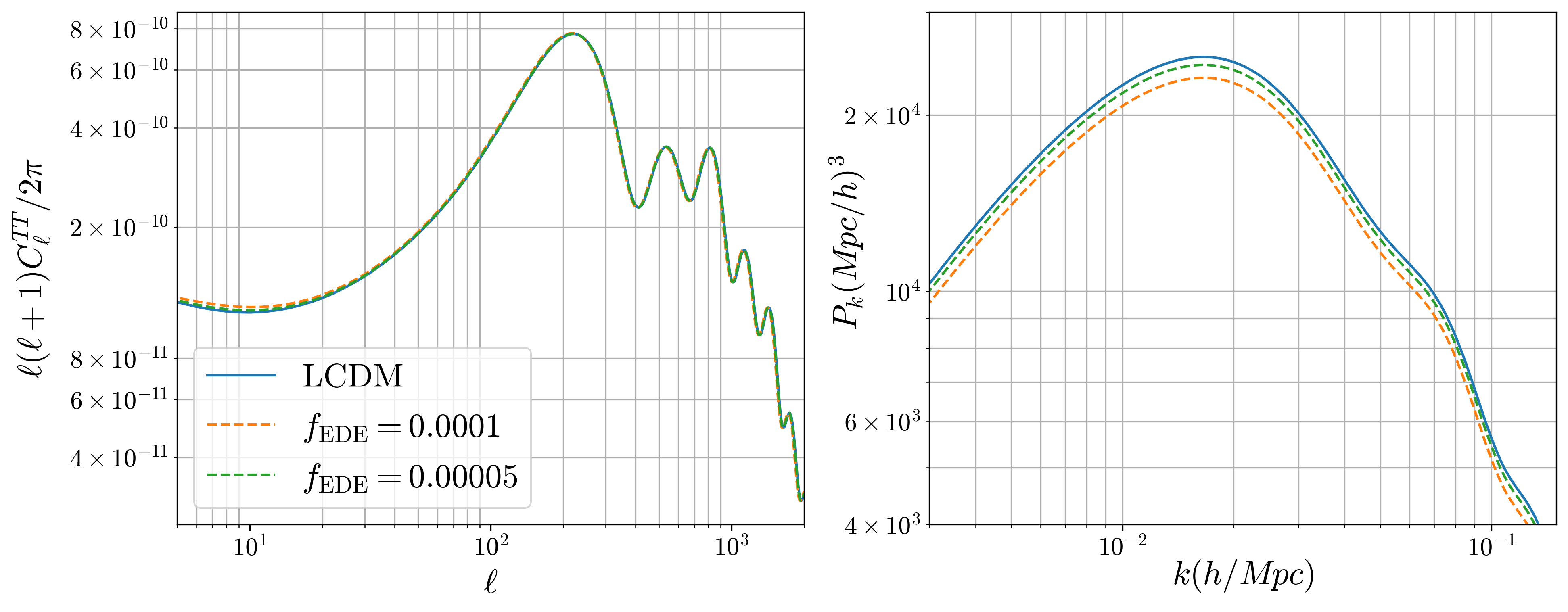}
     \end{subfigure}
     \hfill
     \begin{subfigure}[b]{0.99\textwidth}
         \centering
         \includegraphics[width=\textwidth]{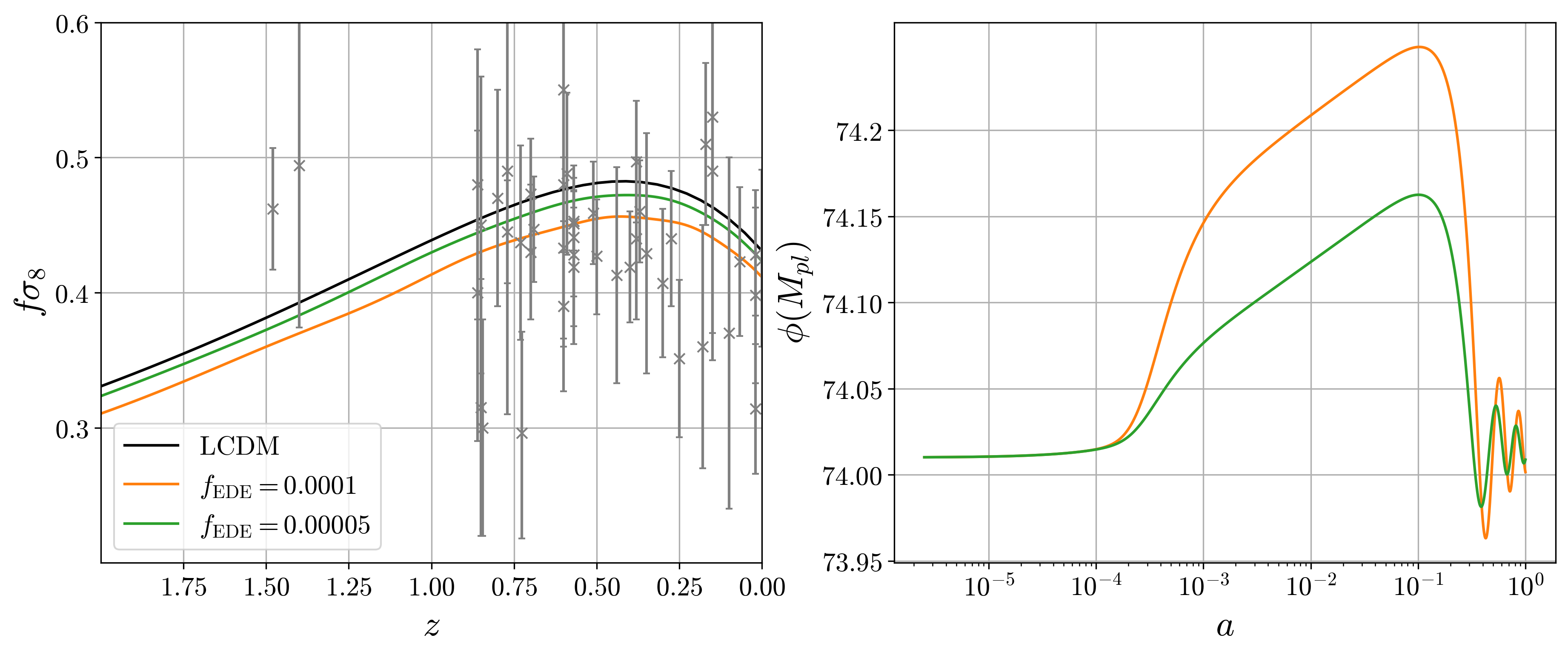}
     \end{subfigure}
     \hfill
     
        \caption{The case of a Yoga potential $V$ for the dilaton and a quadratic $W$ which satisfies screening. Top row shows the background evolution for $m_\mfa = 2\times10^{-15}\,\rm eV$,  
        %$\aplus-\aminus
        $\aminus  = 1.4\times10^4 \,\rm GeV$, $\Lambda_{m} = 5\times10^5 \,\rm GeV$, $\Lambda_\phi = 3.2\times10^3 \,\rm GeV$. Middle row shows the angular and matter power spectra and bottom row shows $f\sigma_8$ and the evolution of the dilaton field for $\aminus = 1.4\times10^4 \,\rm GeV$ and $1.0\times10^4 \,\rm GeV$ in orange and green respectively. In all plots $\beta = 10^{-2}$. Finally, in all cases $\Lambda_e = 10^{11} \text{GeV}$.}
        \label{fig:screened quadratic well}
\end{figure}

The results of this are shown in Fig.\,\ref{fig:screened quadratic well}. Although the tachyonic instability does not completely disappear -- as can be seen in the top left panel -- it is reduced by the presence of the quadratic well holding the dilaton in place. However, increasing the early dark energy fraction results in larger and larger excursions of the dilaton within its potential well, as shown in the bottom right panel, increasing the dark energy density during this excursion and in turn increasing the effective Hubble friction acting on structure growth. This causes a reduction in $f\sigma_8$ as can be seen in the bottom left panel, but also also means that increasing the early dark energy fraction beyond 0.25$\%$ results in larger and larger dilaton excursions increasing the axion equation of state greater than $\omega_\mfa>-1$ around the transition, ruining the early dark energy effect.

Such dilaton excursions can also be triggered around the start of matter domination by stronger dilaton-matter couplings brought on by increased $\beta$, as studied in \cite{Smith:2024ayu}, meaning $\beta$ needs to be kept small to allow for a non-negligible early dark energy fraction. Achieving a dark energy fraction of around $\sim 0.25\%$ requires $\beta\lesssim10^{-2}$ to keep the dilaton from being drawn too far from the minimum of its well, which corresponds to the screening bound. These excursions can otherwise be reduced by increasing the depth/steepness of the well beyond the mild quadratic we are considering here, possibly leading to much larger dark energy fraction being viable cosmologically. 

Keeping the dark energy fraction below $\sim 1\%$ also results in almost negligible effects on the angular and matter power spectra, as can be seen in the middle panels. Here the small deviations in the peaks of the angular power spectrum arise because of the dilaton evolution due to the axion-dilaton coupling inducing a change in the electron mass pre-recombination. Similar effects are observed in \cite{Planck:2014ylh}.

\begin{figure}[hbtp!]
     \begin{subfigure}[b]{0.97\textwidth}
         \centering
         \includegraphics[width=\textwidth]{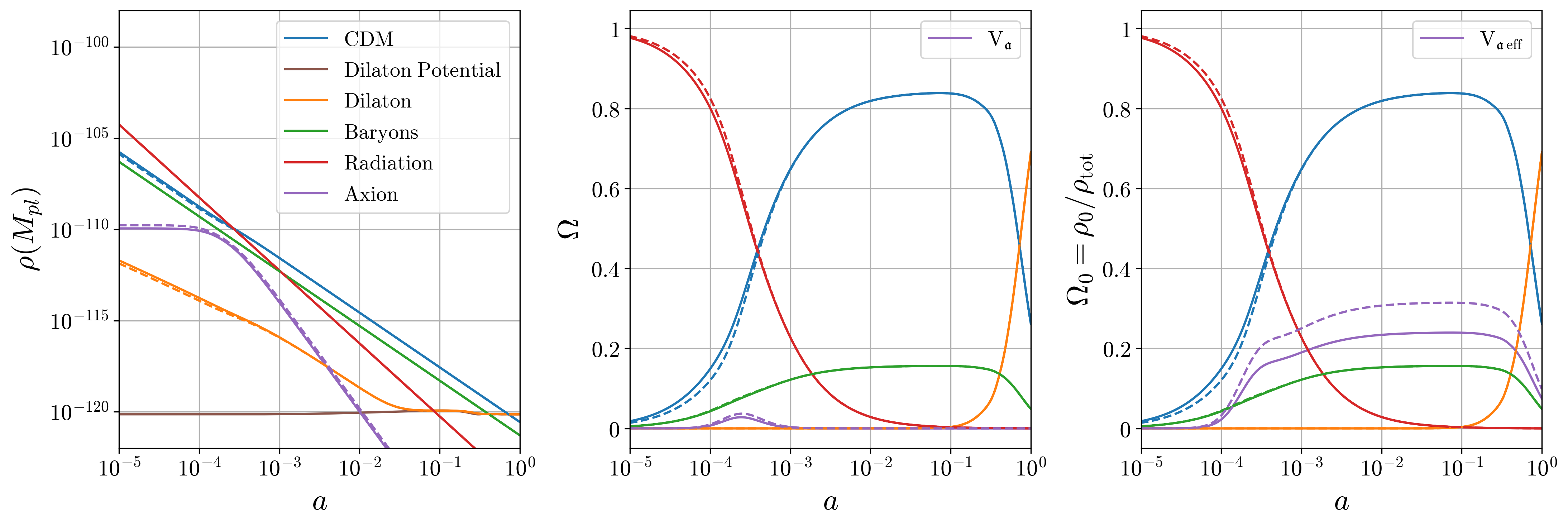}
     \end{subfigure}
     \hfill
     \begin{subfigure}[b]{0.97\textwidth}
         \centering
         \includegraphics[width=\textwidth]{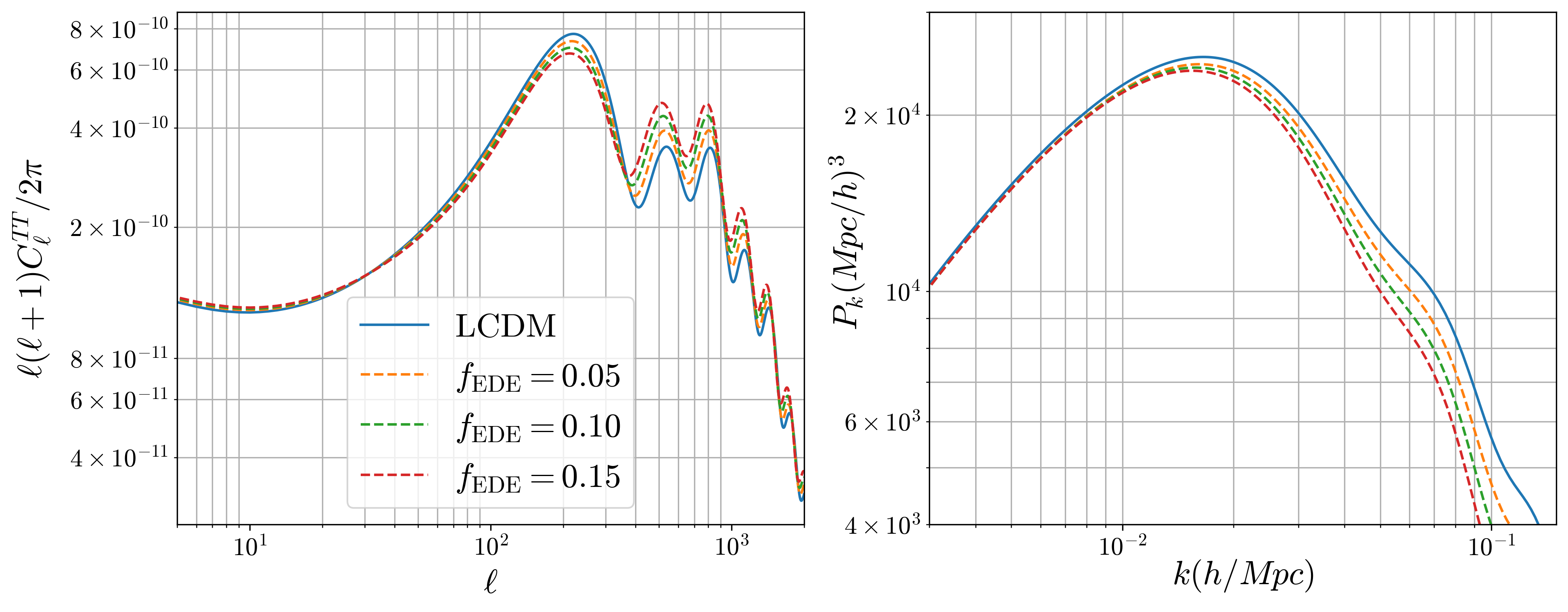}
     \end{subfigure}
     \hfill
     \begin{subfigure}[b]{0.97\textwidth}
         \centering
         \includegraphics[width=\textwidth]{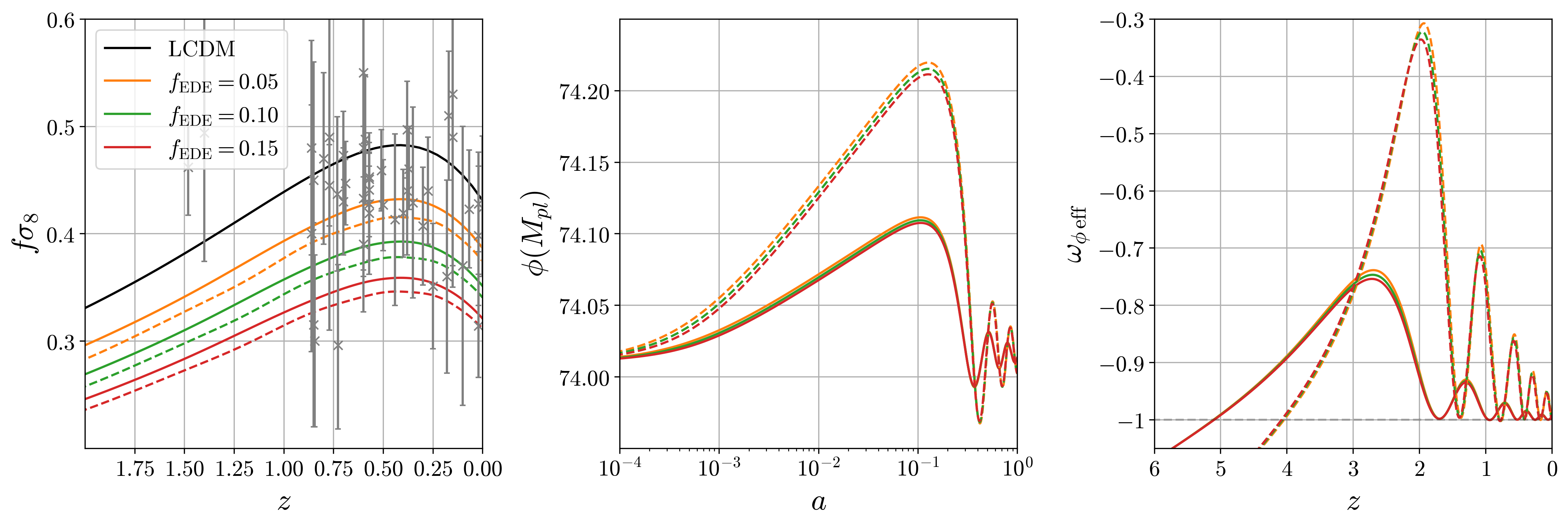}
     \end{subfigure}
     \hfill
     
        \caption{The case of a Yoga  potential $V$ and an exponential $W$ which satisfies screening. Top row shows the background evolution for $m_a = 2\times 10^{-15}\rm eV$, $\Lambda_{m} = 5\times 10^5 \rm GeV$, $\Lambda_\phi = 2\times10^8 \rm GeV$, 
        %$\aplus-\aminus
        $\aminus  = 5.5\times10^5\rm GeV$ and $4.5\times10^5 \rm GeV$ in dashed and solid respectively. Middle row shows the angular and matter power spectra and bottom row shows $f\sigma_8$, the evolution of the dilaton field and the effective dark energy equation of state for 
        %$\aplus-\aminus
        $\aminus  = 5.5\times10^5 \rm GeV$, $4.5\times10^5 \rm GeV$ and $3.2\times10^5 \rm GeV$ in red, green and orange respectively. The dashed lines in the bottom plots correspond to the same parameters except for $\beta=2\times10^{-2}$. In all other cases, $\beta = 10^{-2}$. Finally, in all cases $\Lambda_e = 10^{11} \text{GeV}$.}
        \label{fig:Yoga screened}
\end{figure}

\subsection{Yoga models}\label{Yoga_section}

Saving the best till last, our final case is to consider an axio-dilaton setup with an exponential $W(\phi)$ to illustrate how such a scenario can arise as the best of both worlds, allowing for a significant phase of early dark energy while still satisfying solar system screening constraints. Here we take (recall \pref{Uquaddef})
\begin{align}
    V(\phi) =V_0\left[1-u_1\phi+u_2\phi^2/2\right]e^{-4\zeta\phi/M_{\rm p}} \qq{and}  W(\phi) = e^{-{\phi}/({2\Lambda_\phi)}},
\end{align}
with $\zeta = \sqrt{2/3}$, and $\Lambda_\phi = 2\times10^8\rm GeV$ to satisfy screening. The results are shown in Fig.\,\ref{fig:Yoga screened}, where as predicted in \S\ref{Dilaton Tachyonic Instability} we can see the exponential $W$ causing the dilaton's tachyonic instability disappearing even for a significant fraction of early dark energy present in the background evolution.

The absence of this tachyonic instability means that the dilaton does not receive large kicks from its axion coupling, removing the reduced $f\sigma_8$ effect occurring in the previous case due to dilaton evolution. We however see reduced structure growth due to two different effects in this case. The first being the same reduction in power in the matter power spectrum on small scales caused by the axion's transition causing matter-radiation equality to occur later that we observed in \S\ref{cosmological constraints}. The second effect comes from larger Brans-Dicke coupling between the dilaton and matter species causing a dilaton excursion during matter domination, as can be seen in the bottom right panel. 

In \S\ref{cosmological constraints} this second effect caused an increased structure growth due to the larger fifth forces associated with a larger Brans-Dicke coupling strength, but here we see a {\it reduction} due to the presence of the dilaton's local minimum. This difference arises because the local minimum ensures that when the dilaton goes on its temporary excursion its potential energy goes up, and this in turn raises the Hubble rate,  dampening structure growth. This suggests that we should therefore expect reduced $f\sigma_8$ as the dilaton-matter coupling strength is increased, as is indeed shown by the dashed lines in the bottom left panel, which doubles $\beta$ to $\beta =2\times10^{-2}$. We also observe the same shifting of power from the first peak of the angular power spectrum to higher peaks discussed in \S\ref{cosmological constraints} for this case.

This shows that in the case best motivated by fundamental physics, where we choose the functional form of the dilaton's potential $V(\phi)$ and axion coupling $W(\phi)$ to correspond to those expected from extra-dimensional physics, it is possible to  achieve simultaneously a significant fraction of early dark energy and reduced structure growth generically, while still satisfying solar system constraints on matter couplings. 

Another interesting feature of the dilaton's local minimum is the resulting behaviour of the effective equation of state plotted in the bottom right panel. Given that this feature allows for the dilaton to travel to larger values during matter domination before returning to smaller values at later times, as shown in the bottom-middle panel, the denominator of (\ref{exp eos}) will be $< 1$ during matter domination, causing the effective equation of state to generically cross the phantom divide line. At later times we then observe an oscillating equation of state as the dilaton oscillates around its local minimum.

\section{Conclusion}

Dynamical dark energy requires the existence of very light fields since it is only scalars with masses smaller than the Hubble scale $H$ that move over cosmological timescales, as would be required to be responsible for the time-dependence. On the other hand, it is also generic that a dimensionless scalar field that is gravitationally coupled to a scalar potential of size $V = v^4 \mfu(\phi)$ -- where $\mfu(\phi)$ is a generic order-unity function -- acquires a mass of order $m \sim v^2/\MPL \sim H$. So given the fact that the Dark Energy density is so small, it perhaps might not be a surprse to find very light scalars at play in the recent history of the universe.\footnote{{\it Why} the Dark Energy density is so small in the first place is another question, though the hope is that Yoga models \cite{Burgess:2021obw} are a step in the right direction for understanding this.} 

Of theories with light scalar fields those with only one scalar are special because only for these is it impossible to have two-derivative sigma-model interactions. Two-derivative interactions are special because they scale with energy in the same way as do the two-derivative interactions of GR and this allows them to compete with GR without also jeopardizing the low-energy expansion that justifies working within the semiclassical approximation. The axio-dilaton models we study here provide a minimal, well-motivated example of what the dynamics of such fields can look like.

If such scalars couple to matter with gravitational strength, they generically lead to large deviations from GR in the solar system and so would be ruled out by observations. This can be prevented either by having the scalars couple more weakly or by invoking screening mechanisms whereby the effective coupling of the light fields to matter is dynamically reduced for the macroscopic objects like the Sun or the Earth. A screening mechanism of this sort is known to exist for the kinds of axio-dilaton models we consider \cite{Brax:2023qyp}, that uses the same kinetic sigma-model couplings between axion and dilaton that make the models interesting in the first place. Screening proceeds by coupling matter to the axion so that the axion field shifts from its vacuum value in the presence of matter. This implies that the axion field necessarily has gradients when going from outside to inside dense bodies, and it is the interaction of the dilaton with this axion gradient that drives the effective coupling of the dilaton to matter to small values. 

In this paper we explore what the cosmology of these models looks like in a scenario where the dilaton of an axio-dilaton pair is both responsible for dark energy and is screened locally. Although the axion has a scalar potential and couples to ordinary matter it is not itself the Dark Matter, which we assume also couples to the axion. As might be expected. the presence of matter-axion and Dark Matter-axion couplings can change cosmology fairly dramatically because of axion interactions with the ambient matter and Dark Matter environments encountered in the early universe. We find in particular that an unexpected bonus of the axion-matter couplings is that the axion energy density behaves like a temporary cosmological constant at high redshift, but one that naturally evaporates once the local matter and Dark Matter density become sufficiently small.

The fact that axion cosmology with a matter-dependent potential can resemble Early Dark Energy (EDE) is interesting because early dark energy has been proposed as a possible alleviation mechanism for the Hubble tension \cite{Schoneberg:2021qvd, Poulin:2023lkg}. Although it is not completely clear if this tension needs solving ({\it i.e.}~whether it is physical or due to some uncontrolled astrophysical error), the possibility of linking screening of the light dark energy field -- here the dilaton -- to the existence of early dark energy generated by its partner the axion seems worthy of note. We find the amount of EDE produced in this way cannot be very large, being limited for some choices of model parameters by a tachyonic instability in the dilaton sector triggered by the kinetic coupling to the axion, or by possible indirect imprints on the CMB or the growth of structure through the specific matter couplings generating the EDE. 

Of the options we explore it is Dark Energy potentials of the Albrecht-Skordis type that allow the largest contributions of EDE as a fraction of the total energy of the cosmic fluids, of order 5\%. For these kinds of potentials we find that the growth of structure can be {\it reduced} when increasing the coupling to matter of the dilaton. This follows from the fact that the coupling to matter displaces the dilaton field from the vicinity of the minimum of its potential, in the process increasing the Hubble rate during structure formation and therefore the friction term in the growth equation. This is a background effect which counterbalances the natural tendency of scalar-tensor models to increase the growth by the presence of an attractive scalar interaction between CDM particles, as observed in e.g. \cite{vandeBruck:2022xbk, Brax:2023tls}. This also could have interesting phenomenological consequences (though we here leave these for future work). 

We do not here try to match the hints for a time-dependent equation of state $\omega_0$ and its time drift $\omega_a$ described by the DESI collaboration \cite{DESI:2024mwx}, though we do observe that the dilaton-DM interactions these models have can easily appear to give a `phantom' equation of state parameter, $\omega_0 < -1$, if its dynamics are interpreted as being due to vanilla Dark Matter plus a single-field quintessence model.\footnote{The same is true for the more minimal model of \cite{Smith:2024ibv} in which the axion of the axio-dilaton combination is itself regarded as the Dark Matter.} Trying to obtain the parameters $\omega_0$ and $\omega_a$ would require modifications to the dilaton potential without tampering with the axio-dilaton screening mechanism. If such a model could be built, some of the features of the model presented here such as the existence of early dark energy would remain. On the other hand, features such as a decrease in structure growth would depend on the dilaton potential which could for instance be taken of the thawing type instead of the frozen kind. Again, this is left for future work. 

The key constraint on the size of early dark energy fractions shown in \S\ref{five} is not a physical constraint but calculational issue that limits us to only considering small axion-matter coupling potentials, $\cU_m \ll 1$. A consideration of the full form of the axion's effective potential in (\ref{Veffderiv}) that does not rely on this limit would also be of great interest to see if the effects of larger early dark energy fractions on the cosmology.

Finally although the choices made for the functional forms of potentials and couplings are inspired by string constructions,\footnote{The cases considered in sections \ref{Yoga_quadratic_section} and \ref{Yoga_section} are the same as those derived in \cite{Burgess:2021obw} however with different parameter ranges.} we have not attempted to embed our model into a UV completion from first principles, such as from an extra-dimensional or string point of view. Of course this would be a step forward towards a better physical understanding of dark energy and its sensitivity to ultra-violet physics, and from this point of view its natural roots as the low-energy limit of models with supersymmetric large extra dimensions \cite{Brax:2022vlf} is suggestive, given the progress such models allow on understanding the UV side of the cosmological constant problem \cite{Aghababaie:2003wz,Burgess:2015lda} (see \cite{Burgess:2013ara} for a brief review). Much here is also  left for future work. 

\appendix
\addtocontents{toc}{\setcounter{tocdepth}{1}}

\section*{Acknowledgements}

We thank Jeon Han Kim and Maria Mylova for helpful and illuminating discussions and Elsa Teixeira for useful numerical resources. This work evolved out of discussions at the Astroparticle Symposium at the Institut Pascal. CB, CvdB and ACD thank the Institut Pascal for their hospitality during their 2023 programme. AS is supported by the W.D. Collins Scholarship. CvdB is supported by the Lancaster–Sheffield Consortium for Fundamental Physics under STFC grant: ST/X000621/1. ACD is partially supported by the Science and Technology Facilities Council (STFC) through the STFC consolidated grant ST/T000694/1. CB's research was partially supported by funds from the Natural Sciences and Engineering Research Council (NSERC) of Canada. Research at the Perimeter Institute is supported in part by the Government of Canada through NSERC and by the Province of Ontario through MRI.

\section{The phase of the axion field}
\label{thephase}
Let us come back to the axion field and its variation in the presence of perturbations. 
At the background level we have
\be 
\dot S = 0 .
\ee
This implies that at leading order we can write
\be
\bar \mfa(t)= \bar \mfa(\rho) + \frac{\sqrt{2\bar \rho_\mfa}}{ \bar m(t)}\cos(\int dt\, \bar m(t)- S_0),
\ee
up to an irrelevant constant $S_0$. We have emphasized the dependence on the background cosmology by putting a bar on the mass for instance. 

When including linear perturbations and neglecting gravity first, the phase of the axion field becomes
\be 
\int \bar m(t)dt - \int \dot S dt= \int \left(\bar m(t) +\frac{1}{2}\frac{\delta m^2}{\bar m^2}\right)dt= \int m dt,
\ee
where $m^2=\bar m^2 +\delta m^2$ and we have emphasized that we expand around the background axion field determined by its background mass $\bar m$,
i.e. the behaviour of $\dot S$ given by the Hamilton-Jacobi equation transforms the phase of the axion field from the background one to the full phase including the perturbation of the mass. When including gravity and the effect of the axion velocity field in (\ref{HamiltonJacobi}) we have a phase
\be 
\int \bar m(t)dt - \int \dot S dt= \int E_{\rm axion} dt,
\ee
i.e. the phase depends on the energy of the axion particles along the fluid lines where
\be 
E_{\rm axion}= m(t)\left(1+ \frac{v_\mfa^2}{2}+ \Phi\right),
\ee
as well known from the axion behaviour in the galactic halo where velocities follow a Maxwell-Boltzmann distribution.

\section{The Bianchi Identity}\label{the bianchi identity}

The Bianchi identity implies the non-conservation equation
\be 
D^\mu T^{\tot}_{\mu\nu}= - D^\mu (T^\phi_{\mu\nu}+ T^\mfa_{\mu\nu}).
\ee
Explicitly we have
\be 
D^\mu T^\phi_{\mu\nu}= \left(\frac{\beta}{\MPL } T^m + \frac{1}{2} \frac{\partial W^2(\phi)}{\partial \phi} (\partial \mfa)^2\right) \partial_\nu \phi,
\ee
and 
\be 
D^\mu T^\mfa_{\mu\nu}= -\frac{\partial U(\mfa)}{\partial \mfa}\partial_\nu \mfa T^m - \frac{1}{2} \frac{\partial W^2(\phi)}{\partial \phi}(\partial \mfa)^2 \partial_\nu \phi,
\ee
from which we obtain 
\be 
D^\mu T^\tot_{\mu\nu}= -\frac{\beta}{\MPL } T^\tot \partial_\nu \phi +\frac{\partial \cU_\C(\mfa)}{\partial \mfa}\partial_\nu \mfa T^\C +\frac{\partial \cU_\B(\mfa)}{\partial \mfa}\partial_\nu \mfa T^\B.
\ee
Notice the absence of the mixing term $W^2(\phi)$. Indeed, the kinetic mixing leads to the exchange of energy between the axion and the dilaton. When considering the total energy momentun tensor the two scalars, this exchange cancels out.  
We now average over the fast axion oscillations and get
\begin{equation} 
\left\langle \partial_\mu \mfa\frac{\partial \cU_i(\mfa)}{\partial \mfa}\right\rangle=
\frac{(\bar \mfa(\rho) -\aminus )}{\Lambda_{i}^2} \partial_\mu \bar \mfa + \frac{1}{2\Lambda_{i}^2}\partial_\mu \vert \psi\vert^2.
\end{equation}
Using the fluid description for the axion fluctuations we have
\begin{equation} 
\left\langle \partial_\mu \mfa\frac{\partial \cU_i(\mfa)}{\partial \mfa}\right\rangle=
\frac{(\bar \mfa(\bar\rho) -\aminus )}{\Lambda_{i}^2} \partial_\mu \bar \mfa + \frac{1}{2\Lambda_{i}^2}\partial_\mu \left( \frac{\rho_\mfa}{m^2(t)}\right),
\end{equation}
We can split the non-conservation equation into the CDM and the baryonic equations as $D^\mu T^\tot_{\mu\nu} = D^\mu T^\C_{\mu\nu} + D^\mu T^\B_{\mu\nu}$ where
\begin{equation}
    D^\mu T^\C_{\mu\nu}  = -\frac{\beta}{\MPL } T^\C \partial_\nu \phi +\frac{\partial \cU_\C(\mfa)}{\partial \mfa}\partial_\nu \mfa T^\C,
\end{equation}
and
\begin{equation}
    D^\mu T^\B_{\mu\nu}  = -\frac{\beta}{\MPL } T^\B \partial_\nu \phi +\frac{\partial \cU_\B(\mfa)}{\partial \mfa}\partial_\nu \mfa T^\B.
\end{equation}
These equations will give us the non-conservation of matter and the geodesic equations leading to the Euler equation for the baryon and CDM fluids as determined in the main text.

\bibliographystyle{JHEP}
\bibliography{bibliography}

\end{document}